\documentclass[prd,preprint,superscriptaddress,preprintnumbers,eqsecnum,showpacs,nofootinbib,nobibnotes]{revtex4-1}
\usepackage{amsfonts,amsmath,amssymb,bm,natbib}
\usepackage{graphicx} 
\usepackage{pstricks}
\usepackage{slashed}

\newcommand{\be}{\begin{equation}}
\newcommand{\bea}{\begin{eqnarray}}
\newcommand{\ee}{\end{equation}}
\newcommand{\eea}{\end{eqnarray}}

\def\s#1{{\scriptscriptstyle #1}}

\def\1eq#1{Eq.(\ref{#1})}

\def\2eqs#1#2{Eqs.~(\ref{#1}) and~(\ref{#2})}
\def\3eqs#1#2#3{Eqs.~(\ref{#1}),~(\ref{#2}) and~(\ref{#3})}

\def\ie{{\it i.e.}, }
\def\eg{{\it e.g.}, }

\def\n#1{({\it #1}\,)}

\def\qslash{q\hspace{-0.19cm}\slash}

\def\ps#1{p\hspace{-0.18cm}\slash_#1}
\def\tslash{t\hspace{-0.18cm}\slash}

\begin{document}

\title{Non-Abelian Ball-Chiu vertex for arbitrary Euclidean momenta}

\author{A.~C. Aguilar}
\affiliation{University of Campinas - UNICAMP, 
Institute of Physics ``Gleb Wataghin'',
13083-859 Campinas, SP, Brazil}

\author{J. C. Cardona}
\affiliation{University of Campinas - UNICAMP, 
Institute of Physics ``Gleb Wataghin'',
13083-859 Campinas, SP, Brazil}

\author{M. N. Ferreira}
\affiliation{University of Campinas - UNICAMP, 
Institute of Physics ``Gleb Wataghin'',
13083-859 Campinas, SP, Brazil}

\author{J. Papavassiliou}
\affiliation{\mbox{Department of Theoretical Physics and IFIC, 
University of Valencia and CSIC},
E-46100, Valencia, Spain}

\begin{abstract}

We determine the non-Abelian version of the four non-transverse form factors 
of the  quark-gluon vertex, using exact expressions derived from the Slavnov-Taylor identity 
that this vertex satisfies. In addition to the quark and ghost propagators,
a key ingredient of the present approach is 
the quark-ghost scattering kernel, which is computed  
within the one-loop dressed approximation.  
The vertex form factors obtained from this procedure 
are evaluated for arbitrary Euclidean momenta, and display  
features not captured by the well-known Ball-Chiu vertex, deduced from the 
Abelian (ghost-free) Ward identity.
 Particularly interesting in this analysis is the so-called soft gluon limit,
which, unlike other kinematic configurations considered, 
is especially sensitive to the approximations employed
for the vertex entering in the quark-ghost scattering kernel, and
  may even be affected by a subtle numerical instability.
As an elementary application of the results obtained, we evaluate and compare certain 
renormalization-point-independent combinations, which contribute to  
the interaction kernels appearing in the 
standard quark gap and Bethe-Salpeter equations.  In doing so, even though
  all form factors of the quark-gluon vertex, and in particular the transverse ones which are unconstrained by our procedure, enter non-trivially in the aforementioned kernels,
  only the contribution of a single form factor, corresponding to the classical (tree-level)
  tensor, will be considered.

\end{abstract}


\maketitle

\section{\label{sec:intro} Introduction}


Despite the fact that the quark-gluon vertex,  $\Gamma_{\mu}^{a}$, has been the focal point of 
countless theoretical and phenomenological studies that span at least two decades,
a complete understanding of its structure and properties still eludes us. 
Given the central role that this particular vertex plays in  
some of the most important nonperturbative
phenomena of QCD, such as dynamical chiral symmetry breaking,
the generation of constituent 
quark masses~\mbox{\cite{Roberts:1994dr,Maris:2003vk,Fischer:2003rp,Aguilar:2010cn,Cloet:2013jya}},  
and  the formation of  bound states~\cite{Bender:1996bb,Maris:1999nt,Bender:2002as,Bhagwat:2004hn,Holl:2004qn,Chang:2009zb,Williams:2014iea,Williams:2015cvx,Eichmann:2016yit,Sanchis-Alepuz:2015qra}, its systematic scrutiny 
constitutes one of the main challenges of contemporary hadron physics.
In fact, the level of complexity may require the skillful combination of 
ingredients obtained from diverse approaches and frameworks,
such as Schwinger-Dyson equations (SDEs)~\cite{Williams:2015cvx,Bhagwat:2004kj,Bender:1996bb,Sanchis-Alepuz:2015qra,Hopfer:2013np,LlanesEstrada:2004jz,Williams:2014iea,Alkofer:2008tt,Matevosyan:2006bk,Aguilar:2013ac,Rojas:2013tza,Fischer:2006ub}, gauge-technique inspired Ans\"atze~\cite{Salam:1963sa,Salam:1964zk,Delbourgo:1977jc,Delbourgo:1977hq,Ball:1980ay,Curtis:1990zs,Bashir:1997qt,Kizilersu:2009kg,Aguilar:2013ac,Aguilar:2014lha,Heupel:2014ina}, 
functional renormalization group~\cite{Braun:2014ata,Mitter:2014wpa},
and lattice simulations~\cite{Skullerud:2002sk,Skullerud:2002ge,Skullerud:2003qu,Skullerud:2004gp,Lin:2005zd,Kizilersu:2006et,Oliveira:2016muq,Sternbeck:2017ntv}, before a fully satisfactory nonperturbative 
picture could emerge\footnote{In perturbation theory, a complete study has been 
carried out at the one- and two-loop level in arbitrary linear covariant gauges,  dimensions and kinematics in Refs.~\cite{Davydychev:2000rt} and~\cite{Gracey:2014mpa} respectively. In addition, Refs.~\cite{Bermudez:2017bpx, Chetyrkin:2000fd,Chetyrkin:2000dq} contain results at  the one-, two- and three-loop order  for specific gauges and kinematic limits.}.

In the linear covariant ($R_\xi$) gauges,
the full vertex $\Gamma_{\mu}^{a}(q,p_2,-p_1)$, when contracted by the gluon momentum $q^{\mu}$, 
satisfies a non-linear Slavnov-Taylor identity (STI), imposed by the Becchi-Rouet-Stora-Tyutin (BRST) 
symmetry of the theory. This STI is the non-Abelian equivalent of the   QED
Ward-Takahashi identity (WTI), \mbox{$q^{\mu} \Gamma_{\mu}(q,p_2,-p_1) = S_e^{-1}(p_1) - S_e^{-1}(p_2)$},  
which relates the photon-electron vertex with the electron propagator $S_e$. 
The non-Abelian nature of the STI manifests itself through the presence of multiplicative 
contributions originating from the ghost-sector of the theory, and in particular the ``ghost dressing function'', $F(q^2)$,   
and the ``quark-ghost scattering kernel'', $H$, together with its ``conjugate'', $\overline H$.

Exactly as happens with the QED vertex, the Lorentz decomposition of 
$\Gamma_{\mu}^{a}$ consists of twelve linearly independent tensorial structures, 
which are most conveniently expressed in the well-known Ball-Chiu (BC) basis~\cite{Ball:1980ay}; 
the corresponding form factors are functions of three kinematic variables, 
\eg the moduli of $p_1$ and $p_2$, and their relative angle $\theta$. The actual form of the BC basis 
is inspired by the aforementioned STI, being explicitly 
separated into two distinct pieces: 
\n{i} the ``{\it non-transverse part}'', which saturates the STI, 
and is composed of four tensors that 
are not annihilated upon contraction by  $q^{\mu}$, and \n{ii} 
the  purely ``{\it transverse}'' (automatically conserved) part, which is composed of the remaining eight 
elements of the BC basis, all of which vanish identically when contracted by $q^{\mu}$. 

Evidently, the STI imposes stringent constraints on the non-transverse form factors, 
denoted by $L_1$, $L_2$, $L_3$, and $L_4$; in fact, as has been 
demonstrated in detail in~\cite{Aguilar:2010cn}, 
these four quantities are {\it fully determined} in terms of closed formulas involving the components 
of $S$, $F$, $H$, and $\overline H$.  
In the Abelian limit, \ie when the ghost-related contributions are set to their 
tree-level values, these expressions reduce to the known 
``BC vertex'', with the corresponding form factors denoted by $L_1^{\rm{BC}}$, $L_2^{\rm{BC}}$, 
$L_3^{\rm{BC}}$, and $L_4^{\rm{BC}}$ (note that they depend only on the moduli of 
$p_1$ and $p_2$, and that $L_4^{\rm{BC}}$ vanishes identically).  
The BC vertex has been extensively employed in the literature, both in QED, where it 
captures the non-transverse part of the photon-electron vertex exactly,
as well as in QCD, where it is clearly approximate, but serves as a starting point towards a systematic 
improvement over the  rainbow-ladder truncation~\cite{Roberts:1994dr}. 
Instead, the approach put forth in~\cite{Aguilar:2010cn} permits, at least in principle, 
the complete {\it non-Abelian}  conversion of the BC vertex, namely the reconstruction 
of the  part of $\Gamma_{\mu}^{a}$ that satisfies the 
{\it exact} STI, as dictated by the BRST symmetry. 

The practical implementation of this particular approach 
requires the evaluation of $H$ and $\overline H$ by means of their own dynamical equations, 
rather than the more cumbersome treatment of the typical SDE 
that controls the dynamics of the form factors of $\Gamma_{\mu}^{a}$. 
The equations that govern $H$ and  $\overline H$ are also of the SDE-type, but, unlike the vertex SDE, 
their one-loop dressed approximation involves a single Feynman diagram. 
Actually, a considerable simplification stems from the 
fact that the three-gluon vertex, a well-known source of technical complexity, 
does not appear in this particular diagram, and becomes relevant only at the next order 
of the loop expansion. Even so, the dependence of $H$ and $\overline H$ on three kinematic variables 
has been a limiting factor in the numerical treatment presented in~\cite{Aguilar:2010cn},  
where only certain special kinematic configurations, involving a single momentum variable,  were considered (see also~\cite{Rojas:2013tza} for a related study). 

In the present work we compute the general form of $L_1$, $L_2$, $L_3$, and $L_4$ by evaluating the 
one-loop dressed version of the dynamical equations for the components of   
$H$ and  $\overline H$,  for arbitrary Euclidean momenta, in the {\it Landau gauge}.  
These equations contain the following main ingredients: \n{a} the gluon propagator, $\Delta(q^2)$; 
\n{b} the ghost propagator, $D(q^2)$ or, equivalently, its dressing function, $F(q^2)$;  
\n{c} the two standard Dirac components of the quark propagator, $A(q^2)$ and $B(q^2)$, 
introduced in  Eq.~\eqref{qAB};
\n{d} the full ghost-gluon vertex and 
the full quark-gluon vertex $\Gamma_{\mu}^{a}$, both nested inside the one-loop dressed diagram.

Ideally, the above quantities ought to be determined self-consistently 
from their own dynamical equations, which would be solved simultaneously together with the 
equations determining $H$ and  $\overline H$, thus forming an extended system of coupled integral equations. 
However, given the complexity of such an endeavor, in the present work we have opted for a 
simpler procedure. In particular, for the Landau gauge 
$\Delta(q^2)$ and $F(q^2)$ we use directly the results of the large-volume lattice simulations of~\cite{Bogolubsky:2007ud},
whereas two different sets of $A(q^2)$ and $B(q^2)$ are obtained from the 
solution of two standard forms of the quark gap equation in the same gauge.
The main ingredients composing the kernels of these gap equations are again  
the aforementioned lattice results for the $\Delta(q^2)$ and $F(q^2)$, 
with judicious modeling of the $\Gamma_{\mu}^{a}$ entering in them.
As far as the one-loop dressed diagram describing $H$ 
is concerned, we will use for the internal propagators again the same $\Delta(q^2)$ and $F(q^2)$,
for the ghost-gluon vertex its tree-level expression,
while for the $\Gamma_{\mu}^{a}$ we will only keep its component of $L_1^{\rm{BC}}$.

The main results obtained from our analysis may be briefly summarized as follows.

\begin{enumerate}

\item All four form factors are finite within the entire range of Euclidean momenta. 

\item $L_1$ displays a smoother and more enhanced structure compared to $L_1^{\rm{BC}}$.  

\item $L_2$ has a rather intricate structure, whose details  depend strongly  on the particular shape of $A(p^2)$,  but is, in general,  considerably different  from $L_2^{\rm{BC}}$.

\item $L_3$ exhibits practically the same qualitative 
behavior as its BC counterpart, with mild differences in the deep infrared.

\item $L_4$ is non-vanishing but extremely suppressed in the entire range of momenta, with 
its maximum value being only $0.027\,\mbox{GeV}^{-1}$.

\item In general, the dependence of  $L_2$, $L_3$, and $L_4$ on the angle $\theta=0$ is rather mild; $L_1$ is also rather insensitive to changes in $\theta$. However, when  $\theta=0$ and $p_1=p_2$, it develops a more intricate behavior which requires a delicate analysis.  In that sense, the form factors obtained depend mainly on the moduli of 
$p_1$ and $p_2$, exactly as happens with their BC counterparts, even though their corresponding  functional dependences are in general different.

\item  For all values of $\theta$, $L_2$, $L_3$, and $L_4$ suffers only quantitative changes when the Ansatz for the quark-gluon vertex entering in the calculation 
of the quark-ghost scattering kernel is modified; $L_1$, is also quite insensitive to  the Ansatz chosen, except when  $\theta=0$ and $p_1=p_2$, where a particularly strong dependence is observed.

\end{enumerate}

We end this introductory section by 
emphasizing that the method presented here, being a 
variant of the ``gauge-technique''~\cite{Salam:1963sa,Salam:1964zk,Delbourgo:1977jc,Delbourgo:1977hq}, 
leaves the ``transverse'' part of the vertex completely undetermined. The proper inclusion of this 
part in SDE studies is essential because it enforces the multiplicative
renormalizability of the electron and quark gap equations~\mbox{\cite{Bashir:1997qt,Curtis:1990zs,Kizilersu:2009kg,Aguilar:2010cn}}.
Moreover, it affects considerably the amount of dynamical chiral symmetry breaking obtained~\cite{Bashir:2011dp},
and is involved in the dynamics of various emerging nonperturbative phenomena~\cite{Chang:2010hb,Chang:2012cc,Sanchis-Alepuz:2015qra,Williams:2014iea,Mitter:2014wpa, Hopfer:2013np, Williams:2015cvx}.  
Even though the transverse part is only partially determined by the so-called ``transverse'' WTI~\cite{Takahashi:1985yz,Kondo:1996xn,He:2000we,Pennington:2005mw,He:2006my,Qin:2013mta},  
a recent detailed study reveals that the imposition of certain crucial physical requirements severely restricts its allowed form and strength~\cite{Binosi:2016wcx}.


The article is organized as follows. In section~\ref{sec:Formalism} we introduce the notation and  
set up the theoretical framework of this work. 
In section~\ref{sec:scatter} we derive the  equations that govern 
the behavior of the form factors of $H$ for arbitrary momenta,  and discuss certain
  phenomenological subtleties related with the choice of the non-transverse basis for the quark gluon vertex.
Our main results are presented in section~\ref{3d_analysis},
where we first obtain the numerical solution for the various $X_i$ for general values of the Euclidean momenta, and then 
determine the  quark-gluon form factors which satisfies the 
{\it exact} STI for arbitrary momenta. 
A considerable part of our study focuses on the dependence of the $L_i$
on the value of the quark mass, and the shape (presence or absence of minimum) 
of the inverse quark wave functions.
In  section~\ref{special} we take a closer look at the form factors $L_i$ in five special kinematic configurations. We pay particular attention to the case of the soft-gluon limit,
  whose numerical evaluations appears to be particularly delicate, and,
  even though subjected to an extensive number of checks, may still contain a certain amount of
 imprecision. 
In section \ref{RGIQ} we use some of the results derived in the previous section in order 
to construct certain renormalization-group invariant (RGI) combinations that 
serve as natural candidates for describing the effective strength of the quark interaction.
Finally, in section~\ref{concl} we draw our conclusions, and    
in the Appendix~\ref{app:taylor} present the Taylor expansions
needed in the derivation of the special kinematic limits discussed in the  section~\ref{special}.

\section{\label{sec:Formalism} General framework}

In this section we set up the notation and conventions that 
will be employed throughout this article, and review the general theoretical 
framework together with the fundamental equations that will be central to our
subsequent analysis.  
 
\begin{figure}[!ht]
\begin{center}
\includegraphics[scale=0.6]{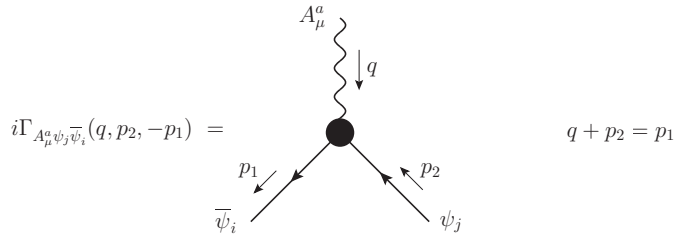}
\end{center}
\caption{The full quark-gluon vertex and the flow of momenta we employ.}
\label{fig_vqg1v}
\end{figure}

 Our starting point is the definition of the quark-gluon vertex, shown in Fig.~\ref{fig_vqg1v}, 
written as 
\begin{align}\label{verc6}
\Gamma^{a}_\mu(q,p_2,-p_1)=gt^a\Gamma_\mu(q,p_2, -p_1)\,,
\end{align} 
where $t^a=\lambda^a/2$ are the $SU(3)$ group generators in the 
fundamental representation, with $\lambda^a$ the Gell-Mann matrices,   
$q$ and $p_2$  are the incoming gluon and quark momenta, $p_1=q+p_2$  is the outgoing  anti-quark momentum.  At tree level, the vertex  reduces to 
$\Gamma_\mu^{[0]}(q,p_2,-p_1) = \gamma_{\mu}$. 
 
\begin{figure}[!t]
\begin{center} 
\includegraphics[scale=0.7]{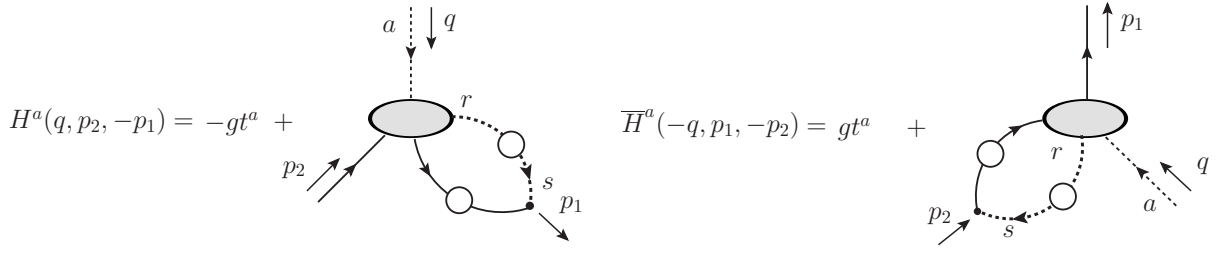}
\caption{\label{H-functions}  Diagrammatic representation of 
the quark-ghost kernels $H$ and $\overline{H}$; 
their  tree-level expressions are $-gt^a$ and $gt^a$, respectively.
The gray oval-shaped 
blob represents the {\it connected part} of the four-point quark-ghost scattering amplitude.}  
\end{center}
\end{figure}

 In addition, $\Gamma_{\mu}$  satisfies  the standard STI given by 
\be
q^\mu\Gamma_{\mu}(q,p_2,-p_1)=F(q^2)\left[S^{-1}(p_1)H(q,p_2,-p_1)-\overline{H}(-q,p_1,-p_2)S^{-1}(p_2)\right],
\label{STI}
\ee
where $F(q^2)$ is the ghost dressing function appearing 
in the definition of the complete ghost propagator
\begin{equation}
D(q^2)=\frac{iF(q^2)}{q^2} \,,
\end{equation}
and  $S^{-1}(p)$ is the inverse of the full quark propagator
expressed as 
\begin{equation}
S^{-1}(p)=A(p^2)\slashed{p}-B(p^2) \,,
\label{qAB}
\end{equation}
where  $A(p^2)$ is the inverse of the quark wave function  and  $B(p^2)$  
is the scalar component (mass function) of the quark propagator, and ${\mathcal M}(p^2) = B(p^2)/A(p^2)$ 
is the dynamically generated quark constituent mass. Finally, 
$H$ denotes the quark-ghost scattering kernel, and $\overline{H}$ its ``conjugate'',  represented diagrammatically in the Fig.~\ref{H-functions}.

 Turning to these last two quantities, 
notice that $\overline{H}$ may be obtained from $H$ through the application of the 
following operations: \n{i} exchange $-p_1$ with $p_2$: $-p_1\leftrightarrow p_2$; \n{ii} reverse the sign of all external momenta: $q,-p_1,p_2\leftrightarrow-q,p_1,-p_2$; \n{iii} take the hermitian conjugate of the resulting amplitude, and use that 
\begin{equation}
\overline{H} := \gamma^{0}H^{\dagger}\gamma^{0} \,.
\label{conj}
\end{equation}
The Lorentz decomposition of $H(q,p_2,-p_1)$ is given by~\cite{Davydychev:2000rt} 
\begin{align}
H &=X_0(q^2,p^2_2,p^2_1) \mathbb{I}+X_1(q^2,p^2_2,p^2_1) \slashed{p}_1+X_2(q^2,p^2_2,p^2_1) \slashed{p}_2 + X_3(q^2,p^2_2,p^2_1) \widetilde{\sigma}_{\mu\nu}p_1^{\mu}p_2^{\nu}\,,  
\label{Hdecomp}
\end{align}
where  $\widetilde{\sigma}_{\mu\nu}=\frac{1}{2}[\gamma_\mu,\gamma_\nu]$ (notice the $i$ difference with respect to the conventional definition of this quantity). 
At tree-level, $X^{(0)}_0=1$ and $X^{(0)}_1=X^{(0)}_2=X^{(0)}_3=0$, while the complete one-loop expressions have been presented in~\cite{Davydychev:2000rt}.

The corresponding decomposition for $\overline{H}$ may
be easily deduced from Eq.~\eqref{Hdecomp} through 
the direct application of the aforementioned operations \n{i}--\n{iii}, using subsequently Eq.~\eqref{conj}. Thus, given that $(\gamma^{\mu})^{\dagger}= \gamma^{0}\gamma^{\mu}\gamma^{0}$ and  $\widetilde{\sigma}_{\mu\nu}^{\dagger} =\gamma^{0}\widetilde{\sigma}_{\nu\mu}\gamma^{0}$, one obtains 
that $\overline{H}(-q,p_1,-p_2)$
\begin{align}
\overline{H} &= {X}_0(q^2,p^2_1,p^2_2) \mathbb{I} + {X}_2(q^2,p^2_1,p^2_2)  \slashed{p}_1+
{X}_1(q^2,p^2_1,p^2_2) \slashed{p}_2 + {X}_3(q^2,p^2_1,p^2_2) \widetilde{\sigma}_{\mu\nu}p_1^{\mu}p_2^{\nu}\,. 
\label{Hdecomp1}
\end{align}
For the sake of notational compactness, in what follows we will employ the definitions 
\begin{align}
X_i := X_i(q^2,p_2^2,p_1^2)\,, \qquad
\overline{X}_i:= X_i(q^2,p_1^2,p_2^2)\,.
\label{sym}
\end{align}

  On the other hand, the tensorial structure of the full quark-gluon vertex, $\Gamma_{\mu}$, consists of  12 independent vectors~\cite{Ball:1980ay}. It is common to divide the vertex into
  a part that ``saturates'' the STI of Eq.~(\ref{STI}), denoted here by $\Gamma_\mu^{\rm{(ST)}}$,
  and a ``transverse part'', denoted by $\Gamma_\mu^{\rm{(T)}}$
which is automatically conserved, 
\be\label{jcc1}
q^{\mu}\Gamma_\mu^{\rm{(T)}}(q,p_2,-p_1)=0\,.
\ee
Thus, 
\be
\Gamma_\mu(q,p_2,-p_1)=\Gamma_\mu^{\rm{(ST)}}(q,p_2,-p_1)+\Gamma_\mu^{\rm{(T)}}(q,p_2,-p_1)\,.
\label{decomp1}
\ee
Evidently, the above decomposition is not unique, given that a ``transverse'' structure  may be
removed from $\Gamma_\mu^{\rm{(T)}}$ and  be reassigned to $\Gamma_\mu^{\rm{(ST)}}$. This
ambiguity introduces a corresponding arbitrariness at the level of the tensorial 
basis used to span $\Gamma_\mu^{\rm{(ST)}}(q,p_2,-p_1)$ and 
$\Gamma_\mu^{\rm{(T)}}(q,p_2,-p_1)$. 
One of the most standard choices for the decomposition of the ST part, is
 the so-called BC basis~\cite{Ball:1980ay}, given by  
\be
\Gamma_{\mu}^{\rm{(ST)}}(q,p_2,-p_1) = \sum^4_{i=1} L_i(q, p_2,-p_1)\lambda_{i,\mu}(p_1,p_2)\,,
\label{sum_long}
\ee 
with 
\bea
\lambda_{1,\mu}&=& \gamma_{\mu} \nonumber\,, \\
\lambda_{2,\mu}&=& (\slashed{p}_1 + \slashed{p}_2)(p_1+p_2)_\mu \nonumber \,, \\
\lambda_{3,\mu}&=&(p_1+p_2)_\mu  \nonumber \,, \\
\lambda_{4,\mu}&=& \widetilde{\sigma}_{\mu\nu}(p_1+p_2)^\nu\,,
\label{tens_long}
\eea
where $L_i(q, p_2, -p_1)$ are the form factors.

For the transverse part, $\Gamma_\mu^{\rm{(T)}}$, one may use   the   basis proposed in Ref.~\cite{Kizilersu:1995iz} 
\be
\Gamma_{\mu}^{\rm{(T)}}(q,p_2,-p_1) = \sum^8_{i=1} T_i(q, p_2,-p_1)\tau_{i,\mu}(p_1,p_2)\,,
\label{sum_long2}
\ee 
where $T_i(q, p_2, -p_1)$ are the form factors and the set of independent tensors $\tau_i$ are given by
\begin{align}
\tau_{1,\mu}&=p_{2\mu}(p_1\cdot q)-p_{1\mu}(p_2\cdot q);&
\tau_{2,\mu}&=\tau_{1\mu}\tslash;\nonumber\\
\tau_{3,\mu}&=q^2\gamma_\mu-q_\mu\qslash;&
\tau_{4,\mu}&= q^2[\gamma_{\mu}\tslash - t_{\mu}]
-2q_{\mu}\widetilde{\sigma}_{\nu\lambda}p_1^\nu p_2^\lambda;\nonumber\\
\tau_{5,\mu}&=\widetilde{\sigma}_{\mu\nu}q^\nu;&
\tau_{6,\mu}&=\gamma_\mu (q\cdot t) -t_\mu\qslash;\nonumber\\
\tau_{7,\mu}&=\frac12(q\cdot t)\lambda_{4,\mu}-t_\mu\widetilde{\sigma}_{\nu\lambda}p_1^\nu p_2^\lambda;&
\tau_{8,\mu}&=\gamma_{\mu}\widetilde{\sigma}_{\nu\lambda}p_1^\nu p_2^\lambda+p_{2\mu}\ps{1}-p_{1\mu}\ps{2} \,,
\label{theTs}
\end{align}
with  $q=p_1-p_2$ and $t=p_1+p_2$. Note that Eq.~\eqref{theTs} not
only explicitly satisfies Eq.~(\ref{jcc1}), but also guarantees that 
$\tau_{i,\mu}(p_1,p_1) =0$.  

It is important to mention that the above decomposition for $\Gamma_{\mu}^{\rm{(T)}}(q,p_2,-p_1)$ 
is slightly different from the one first employed by Ball-Chiu~\cite{Ball:1980ay}. The modification, 
proposed in Ref.~\cite{Kizilersu:1995iz} guarantees that the corresponding form factors are free of kinematic singularities in all covariant gauges~\cite{Kizilersu:1995iz,Bermudez:2017bpx}. In addition, this basis also permits one to establish a more transparent relation between the ST and the transverse parts of the vertex. More specifically, when we contract the tensors  defining the  ST part  with the transverse projector, \mbox{$P_{\mu\nu}(q) = g_{\mu\nu} - \frac{q_{\mu}q_{\nu}}{q^2}$}, we obtain~\cite{Skullerud:2002ge} 
\begin{align}
P_{\mu\nu}(q)\lambda_1^{\nu} &=  \frac{1}{q^2}\tau_{3,\mu} \,; 
&P_{\mu\nu}(q)\lambda_2^{\nu} &=  \frac{2}{q^2}\tau_{2,\mu} \nonumber\,; \\
P_{\mu\nu}(q)\lambda_3^{\nu} &=  \frac{2}{q^2}\tau_{1,\mu}\,;
&P_{\mu\nu}(q)\lambda_4^{\nu} &=  \frac{1}{q^2}\tau_{4,\mu}\,. 
\label{projtra}
\end{align}
which clearly shows that $\Gamma_\mu^{\rm{(ST)}}$  contains transverse contributions,
or else the above contractions would have vanished.

It  is interesting  to  notice that  Eq.~\eqref{projtra}  has a  great
impact on the  applications of the quark-gluon vertex in  the study of
chiral symmetry  breaking and  the formation of bound states.   To see
that,  we  recall that  in  the  dynamical equations  describing  both
phenomena, one  of the relevant  quantities appearing on them is the
contraction  of  the  full  quark-gluon  vertex  with  the  transverse
projector,   namely  $P_{\mu}^{\nu}(q)\Gamma_\nu(q,p_2,-p_1)$   [see  for
  example  Fig.~\ref{fig_rgi}   of  the   section~\ref{RGIQ}].   Using
Eqs.~(\ref{jcc1}), (\ref{decomp1}),   and (\ref{projtra}) we  conclude 
that the aforementioned contraction will produce
\begin{align}
& P_{\mu}^{\nu}(q)\Gamma_\nu(q,p_2,-p_1) = P_{\mu}^{\nu}(q)\Gamma_{\nu}^{\rm{(ST)}}(q,p_2,-p_1) + \Gamma_{\mu}^{\rm{(T)}}(q,p_2,-p_1) \nonumber \\
& \qquad \qquad\qquad = \frac{1}{q^2}[L_1^{\prime}\tau_{3,\mu}(p_1,p_2) +2 L_2^{\prime}\tau_{2,\mu}(p_1,p_2) + 2L_3^{\prime}\tau_{1,\mu}(p_1,p_2) + L_4^{\prime}\tau_{4,\mu}(p_1,p_2) ] \nonumber \\
&\qquad \qquad\qquad + \sum^8_{i=5} T_i\tau_{i,\mu}(p_1,p_2) \,,
\label{cont_gap}
\end{align} 
where the form factors $L_i$ get entangled with the $T_i$, generating the following  modified
(effective) form factors~\cite{Skullerud:2002ge}
\begin{align}
L_1^{\prime} = L_1 + q^2T_3\,; \qquad  \qquad&   L_2^{\prime} = L_2 +\frac{q^2}{2}T_2\,; 
\nonumber \\
L_3^{\prime} = L_3 + \frac{q^2}{2}T_1\,; \qquad  \qquad &   L_4^{\prime} = L_4 +q^2T_4\,.
\label{effL}
\end{align}

It is important to emphasize at this point that, if we had chosen a basis different from that of 
Eqs.~\eqref{tens_long} and~\eqref{theTs}, the above relations would be modified.
In fact, as far as phenomenological applications are concerned (such as those discussed
in section~\ref{RGIQ}), an appropriate choice of basis is necessary in order to
profitably exploit the information encoded  in the STI.

In order to appreciate this point with a concrete example, let us assume that
the basis given in \eqref{tens_long} is modified by changing 
$\lambda_{1,\mu}$ to ${\overline\lambda}_{1,\mu}$
through the addition of a transverse piece, \ie, 
\be
   {\overline\lambda}_{1,\mu} = \gamma_{\mu} + c P_{\mu\nu}(q)  \gamma^{\nu} = 
   \left((1+c)g_{\mu\nu} - c  \frac{q_{\mu}q_{\nu}}{q^2}\right)\gamma^{\nu} ,    
\label{barL}
\ee
where $c$ is an arbitrary real number.  Clearly, in this new basis, the tree level
quark-gluon vertex is decomposed as
\be
\Gamma_\mu^{[0]} = {\overline\lambda}_{1,\mu} - \frac{c}{q^2}\tau_{3,\mu}.
\label{treelincom}
\ee

   Now, given that the difference between $\lambda_{1,\mu}$ and ${\overline\lambda}_{1,\mu}$
   is a purely transverse piece, the change of basis is not felt at the level of the
   STI; therefore, the form factors $L_1$ and ${\overline L}_1$ will be identical,
   $L_1= {\overline L}_1$. On the other hand, the first projection given in Eq.~\eqref{projtra}  
   becomes $c$-dependent, since now 
\be   
 P_{\mu\nu}(q) {\overline\lambda}_{1}^{\nu} =  \frac{1+c}{q^2}\tau_{3,\mu},   
\ee
while the first relation of Eq.~\eqref{effL} becomes
\be
L_1^{\prime} = (1+c)L_1 + q^2T_3 , 
\label{newLprime}
\ee
revealing that, indeed, the answer of the transversely projected vertex depends on
the details of the basis chosen for the $\Gamma_\mu^{\rm{(ST)}}$ part.
In fact, for the special value $c=-1$, for which   
\mbox{${\overline\lambda}_{1,\mu} = q_{\mu}\slashed{q}/{q^2}$}, 
all information furnished by the STI (namely the form of $L_1$), is completely
washed out from the corresponding amplitude.
The above argument may be easily generalized to all remaining 
elements of the basis that spans $\Gamma_\mu^{\rm{(ST)}}$. 

There are two main conclusions that may be
drawn from the discussion presented above. First, a necessary
condition for exploiting the STI in phenomenological applications
is that the basis used for $\Gamma_\mu^{\rm{(ST)}}$ should {\it not} be completely annihilated 
when contracted by the transverse projector. Second, 
the amount of $L_i$ that enters into the
amplitude (in other words, the value of ``$c$'' in the case of $L_1$) {\it depends on the basis} chosen for $\Gamma_\mu^{\rm{(ST)}}$;
we will consider this issue again in section~\ref{RGIQ}.

Returning to Eq.~\eqref{sum_long}, it is clear that 
the  form factors $L_i$ can be related through Eq.~(\ref{STI}) with $A$, $B$, $F$, $X_i$, and $\overline{X}_i$.
Specifically, as was demonstrated in~\cite{Aguilar:2010cn}, the  $L_i$ may be expressed as
\bea
L_1 &=& \frac{F(q)}{2} \left\{
A(p_1)[X_0 - (p_1^2+p_1\!\cdot\!p_2)X_3] 
+ A(p_2)[{\overline X}_0 -(p_2^2 + p_1\!\cdot\!p_2){\overline X}_3]\right\} 
\nonumber\\
&+&
\frac{F(q)}{2} \left\{ B(p_1)(X_2-X_1) + B(p_2)({\overline X}_2-{\overline X}_1)\right\};
\nonumber\\
L_2 &=& \frac{F(q)}{2(p_1^2 - p_2^2)} \left\{
A(p_1)[X_0 + (p_1^2 - p_1\!\cdot\!p_2)X_3] 
- A(p_2)[{\overline X}_0 +(p_2^2-p_1\!\cdot\!p_2){\overline X}_3]\right\}
\nonumber\\
&-&
\frac{F(q)}{2(p_1^2 - p_2^2)} \left\{ B(p_1)(X_1+X_2) - B(p_2)({\overline X}_1+{\overline X}_2)\right\};
\nonumber\\
L_3 &=&  \frac{F(q)}{p_1^2 - p_2^2}
\left\{  
A(p_1) \left( p_1^2 X_1 + p_1\!\cdot\!p_2 X_2 \right)
- A(p_2) \left( p_2^2 {\overline X}_1 +p_1\!\cdot\!p_2 {\overline X}_2\right)
- B(p_1)X_0 + B(p_2){\overline X}_0\right\};
\nonumber\\
L_4 &=&\frac{F(q)}{2} \left\{ 
A(p_1) X_2 - A(p_2) {\overline X}_2 - B(p_1) X_3 + B(p_2){\overline X}_3 
\right\}.
\label{expLi}
\eea

The derivation of the above equation has been carried out in Minkowski space; 
its Euclidean version may be obtained through direct application of the rules given  
in the subsection~\ref{sec:eucl}.

Setting  in Eq.~\eqref{expLi} $X_0 = {\overline X}_0=1$ and $X_i = {\overline X}_i=0$, 
for $i \geq 1$, and  $F(q) =1$, we obtain  
the  following expressions  (still in Minkowski space)
\bea
L_1^{\rm{BC}} &=& \frac{A(p_1)+A(p_2)}{2}\,,\qquad L_2^{\rm{BC}} = \frac{A(p_1)- A(p_2)} {2(p_1^2 - p_2^2)} \,,  \nonumber \\ 
L_3^{\rm{BC}} &=& \frac{B(p_2)- B(p_1)}{p_1^2 - p_2^2} \,, \qquad L_4^{\rm{BC}} = 0 \,.
\label{BC_vertex}
\eea 
which is precisely the well-known BC vertex~\cite{Ball:1980ay}.

\section{\label{sec:scatter} Quark-ghost kernel at the one-loop dressed level}

\begin{figure}[!ht]
\begin{center}
\includegraphics[scale=0.7]{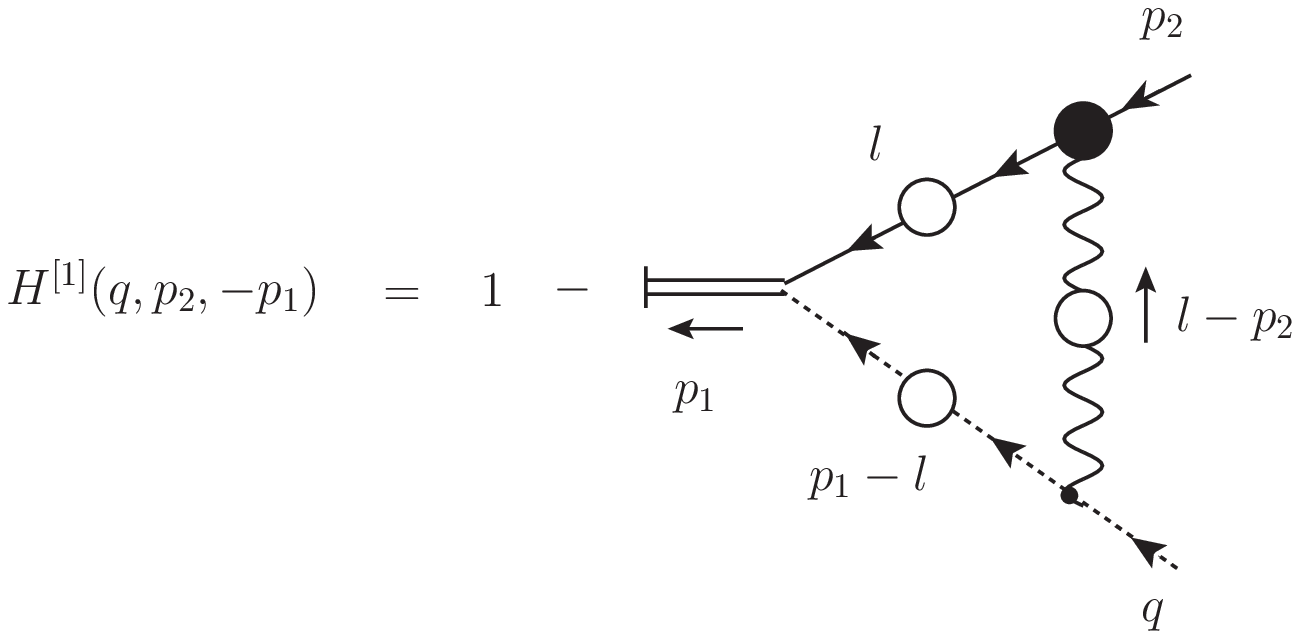}
\end{center}
\caption{\label{honeloopdressed} 
The scattering kernel $H^{[1]}(q,p_2,-p_1)$
at  one-loop dressed approximation.}
\end{figure}

In this section we derive the expressions for the form factors $X_i$  within the 
one-loop dressed approximation. In particular, the four-point quark-ghost scattering 
amplitude, entering in the diagrammatic definition of  
$H^a = -gt^aH$ in Fig.~\ref{H-functions}, is approximated by its lowest order diagram, the one gluon 
exchange term, which is subsequently ``dressed'' as shown in Fig.~\ref{honeloopdressed}.

Using the flow of momenta defined in Fig.~\ref{honeloopdressed} and factoring out
its color structure ($-gt^a$), the expression for $H^{[1]}(q,p_2,-p_1)$ is given by 
\bea
H^{[1]}&=&1- \frac{1}{2}\,i\,C_Ag^2\int_l\Delta^{\mu\nu}(l-p_2)G_{\nu}(p_1-l)D(l-p_1)S(l)\Gamma_\mu(l-p_2,p_2,-l) \,,
\label{Hexp}
\eea
where we have already used the three level expression for the
the quark-ghost kernel, \mbox{$H^{[0]\,a} = -gt^a$}, as indicated in the Fig.~\ref{H-functions}. In addition, $C_A$ is the eigenvalue of the Casimir operator in the adjoint representation and we have defined the
integration measure
\be
\int_{l}\equiv\!\int\!\frac{\mathrm{d}^4 l}{(2\pi)^{4}}; 
\label{dqd}
\ee
it is understood that a symmetry preserving regularization scheme must be employed
(see also subsection~\ref{closer}).
   In addition, $\Delta^{\mu\nu}(k)$ is  the full gluon propagator,
which in the  Landau gauge has the form 
\begin{align}
\Delta_{\mu\nu}(k)=-iP_{\mu\nu}(k)\Delta(k^2),\quad& \qquad P_{\mu\nu}(k)= g_{\mu\nu}-\frac{k_\mu k_\nu}{k^2}.
\label{propagators}
\end{align}

In order to evaluate Eq.~(\ref{Hexp}) further, we will use for the full gluon-ghost vertex its tree-level value\footnote{Evidently,  a more detailed analysis along the lines of the Ref.~\cite{Rojas:2013tza} should be eventually performed, in order to establish the numerical impact of this approximation.}, {\it i.e.}, $G^{abc}_{\nu} = -gf^{abc}(p_1-l)_{\nu}$.

The question of how to approximate the $\Gamma_{\mu}$ that enters in $H^{[1]}$ turns out to 
  be particularly subtle. Evidently, if one were to consider only the non-transverse
  part of this vertex (as we do throughout this work), the most complete treatment would entail
  to replace the $\Gamma_{\mu}$ by the $\Gamma_{\mu}^{\rm{(ST)}}$ of Eq.~(\ref{sum_long}),
  using the $L_i$ given in Eq.~(\ref{expLi}). This choice, however,
  would convert the problem into a system of coupled integral equations
  for the $L_i$, whose solution, unfortunately, lies beyond our present powers.
  Instead, we will reduce the level of technical complexity by 
  employing a simpler expression for $\Gamma_{\mu}$,  
  retaining only the component proportional to $\gamma_\mu$, and approximating its
  form factor $L_1$ using tree-level values for the $X_i$ entering in it.
  With these simplifications, one has  
  \bea
\Gamma_{\mu}(l-p_2,p_2,-l) &=&\frac{F(l-p_2)}{2}\left[A(l)+A(p_2)\right]\gamma_\mu\,. 
\label{vertex2}
\eea
However, as we will discuss in section~\ref{long_vertex}, the 
use of this particular expression 
leads to unnatural results for the form factor
corresponding to the soft gluon kinematics: essentially, the
curve reverses sign, and deviates dramatically from the
expected perturbative behavior in the ultraviolet.
Interestingly enough, the expected ultraviolet behavior is restored if instead of
(\ref{vertex2}) one uses 
\bea
\Gamma_{\mu}(l-p_2,p_2,-l) &=&\frac{1}{2}\left[A(l)+A(p_2)\right]\gamma_\mu \,,
\label{vertex1}
\eea
which is simply $L_1^{\rm{BC}}(l-p_2,p_2,-l)$.

It seems therefore that, depending on the kinematic circumstances, 
the presence of the ghost dressing function $F(l-p_2)$
in this particular part of the calculation 
destabilizes the truncation procedure. In what follows we will present
the results obtained using Eq.~\eqref{vertex1} for all kinematic configurations
other than that of the soft gluon limit; we have checked explicitly that,
for all these cases, the use of Eq.~\eqref{vertex2}
does not affect the answers appreciably, and, in that sense, our results are rather stable.
Instead, for the special case of the soft gluon configuration, 
we will show
the results obtained with both Eqs.~\eqref{vertex2} and \eqref{vertex1} 
(see the panels of  Fig.~\ref{fig:vertexL1angle0} and Fig.~\ref{fig:L1pert}), in order 
to fully appreciate the difference between the two.

Then, we proceed  inserting into Eq.~(\ref{Hexp}) the propagators of Eqs.~\eqref{qAB} and~(\ref{propagators}) together with the  Ansatz 
given by Eq.~(\ref{vertex1}), it is straightforward to derive the following expression 
for $H$ 
\begin{align}
H(q,p_2,-p_1)&=1+\frac{i}{4}C_\mathrm{A}g^2\int_l\mathcal{K}(p_1,p_2,l)f(p_2,q,l)\,,
\label{Hfinal}
\end{align}
where we have introduced  the kernel
\begin{align}
\mathcal{K}(p_1,p_2,l)&=\frac{F(l-p_1)\Delta(l-p_2)[A(l)+A(p_2)]}{(l-p_1)^2[A^2(l)l^2-B^2(l)]}\,,
\label{kernelsH}
\end{align}
with
\begin{align}
f(p_2,q,l)&=A(l)\left[\slashed{l}\slashed{q}- q\cdot(l-p_2)\left(1+ \frac{(\slashed{p}_2\slashed{l}-p_2^2)}{(l-p_2)^2}\right)\right] \nonumber \\
&\hspace{2.5cm}+B(l)\left[\slashed{q}-(\slashed{l}-\slashed{p}_2)\frac{q\cdot (l-p_2)}{(l-p_2)^2}\right] \,.
\end{align}

Notice that if we  had used the Ansatz given by Eq.~(\ref{vertex2}) instead of Eq.~(\ref{vertex1}), the unique difference in the derivation was that the kernel of Eq.~\eqref{kernelsH} would be replaced by  \mbox{$\mathcal{K}(p_1,p_2,l) \to F(l-p_2)\mathcal{K}(p_1,p_2,l)$}.

The next step is to project out of Eq.~\eqref{Hfinal} the individual form factors $X_i$. This is easily accomplished by means of the following 
formulas~\cite{Aguilar:2010cn}    
\begin{align}
X_0 &=\frac{\text{Tr}\lbrace H\rbrace}{4}\,,\nonumber\\
X_1 &=\frac{p_2^2\text{Tr}\lbrace \slashed{p}_1H\rbrace-p_1\cdot p_2 \text{Tr}\lbrace \slashed{p}_2H\rbrace}{4h}\,,\nonumber\\
X_2 &=\frac{p_1^2\text{Tr}\lbrace \slashed{p}_2H\rbrace-p_1\cdot p_2 \text{Tr}\lbrace \slashed{p}_1H\rbrace}{4h}\,,\nonumber\\
X_3 &=-\frac{\text{Tr}\lbrace\tilde{\sigma}_{\alpha\beta}p_1^\alpha p_2^\beta H\rbrace}{4h}\,,
\label{projector}
\end{align}
where we have introduced the function
\be
h = p_1^2p_2^2-(p_1\cdot p_2)^2\,, 
\label{hfunction}
\ee
and the arguments of $X_i$ have been suppressed as before.  

After substituting the Eq.~\eqref{Hfinal} 
into the projectors given by Eqs.~\eqref{projector} 
and taking the appropriate traces, 
we obtain 
\begin{align}
\label{generalx}
X_0 &=1+\frac{i}{4}C_\mathrm{A}g^2\int_l\mathcal{K}(p_1,p_2,l)A(l){\mathcal G}(p_2,q,l)\,,\nonumber\\
X_1 &=\frac{i}{4}C_\mathrm{A}g^2\int_l\frac{\mathcal{K}(p_1,p_2,l)B(l)}{h(p_1,p_2)}\left[p_2^2{\mathcal G}(p_1,q,l)-(p_1\cdot p_2){\mathcal G}(p_2,q,l)\right]\,,\nonumber\\
X_2 &=\frac{i}{4}C_\mathrm{A}g^2\int_l\frac{\mathcal{K}(p_1,p_2,l)B(l)}{h(p_1,p_2)}\left[p_1^2{\mathcal G}(p_2,q,l)-(p_1\cdot p_2){\mathcal G}(p_1,q,l)\right] \,,\nonumber\\
X_3 &=-\frac{i}{4}C_\mathrm{A}g^2\int_l\frac{\mathcal{K}(p_1,p_2,l)A(l)}{h(p_1,p_2)}\left[ p_2^2{\mathcal G}(p_1,q,l) - (p_1\cdot p_2){\mathcal G}(p_2,q,l)  - {\mathcal T}(p_1,p_2,l)  \right]\,,
\end{align}
where we have introduced the functions 
\bea
{\mathcal G}(k,q,l) &=& (k\cdot q)-\frac{[k\cdot(l -p_2)][q\cdot(l-p_2)]}{(l-p_2)^2} \,, \nonumber \\
{\mathcal T}(p_1,p_2,l) &=& (p_2\cdot q)[(p_1\cdot l) - (p_1\cdot p_2)] - (p_1\cdot q)[(p_2\cdot l)-p_2^2] \,.
\label{fgg}
\eea


\subsection{\label{sec:eucl}Passing to the Euclidean space}

Next, we will convert the Eq.~\eqref{generalx} from Minkowski  to Euclidean space. To do that we will employ the following transformation rules, which are valid for two arbitrary momenta $p$ and $q$
\begin{align}
 \qquad\qquad (\slashed{p},\slashed{q}) \to i(\slashed{p}_{\s E}, \slashed{q}_{\s E});
\qquad\qquad &(p^2, q^2, p\cdot q) \to - (p_{\s E}^2, q_{\s E}^2, \,p_{\s E}\cdot q_{\s E})\,.
\label{eucl_rules}
\end{align}

In addition, the measure defined in Eq.~\eqref{dqd} becomes
\begin{align}
\qquad d^4l \to id^4l_{\s E}, \qquad\qquad \int_l \to i\int_{l_{E}} \,,
\end{align}
where we have introduced the Euclidean measure in spherical coordinates, 
\begin{align}
\int_{l_{E}}=\frac{1}{(2\pi)^4}\int d^4l =\frac{1}{32\pi^4}\int_0^\infty dl_{\s E}^2 l_{\s E}^2\int_0^\pi d\varphi_1 \sin^2\varphi_1 \int_0^\pi d{\varphi_2}\sin{\varphi_2}\int_0^{2\pi} d{\varphi_3}\,.
\end{align}

Applying the above rules to the  scalar functions appearing in the definition of the various propagators, lead us  to the following relations
\begin{align}
&A_{\s E}(p_{\s E}^2)=A(-p^2);\qquad \qquad\qquad\qquad B_{\s E}(p_{\s E}^2)=B(-p^2); \nonumber \\
&\Delta_{\s E}(q_{\s E}^2)=-\Delta(-q^2);\qquad \qquad\qquad\qquad  F_{\s E}(q_{\s E}^2)=F(-q^2)\,.
\label{scalar}
\end{align}

Then, it is straightforward to see that, after applying the conversion rules defined  in the  Eqs.~\eqref{eucl_rules} and \eqref{scalar}, the quantities appearing
in Eq.~\eqref{generalx} transform in the following way  
\begin{align}  
&{\mathcal K}(p_1^2,p_2^2,l^2) \to-{\mathcal K}_{\s E}(-p_{1 \s E}^2,-p_{2\s E}^2,-l_{\s E}^2) \,,\nonumber\\ 
&{\mathcal G}(k^2,q^2,l^2) \to - {\mathcal G}_{\s E}(-k_{\s E}^2,-q_{\s E}^2,-l_{\s E}^2) \,,\nonumber\\ 
&{\mathcal T}(p_1^2,p_2^2,l^2) \to {\mathcal T}_{\s E}(-p_{1\s E}^2,-p_{2\s E}^2,-l_{\s E}^2) \,.
\end{align}
In order to avoid notational clutter, from now on we will suppress the  subscript $E$.

Then, we can easily see that, in a general kinematic configuration, the various form factors $X_i$ and $L_i$  are expressed in terms of the Euclidean  scalar products $(p_1\cdot p_2)$, $(p_1\cdot l)$, and $(p_2\cdot l)$. Without loss of generality, a convenient choice for Euclidean four momenta $p_1$ and $p_2$ is
\begin{align}
p_1^\mu=|p_1|\begin{pmatrix}
\cos{\theta}\\\sin{\theta}\\0\\0
\end{pmatrix}\,,  \qquad
p_2^\mu=|p_2|\begin{pmatrix}
1\\0\\0\\0
\end{pmatrix}\,, 
\end{align}
where  $|p_1|$ and $|p_2|$ are the magnitudes of the Euclidean momenta and 
$\theta$ is the angle between them. Notice that the above choices
guarantee that \mbox{$(p_1\cdot p_1)=p_1^2$} and \mbox{$(p_2\cdot p_2)=p_2^2$}.

Similarly,  the integration momentum  $l$ can be written as
\begin{align}
l^\mu=|l|\begin{pmatrix}
\cos{\varphi_1}\\\sin{\varphi_1}\cos{\varphi_2}\\\sin{\varphi_1}\sin{\varphi_2}\cos{\varphi_3}\\\sin{\varphi_1}\sin{\varphi_2}\sin{\varphi_3}
\end{pmatrix} \,.
\end{align}

With the above definitions,  it is evident that the Euclidean scalar products $(p_1\cdot p_2)$, $(p_1\cdot l)$, and $(p_2\cdot l)$, appearing in the Euclidean version of the Eq.~\eqref{generalx}, 
do not display any dependence  on the angle $\varphi_3$, so that the integral over this angle becomes trivial, and  the measure may be cast in the form
\begin{align}
\int_{l_E}=\frac{1}{(2\pi)^4}\int d^4l =\frac{1}{16\pi^3}\int_0^\infty dl^2 l^2\int_0^\pi d\varphi_1 \sin^2\varphi_1 \int_0^\pi d{\varphi_2}\sin{\varphi_2}\,.
\end{align}

Finally, the  Euclidean form of Eq.~\eqref{generalx} becomes
\begin{align}
X_0(p_1,p_2,\theta)&=1+\frac{C_\mathrm{A}g^2}{4}\int_{l_E}\frac{\mathcal{K}(p_1,p_2,l)A(l^2)}{s^2}\left\lbrace p_2^2l^2\sin^2\varphi_1-s^2p_1p_2\cos\theta\right. \nonumber \\ 
&\left.+[p_1l(\cos\theta\cos\varphi_1+\sin\theta\sin\varphi_1\cos\varphi_2)-p_1p_2\cos\theta](p_2 l\cos\varphi_1-p_2^2)\right\rbrace \,, \nonumber
\end{align}
\begin{align}
X_1(p_1,p_2,\theta)&=\frac{C_\mathrm{A}g^2}{4}\int_{l_E}\frac{\mathcal{K}(p_1,p_2,l)B(l^2)}{s^2}\bigg[s^2-l^2\sin^2\varphi_1\cos^2\varphi_2  \bigg.\nonumber\\
&\bigg.+l(l\cos\varphi_1-p_2)\left(\frac{p_2}{p_1}-\cos\theta\right)\frac{\sin\varphi_1\cos\varphi_2}{\sin\theta}\bigg] \,,\nonumber
\end{align}
\begin{align}
X_2(p_1,p_2,\theta)&=\frac{C_\mathrm{A}g^2}{4}\int_{l_E}\frac{\mathcal{K}(p_1,p_2,l)B(l^2)}{s^2}\left[(l\cos\varphi_1-p_2)^2\left(1-\frac{p_1}{p_2}\cos\theta\right) -s^2 \right.\nonumber\\
&+\frac{p_1l^2}{p_2}\cos\theta\sin^2\varphi_1\cos^2\varphi_2-\frac{p_1l}{p_2}\left(l\cos\varphi_1-p_2\right)\sin\theta\sin\varphi_1\cos\varphi_2\nonumber\\
&\left.-l(l\cos\varphi_1-p_2)\cos\theta\left(1-\frac{p_1}{p_2}\cos\theta\right)\frac{\sin\varphi_1\cos\varphi_2}{\sin\theta}\right]\,, \nonumber
\end{align}
\begin{align}
X_3(p_1,p_2,\theta)&=\frac{C_\mathrm{A}g^2}{4}\int_{l_E}\frac{\mathcal{K}(p_1,p_2,l)A(l^2)}{s^2}\left[-s^2\frac{l}{p_1}\frac{\sin\varphi_1\cos\varphi_2}{\sin\theta}\right.\nonumber\\
&+s^2\frac{l}{p_2}\left(\frac{\cos\theta\sin\varphi_1\cos\varphi_2}{\sin\theta}-\cos\varphi_1\right)\nonumber\\
&\left.+l^2\sin^2\varphi_1\cos^2\varphi_2+l(l\cos\varphi_1-p_2)\left(\cos\theta-\frac{p_2}{p_1}\right)\frac{\sin\varphi_1\cos\varphi_2}{\sin\theta}\right] \,,
\label{euclidean_xi}
\end{align}
where, 
in order to simplify the notation, we have defined $p_1 \equiv |p_1|$,  $p_2 \equiv |p_2|$, $l \equiv |l|$, and the variable $s^2=l^2 + p_2^2 -2lp_2\cos{\varphi}_1$.

Note that the functional dependence of the functions $X_i$, in Euclidean space, will be expressed
in terms of the moduli of the momenta $p_1$, $p_2$, and  their relative angle $\theta$, as denoted by Eq.~\eqref{euclidean_xi}.

\section{\label{3d_analysis}Results for general momenta: 3-D plots}

In this section we will  determine numerically the form factors 
$X_i$ given by Eq.~\eqref{euclidean_xi}
for general values of the Euclidean momenta.
All  results  will be presented in the form of 3-D plots\footnote{ Throughout this work, all 3-D numerical data will be generically written as an array ${\mathcal G}(p_1,p_2,\theta)$ with dimensions (96,96,7), {\it i.e.} we compute the function ${\mathcal G}$ for 96 different values for each momentum $p_1$ and $p_2$ and 7  distinct values for the angle $\theta$, Then,  all 3-D plots were  produced using  the Renka-Cline interpolation on the grid.}, where $p_1$ and $p_2$ will be varied, for fixed values of the angle $\theta$. The culmination of this analysis is presented 
at the final step, where the  the numerical solution obtained for the various $X_i$
are fed into the Euclidean version of  Eq.~\eqref{expLi}, giving rise to all  quark-gluon  form factors $L_i$ for arbitrary momenta.

\subsection{\label{ingred} Inputs for the numerical analysis}

The first step in this analysis is to consider the ingredients entering into the evaluation of the form factors $X_i$  and the corresponding $L_i$. The computation of the 
$X_i$ and $L_i$, in a general kinematic configuration, not only require the knowledge of the nonperturbative behavior of the gluon, $\Delta(q^2)$, and ghost, $D(q^2)$, propagators (or equivalently the ghost dressing function, $F$), but the functions $A(k^2)$ and $B(k^2)$, appearing in the decomposition of the full quark propagator of Eq.~\eqref{qAB},  see for example  Eqs.~\eqref{euclidean_xi} and \eqref{expLi}.

\begin{figure}[!ht]
\hspace{-1.0cm}
\begin{minipage}[b]{0.45\linewidth}
\centering
\includegraphics[scale=0.27]{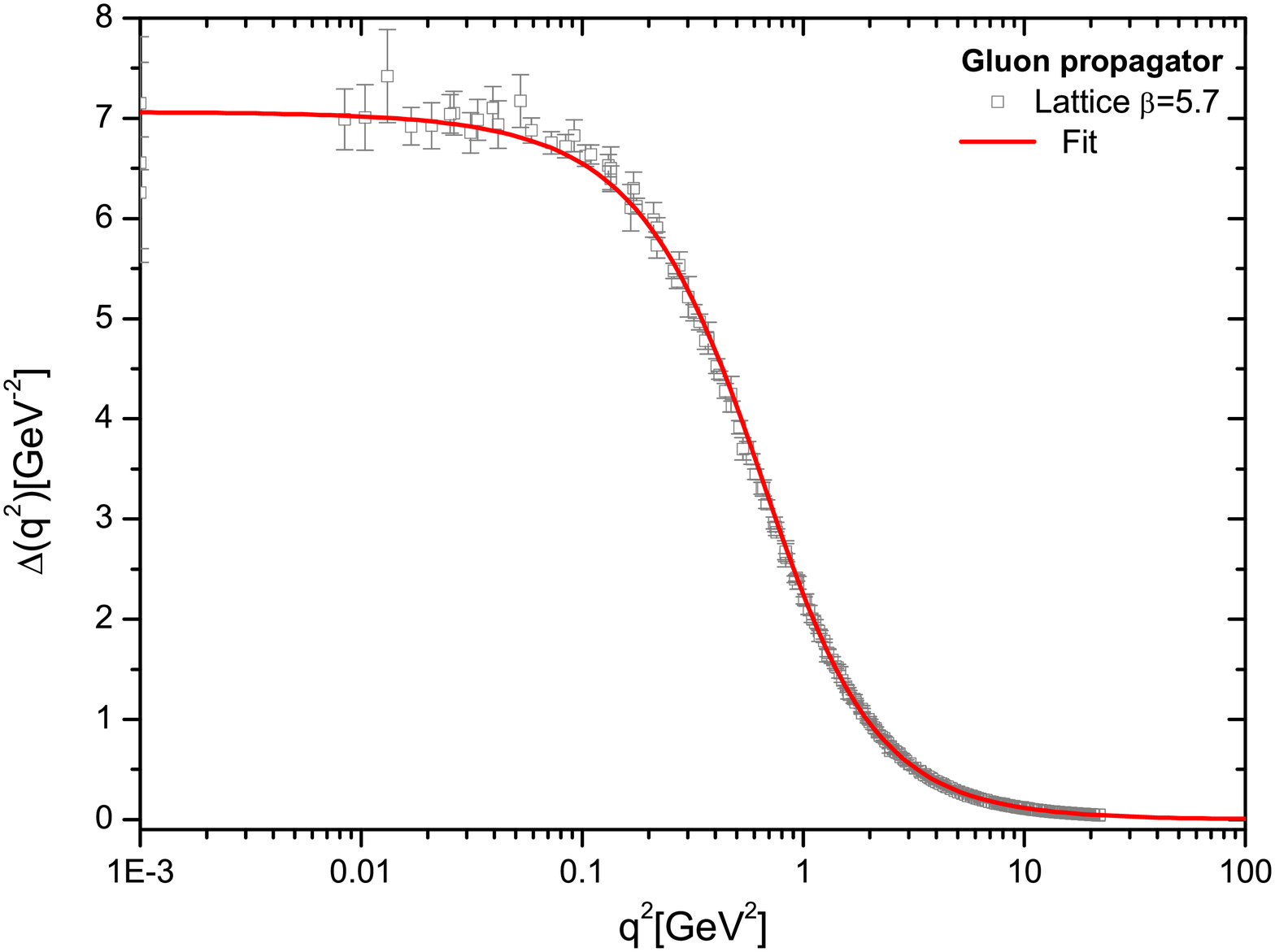}
\end{minipage}
\hspace{0.5cm}
\begin{minipage}[b]{0.50\linewidth}
\includegraphics[scale=0.26]{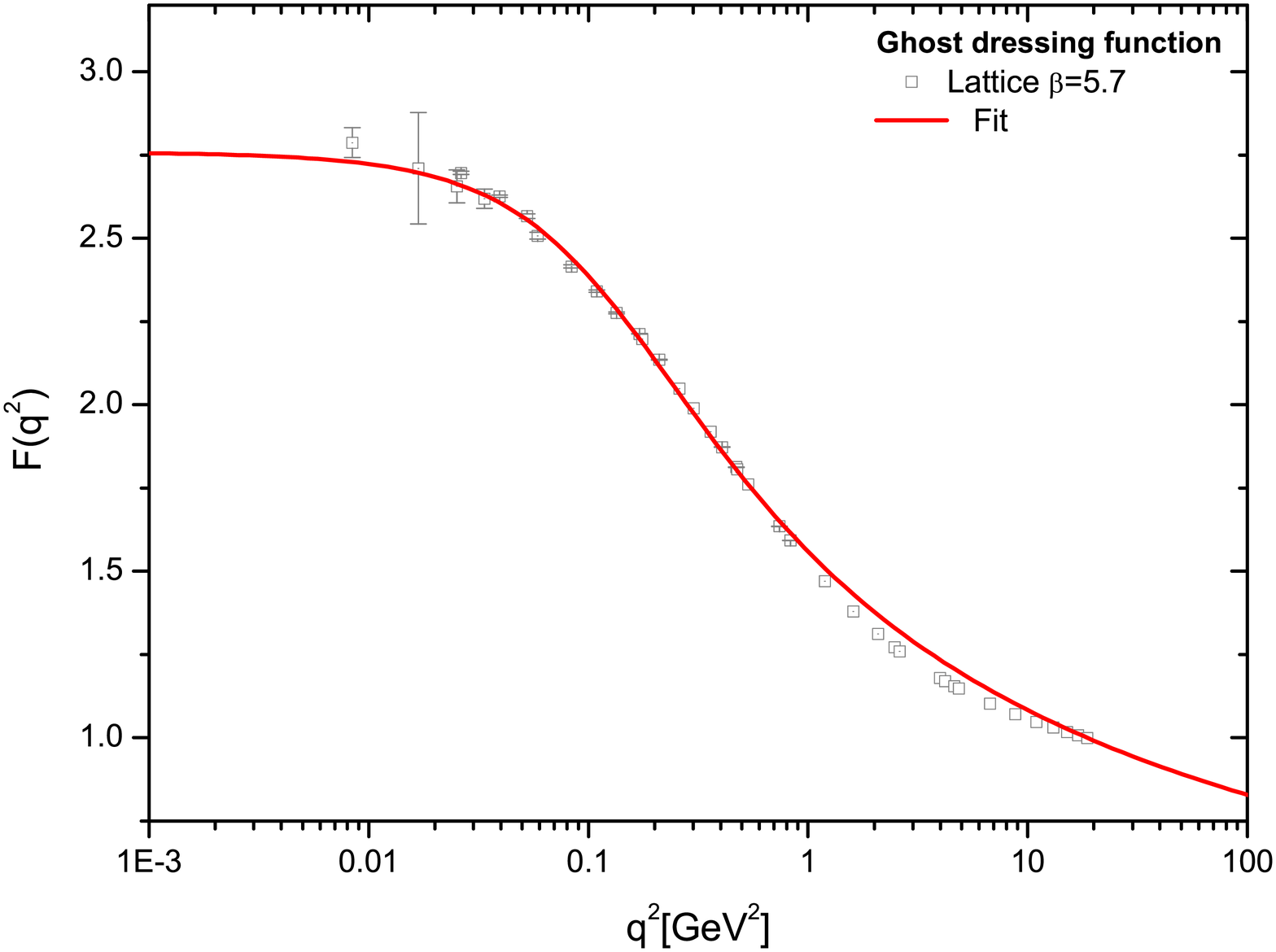}
\end{minipage}
\caption{\label{gluon:plot} The gluon propagator $\Delta(q^2)$ (left panel) and the ghost dressing function $F(q^2)$ (right panel), both renormalized at \mbox{$\mu=4.3$\,GeV}. The lattice data is from Ref.~\cite{Bogolubsky:2007ud}.}
\end{figure}

 Employing the same methodology of previous works~\cite{Aguilar:2013xqa, Aguilar:2012rz,Aguilar:2011ux}, we use  for $\Delta$ and $F$  a fit for the $SU(3)$ lattice data of the Ref.~\cite{Bogolubsky:2007ud}. In Fig.~\ref{gluon:plot} we show the lattice data for $\Delta(q^2)$ and $F(q^2)$ and their corresponding fits, renormalized at \mbox{$\mu=4.3$ GeV}. The explicit functional dependence of 
$\Delta(q^2)$ and $F(q^2)$ may be found 
in a series of  recent articles~\cite{Aguilar:2010cn,Aguilar:2012rz,Aguilar:2011ux}, and their main characteristic is the saturation in 
the deep infrared, associated with the generation of a dynamical 
gluon mass~\cite{Cornwall:1981zr,Aguilar:2008xm}.

\begin{figure}[!t]
\hspace{-1.5cm}
\begin{minipage}[b]{0.45\linewidth}
\centering
\includegraphics[scale=0.33]{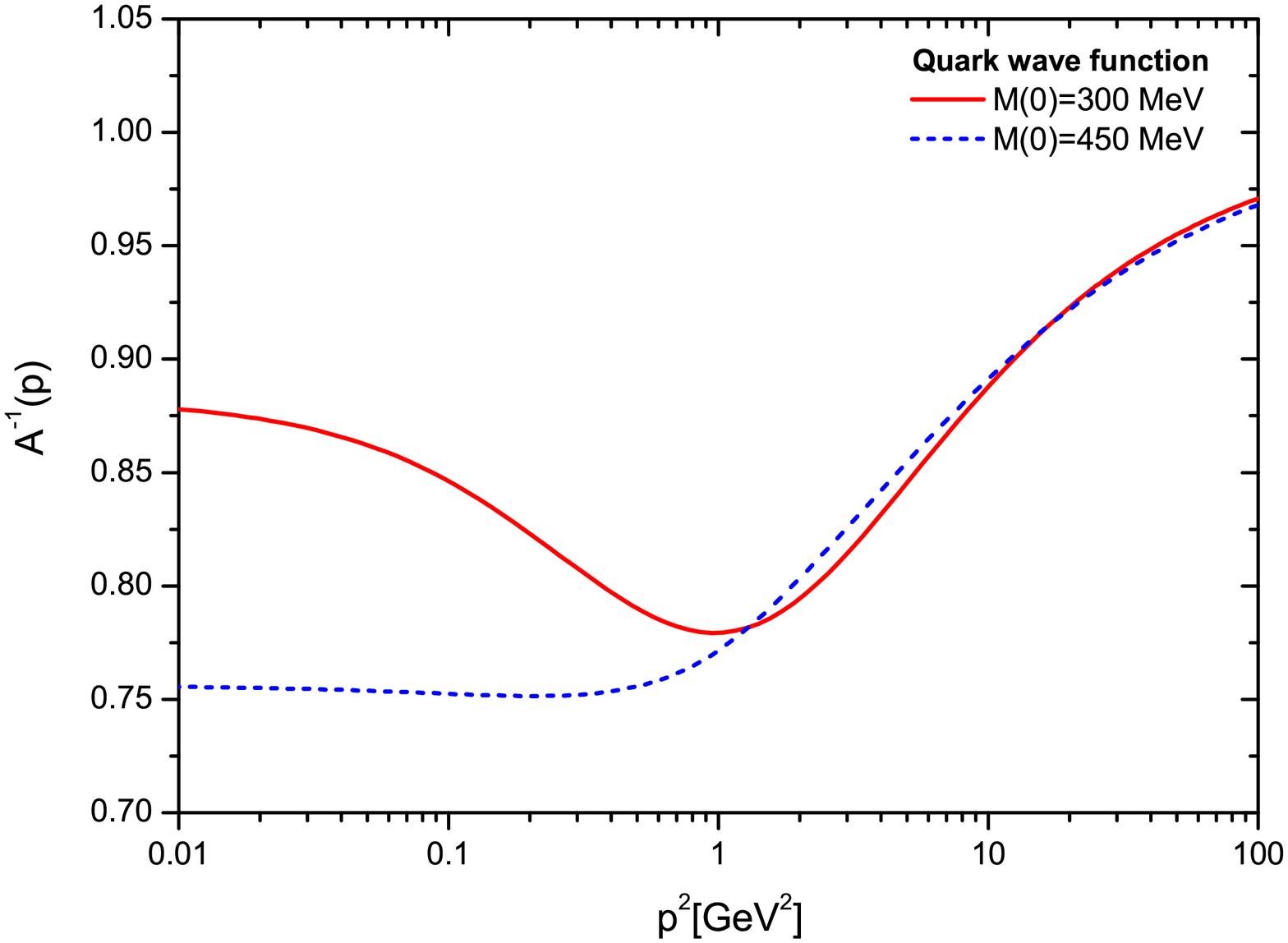}
\end{minipage}
\hspace{0.5cm}
\begin{minipage}[b]{0.50\linewidth}
\includegraphics[scale=0.33]{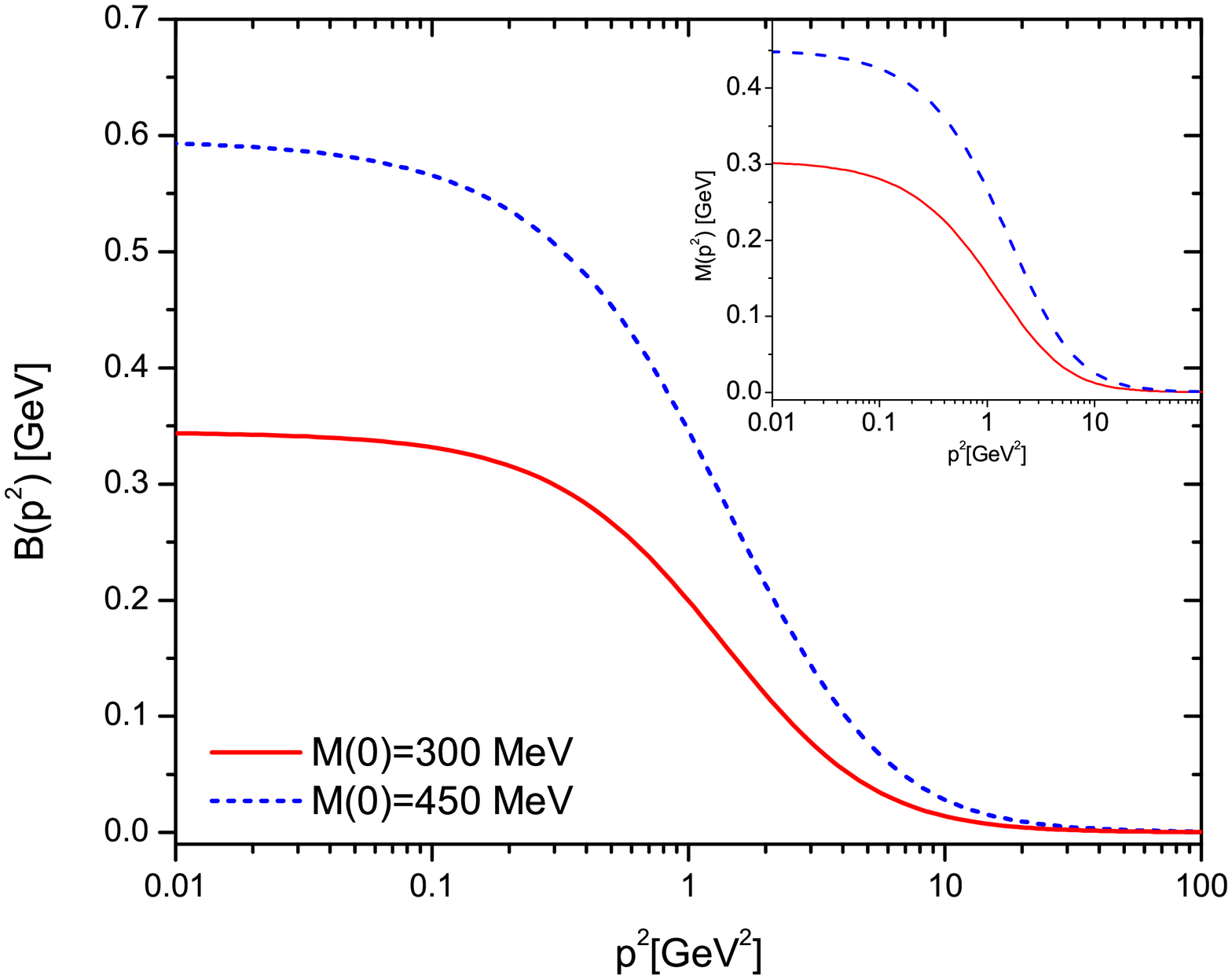}
\end{minipage}
\vspace{-0.8cm}
\caption{\label{AA:plot} The inverse quark wave function, $A^{-1}(p^2)$, (left panel) and the
scalar component of the quark propagator, $B(p^2)$ (right panel). In the inset we show
the  corresponding dynamical quark masses ${\mathcal M}(p^2) = B(p^2)/A(p^2)$.}
\end{figure}
  
The ingredients originating from the quark sector of the theory, namely $A(k^2)$ and $B(k^2)$, will be obtained from
two different versions of the quark gap equation: {\it(i)}
the first one contains the BC vertex, endowed with the minimum  amount of non-Abelian ``dressing''
necessary for achieving compliance with the renormalization group 
[see~\cite{Aguilar:2010cn} and the discussion following Eq.~\eqref{kcsb}], 
and setting $\alpha_s(\mu)=g^2(\mu)/4\pi=0.28$, {\it(ii)} the 
second one employs the Curtis-Pennington vertex~\cite{Curtis:1990zs}, accompanied by a slightly enhanced non-Abelian dressing\footnote{Specifically, in Eq.~\eqref{kcsb} one substitutes $F(q^2)$ by $[1+G(q^2)]^{-1}$; a detailed 
discussion on the properties of the quantity $1+G(q^2)$, 
and its relation to $F(q^2)$, may be found in~\cite{Aguilar:2009nf}.}, 
and $\alpha_s(\mu) =0.29$.  The main practical difference between the two gap equations 
is that they produce  qualitatively different forms of the quark wave function, 
and give rise to distinct constituent quark masses as are shown in Fig.~\ref{AA:plot}. 
In particular, as has been recently pointed out in~\cite{Binosi:2016wcx}, the minimum displayed 
by $A^{-1}(p^2)$ is intimately related to the values of ${\mathcal M}(0)$. Specifically, 
$A^{-1}(p^2)$ maintains its minimum  as long as the corresponding values for ${\mathcal M}(0)$ are relatively low. 
But, when ${\mathcal M}(0)$ exceeds a certain limiting value of 
approximately \mbox{$350$ MeV}~\footnote{The actual value depends, among other things, on the structure and strength 
of the transverse part of the quark-gluon vertex that one uses.}, the aforementioned structure 
is practically eradicated. In order to explore the potential impact of this feature on the 
structure of the $L_i$,  we have obtained a quark mass of \mbox{${\mathcal M}(0) =300$ MeV} with an $A^{-1}(p^2)$ 
with a rather pronounced minimum (red continuous line), and another of  \mbox{${\mathcal M}(0) =450$ MeV}, 
with an $A^{-1}(p^2)$ whose minimum has disappeared (blue dashed line);  note that the corresponding functions $B(k^2)$
are monotonic in both cases.

Note that the $A(p^2)$ and $B(p^2)$ that give rise to a quark dynamical mass of  \mbox{${\mathcal M}(0) =300$ MeV } will be employed in the analysis presented
in the subsections~\ref{3D_scatt},~\ref{long_vertex} and sections~\ref{special} and \ref{RGIQ}, whereas those producing 
\mbox{${\mathcal M}(0) =450$ MeV} will be relevant for the subsection~\ref{CP_results}. 

 Finally, it is important to stress that the inputs used in our calculations (gluon propagator and ghost dressing function) are quenched (no dynamical quarks).
  To be sure, the omission of quark effects while computing the quark-gluon vertex may
  seem a-priori conceptually inconsistent.
However, the effects of ``unquenching'' have been found to be relatively small; in particular, the
estimate provided in~\cite{Aguilar:2014lha} for their relative impact is of the order of $10\%$.

\subsection{\label{3D_scatt} Form factors of the quark-ghost scattering kernel}                                                   

At this point we have all ingredients and shall proceed to determine 
the various $X_i$, given by Eq.~\eqref{euclidean_xi}, for a general kinematic configuration. The main results and observations regarding the $X_i$ obtained using a quark mass with \mbox{$\mathcal{M}(0) =300$ MeV} 
may be summarized as follows.

\begin{figure}[!t]
\begin{minipage}[b]{0.45\linewidth}
\centering
\includegraphics[scale=0.35]{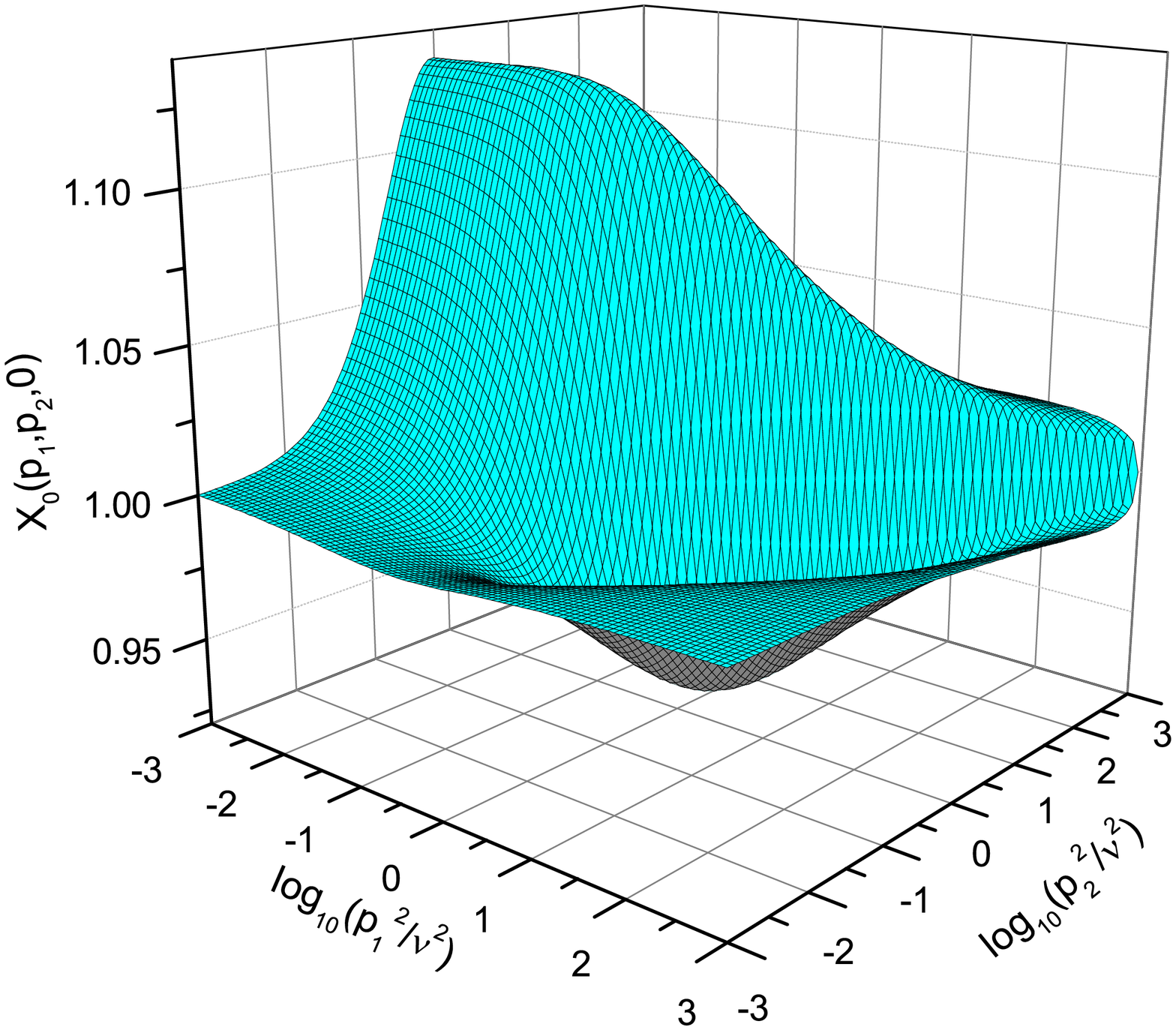}
\end{minipage}
\hspace{0.5cm}
\begin{minipage}[b]{0.50\linewidth}
\includegraphics[scale=0.35]{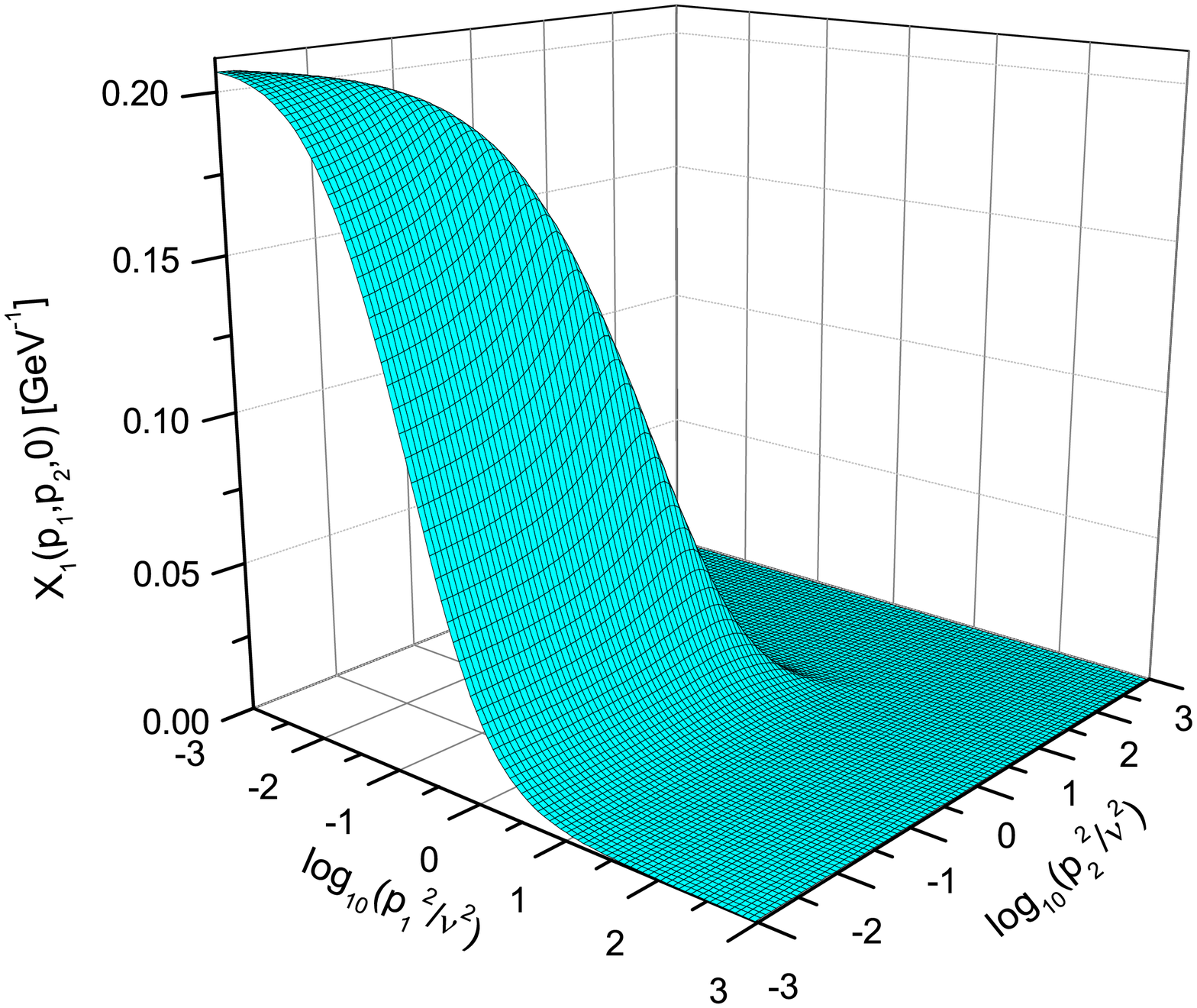}
\end{minipage}
\begin{minipage}[b]{0.45\linewidth}
\centering
\includegraphics[scale=0.35]{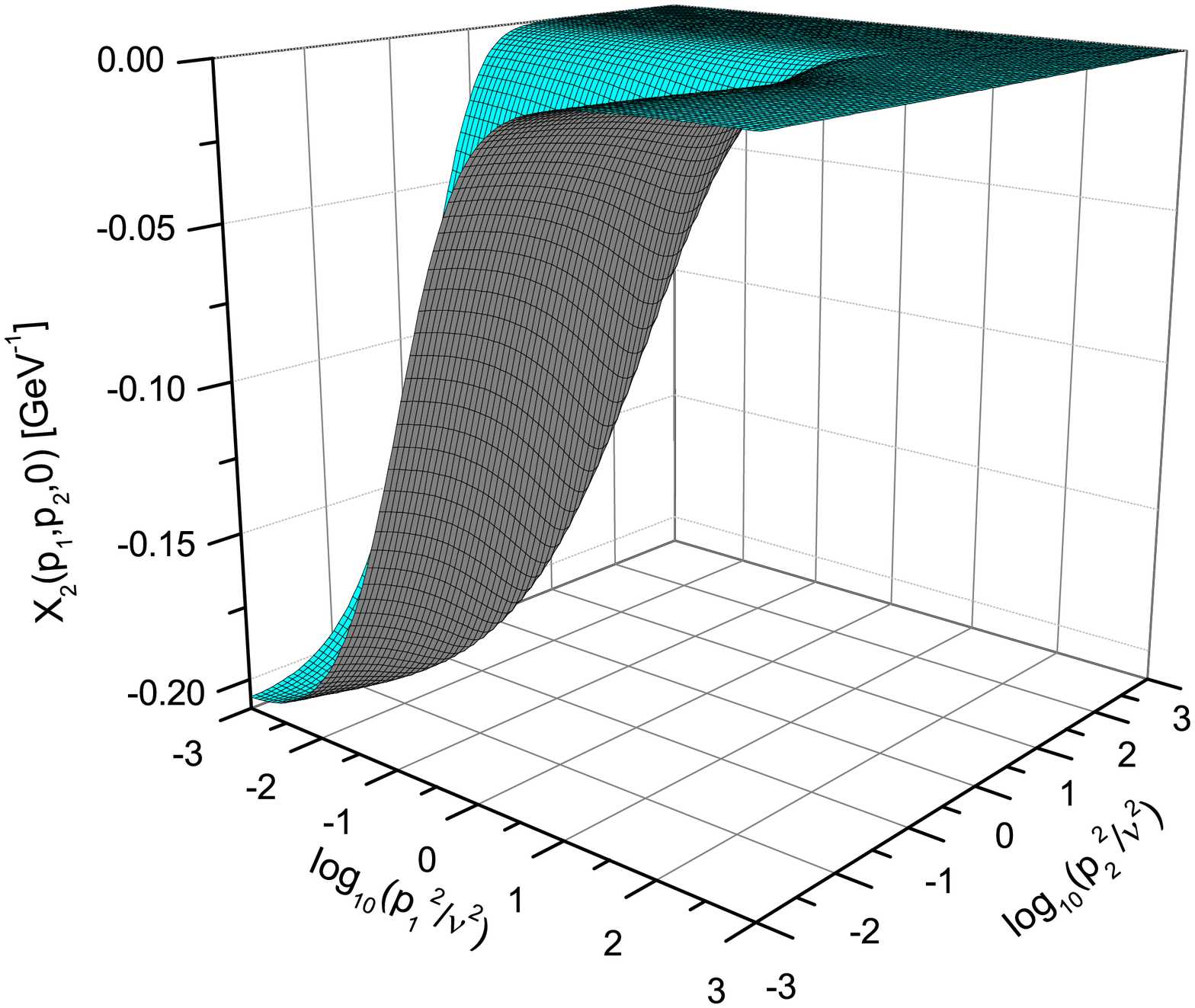}
\end{minipage}
\hspace{0.5cm}
\begin{minipage}[b]{0.50\linewidth}
\includegraphics[scale=0.35]{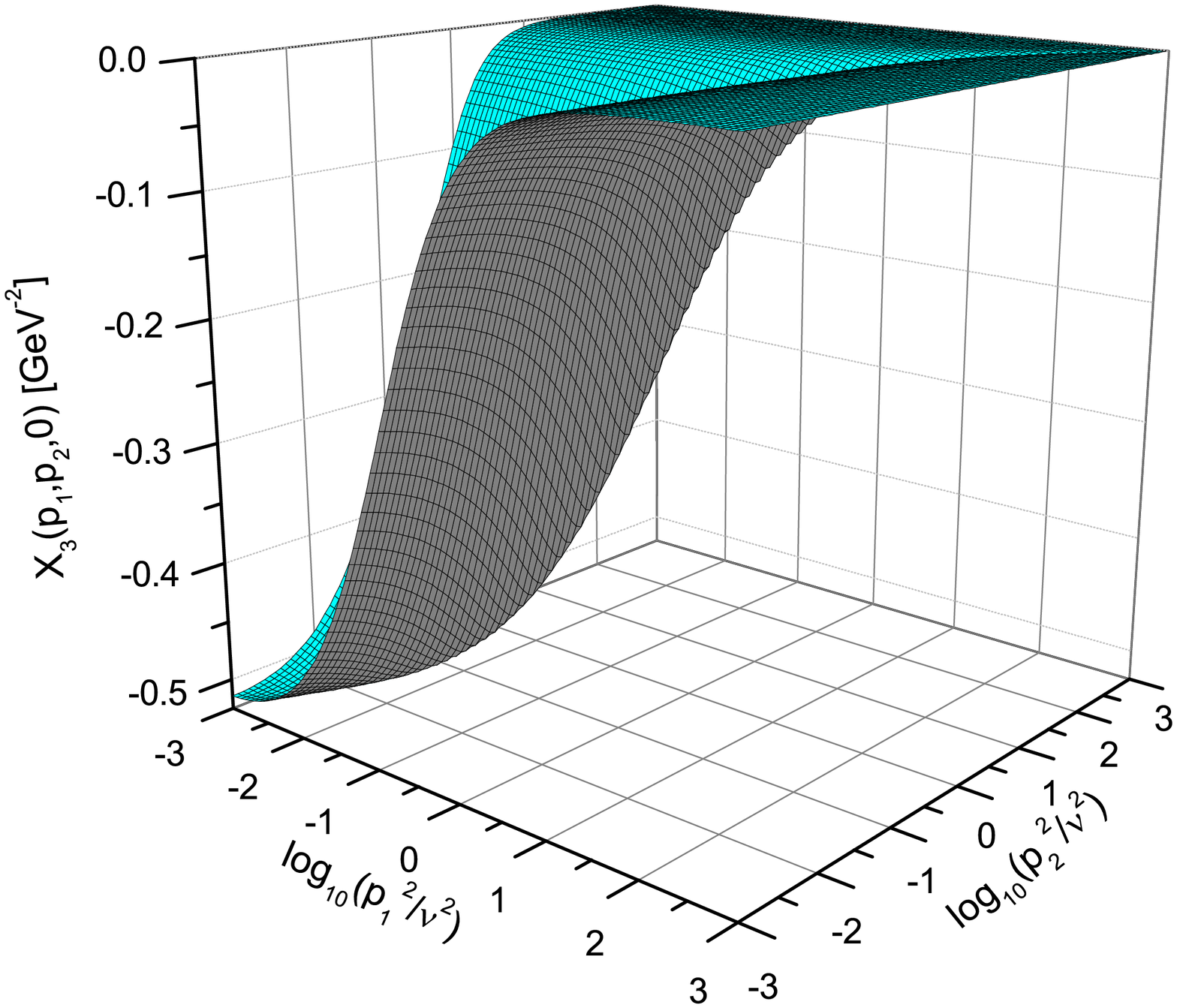}
\end{minipage}
\vspace{-1.25cm}
\caption{\label{Xstheta0:plot} The form factors $X_i(p_1,p_2,0)$ 
 for an arbitrary kinematic configuration given by Eqs.~\eqref{euclidean_xi} when $\theta=0$ and the scale parameter $\nu=1$ GeV.}
\end{figure}

\n{i} In Fig.~\ref{Xstheta0:plot} the 3-D
results  for   all  $X_i(p_1,p_2,0)$   when  $\theta=0$.  
We  observe that all $X_i(p_1,p_2,0)$ are
finite in the infrared, and they recover the correct ultraviolet perturbative behavior.  
More specifically, in the  limit of large momenta ($p_1$ or $p_2$, or both) 
the 3-D curves tend to $X_0=1$ and $X_1=X_2=X_3=0$.

\n{ii} From Fig.~\ref{Xstheta0:plot}  we can  infer the amount by which the $X_i$
depart  from  their tree  level  values.  $X_0$  
deviates $13\%$  from its tree value, while for $X_1$ and $X_2$
the maximum deviation occurs at zero momenta scale, reaching the value
of \mbox{$\pm 0.21\,\mbox{GeV}^{-1}$}. $X_3$ 
displays in the infrared region  the maximum deviation 
value saturating at zero momenta around $-0.52 \,\mbox{GeV}^{-2}$. This
last  observation  indicates  that  the  impact  of $X_3$  on  the
quark-gluon  vertex  may be  quite  sizable.


\n{iii} $X_0(p_1,p_2,0)$ and $X_1(p_1,p_2,0)$ are positive definite 
for all values of $p_1$ and $p_2$, whereas $X_2(p_1,p_2,0)$ and $X_3(p_1,p_2,0)$ are 
strictly negative within the entire range.  

\n{iv} The direct comparison of  the above form factors reveals that  $X_0(p_1,p_2,0)$ 
displays the richest structure, its main features being a pronounced ``peak''  and a shallow ``well''. 
The peak has its maximum  located in the infrared region,  
at \mbox{$p_1^2 =0.009\,\mbox {GeV}^2$} and \mbox{$p_2^2=0.97 \,\mbox {GeV}^2$},   
whereas  the  well has its minimum  around  \mbox{$p_1^2= 3.14\,\mbox {GeV}^2$} and \mbox{$p_2^2 = 0.97 \,\mbox {GeV}^2$}. 

\n{v} We have checked by analyzing a large number of cases that $X_1$, $X_2$ and $X_3$ are quite insensitive
to changes in the value of $\theta$, whereas $X_0$  exhibits certain qualitative changes. 
More specifically,  for the cases where $\theta=\pi/2$ or  $\theta=\pi$, the aforementioned ``well'' disappears. 

As expected from Eq.~\eqref{sym}, the conjugated form factors $\overline{X_i}$, in a general kinematic configuration, display exactly the same behavior shown in Fig.~\ref{Xstheta0:plot}, and for this reason we will omit their explicit derivation here.

\subsection{\label{long_vertex} Form factors of the quark-gluon vertex for ${\mathcal M}(0) = 300$ MeV.}

\begin{figure}[!ht]
\begin{minipage}[b]{0.45\linewidth}
\centering
\includegraphics[scale=0.35]{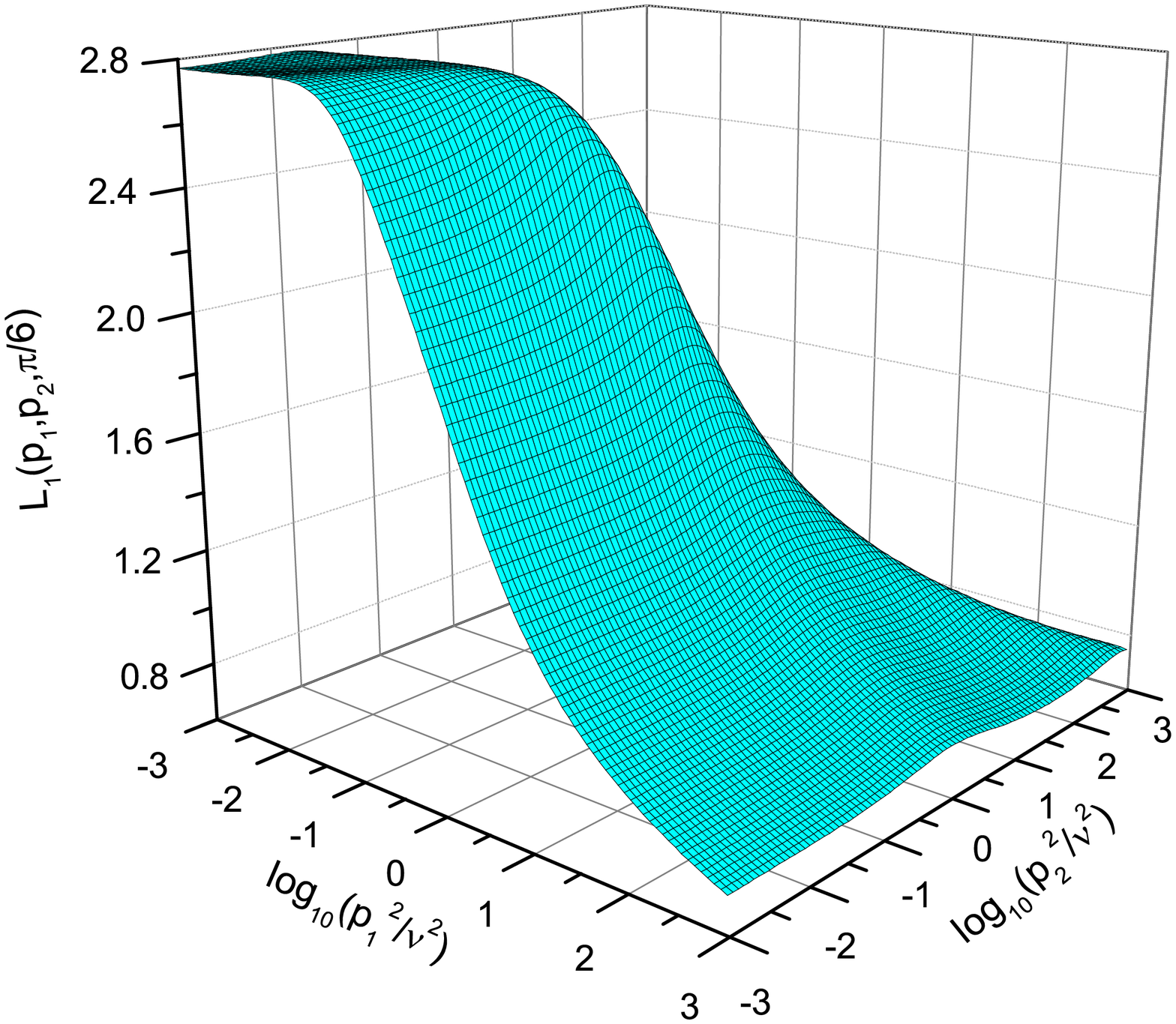}
\end{minipage}
\hspace{0.5cm}
\begin{minipage}[b]{0.50\linewidth}
\includegraphics[scale=0.35]{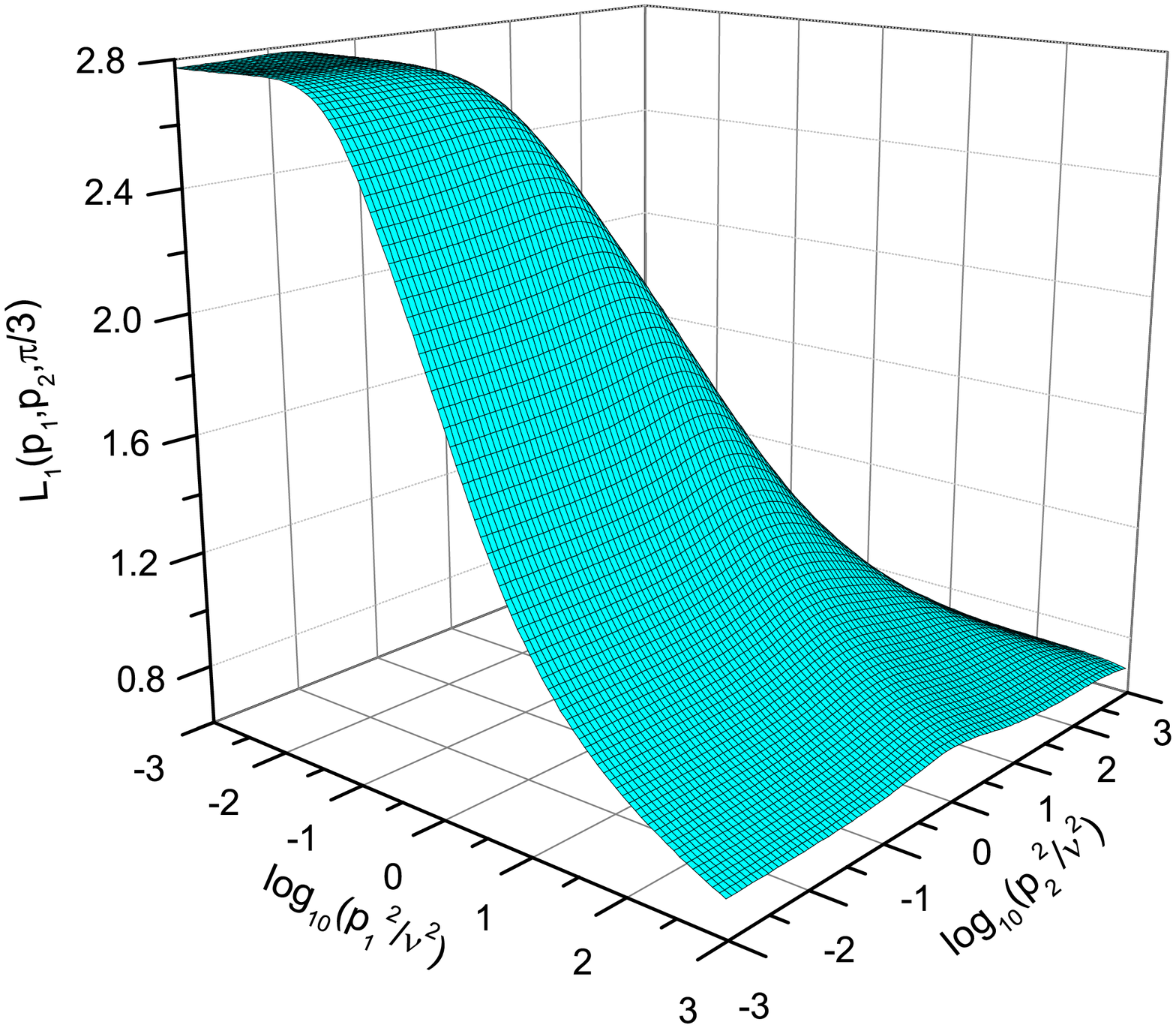}
\end{minipage}
\begin{minipage}[b]{0.45\linewidth}
\centering
\includegraphics[scale=0.35]{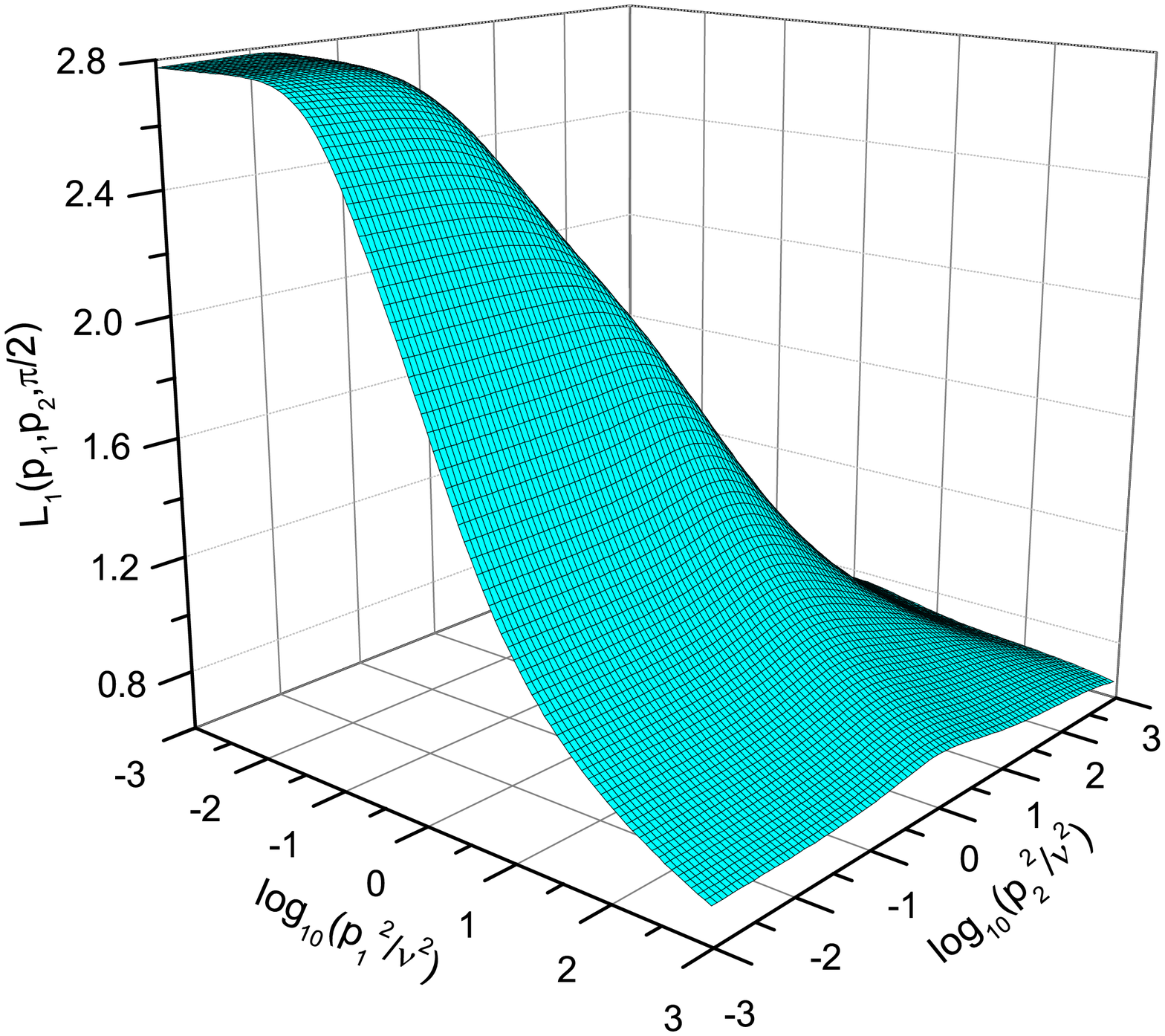}
\end{minipage}
\hspace{0.5cm}
\begin{minipage}[b]{0.50\linewidth}
\includegraphics[scale=0.35]{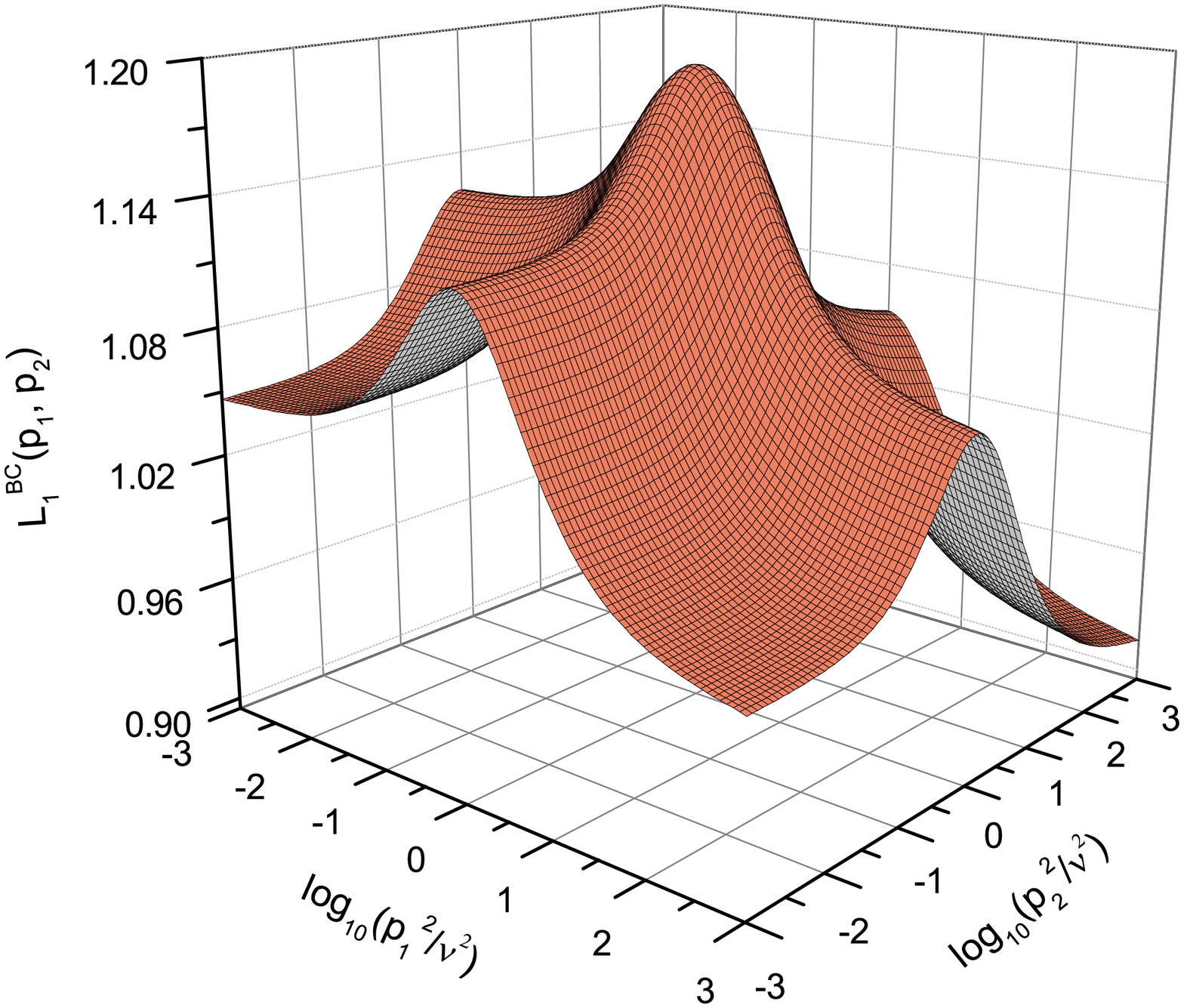}
\end{minipage}
\vspace{-1.25cm}
\caption{\label{fig:vertexL1} $L_1(p_1,p_2,\theta)$ 
for $\theta=\pi/6$, $\theta=\pi/3$, and $\theta=\pi/2$. In the bottom right panel we show  $L_1^{\rm{BC}}$ given by Eq.~\eqref{BC_vertex}.}
\end{figure}

With the $X_i$  for general kinematic configurations at our disposal, 
we may now determine the behavior of the quark-gluon form factors $L_i$ by means of Eq.~\eqref{expLi}.  

Of course, the quark-gluon vertex, and in particular its component $L_1$ (proportional to the tree-level vertex $\gamma^{\mu}$),  
 need to be properly renormalized. The renormalization is implemented as usual, through the introduction of the  
cutoff-dependent constant  $Z_1$, namely   
\begin{align}
\Gamma^{\nu}_R(q,p_2,-p_1;\mu) = Z_1\Gamma^{\nu}(q,p_2,-p_1) \,.
\label{renZ1}
\end{align}
The exact form of $Z_1$ is determined within the momentum-subtraction (MOM) scheme, by 
imposing the condition that,    
at the totally symmetric point, {\it i.e.}, where all squared momenta 
are equal to the renormalization scale $\mu^2$, the quark-gluon vertex recovers its bare value,  \ie
\begin{align}
\Gamma^{\mu}(q,p_2,-p_1)\Big|_{p_1^2=p_2^2=q^2=\mu^2} = \gamma^{\mu} \,. 
\label{cond_mom}
\end{align}

\begin{figure}[!ht]
\begin{minipage}[b]{0.45\linewidth}
\centering
\includegraphics[scale=0.35]{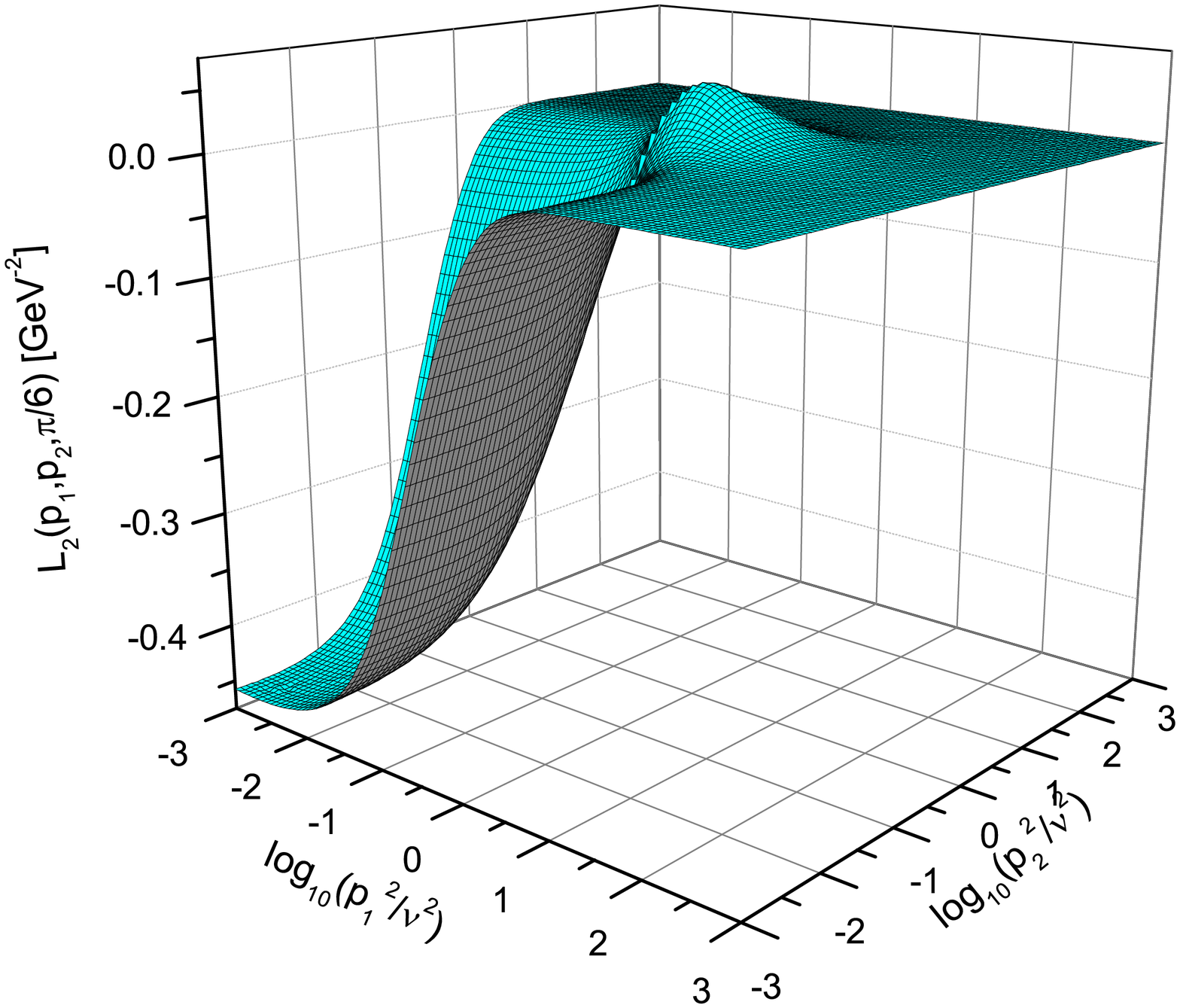}
\end{minipage}
\hspace{0.5cm}
\begin{minipage}[b]{0.50\linewidth}
\includegraphics[scale=0.35]{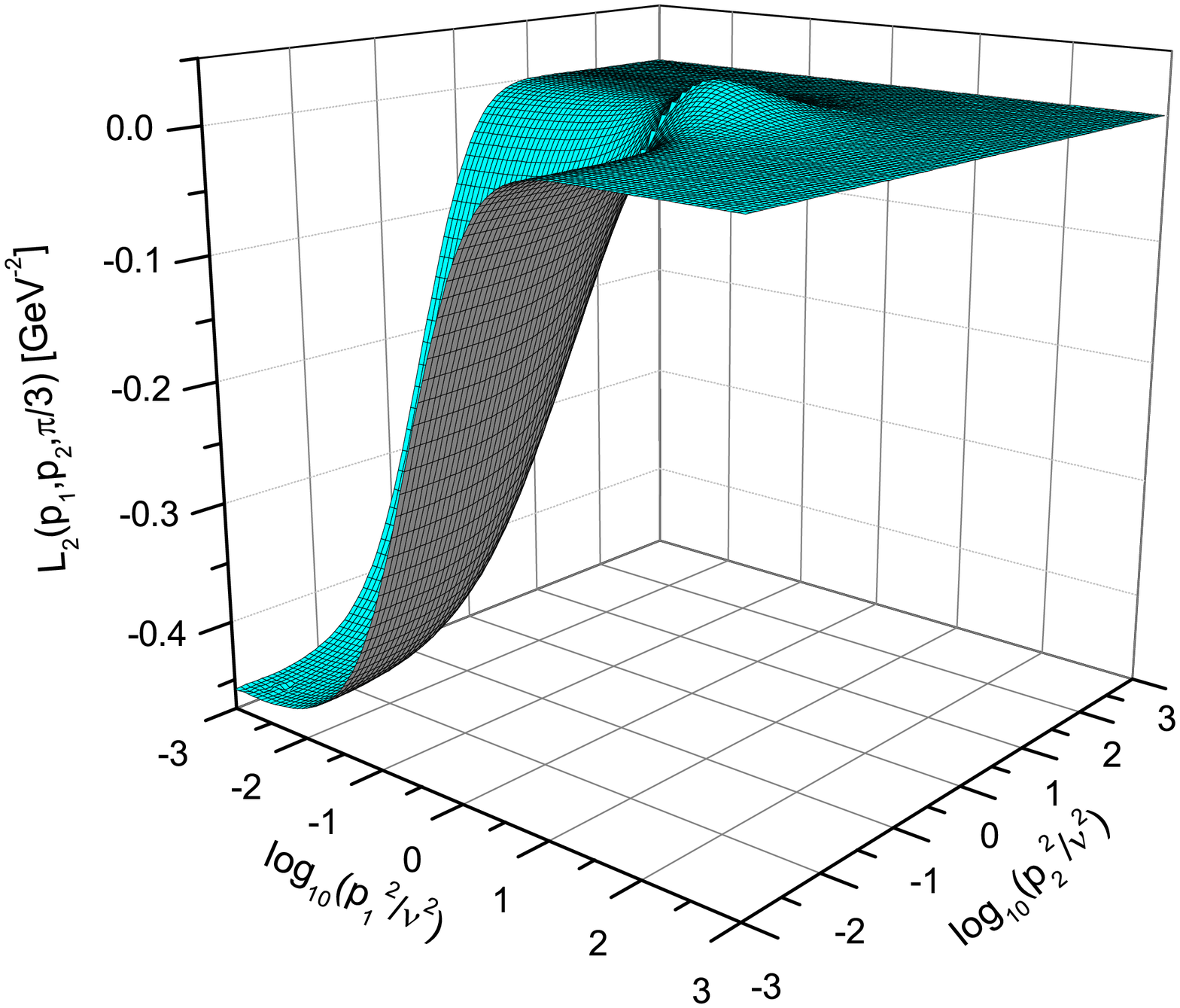}
\end{minipage}
\begin{minipage}[b]{0.45\linewidth}
\centering
\includegraphics[scale=0.35]{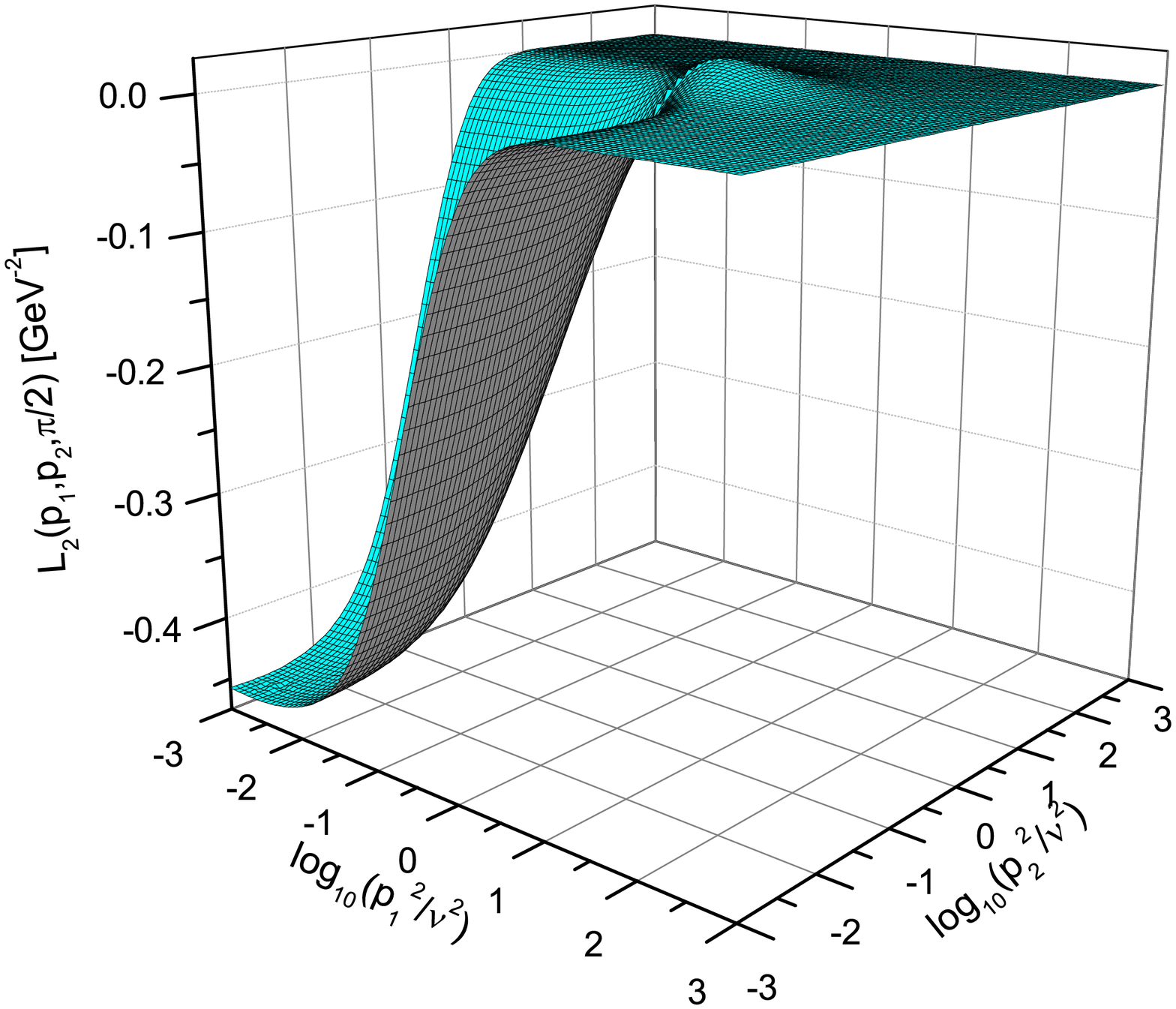}
\end{minipage}
\hspace{0.5cm}
\begin{minipage}[b]{0.50\linewidth}
\includegraphics[scale=0.35]{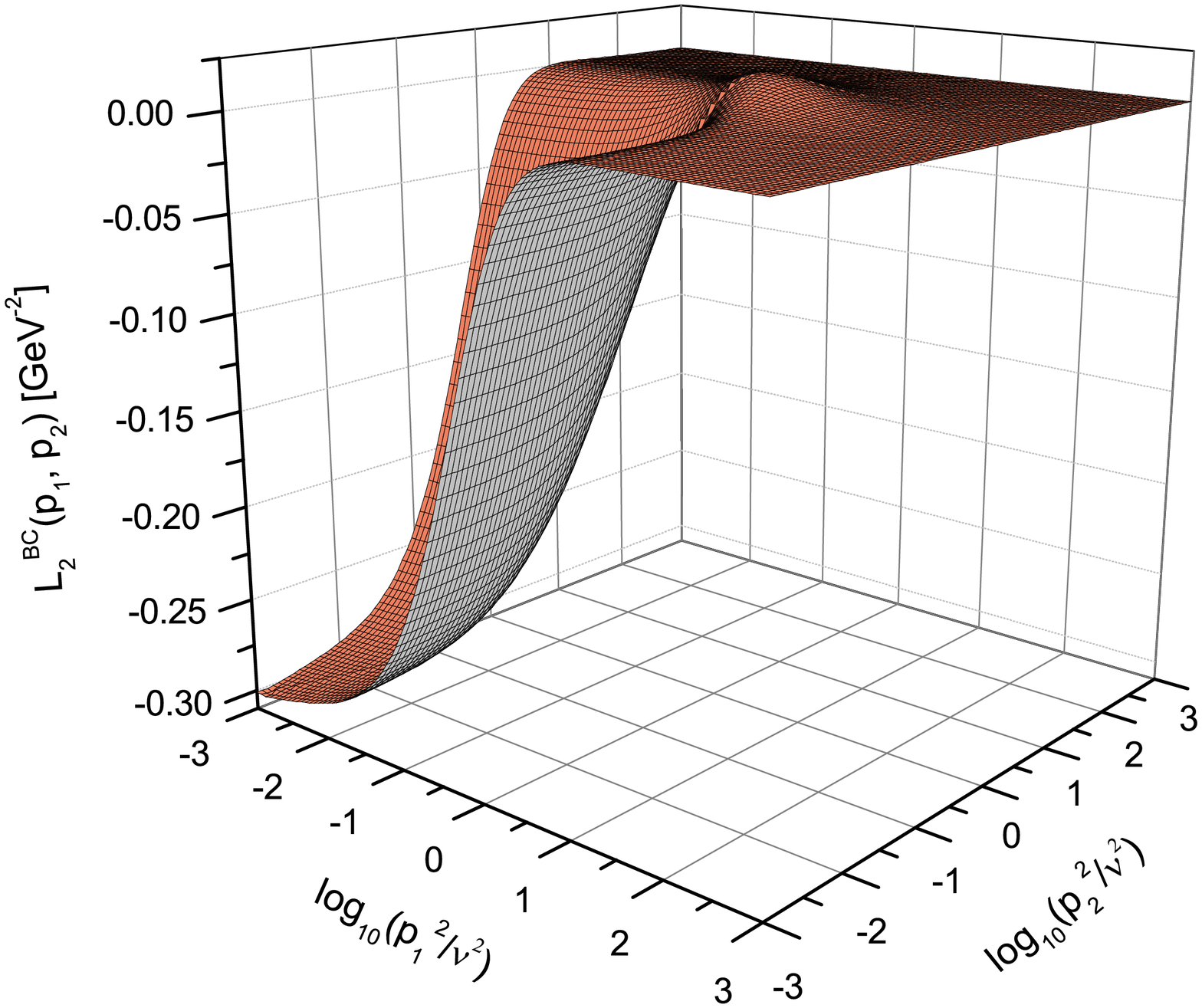}
\end{minipage}
\vspace{-1.25cm}
\caption{\label{fig:vertexL2} $L_2(p_1,p_2,\theta)$ for $\theta=\pi/6$, $\theta=\pi/3$, and $\theta=\pi/2$, together with $L_2^{\rm{BC}}$ (bottom right panel).}
\end{figure}

\begin{figure}[!ht]
\begin{minipage}[b]{0.45\linewidth}
\centering
\includegraphics[scale=0.35]{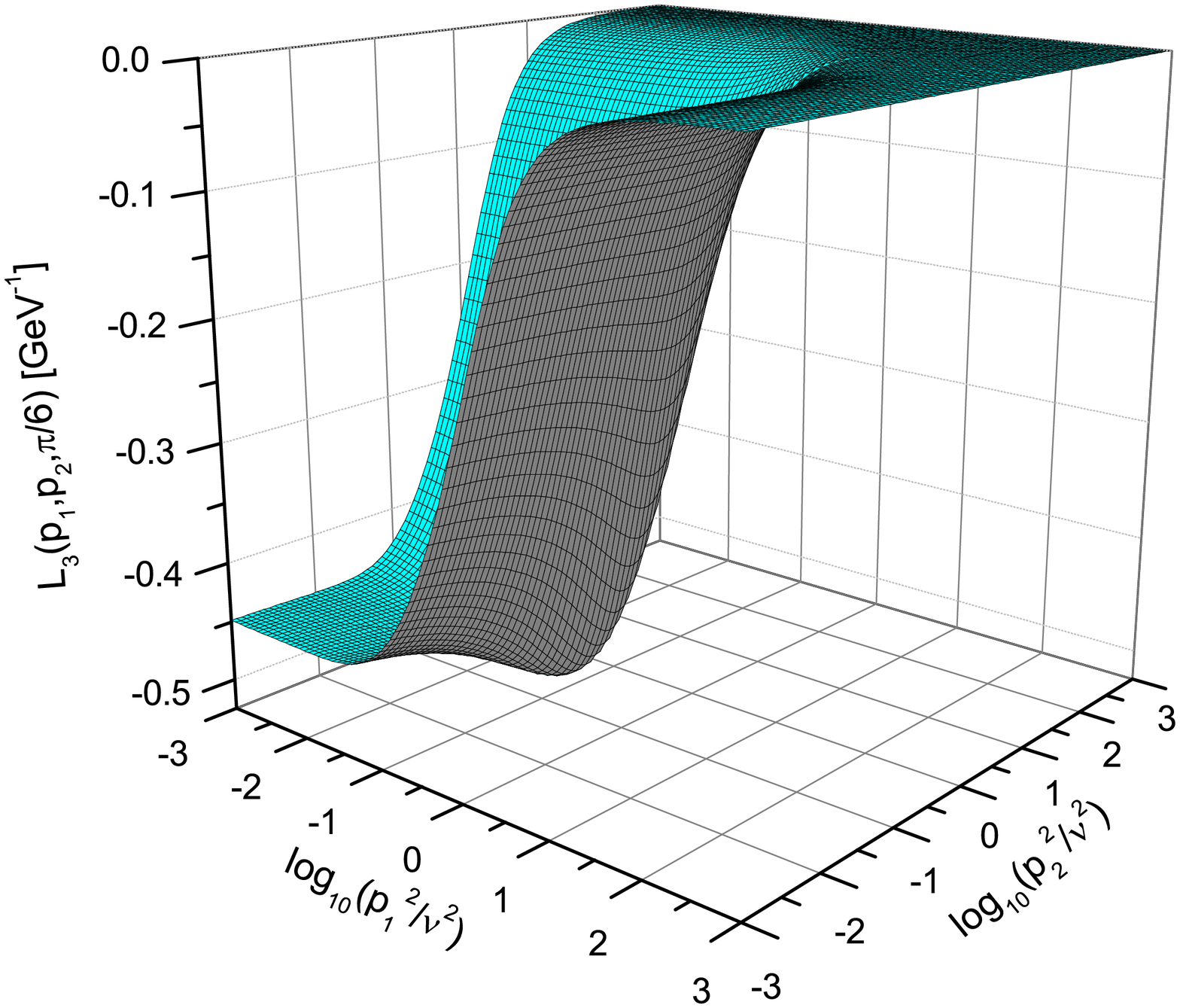}
\end{minipage}
\hspace{0.5cm}
\begin{minipage}[b]{0.50\linewidth}
\includegraphics[scale=0.35]{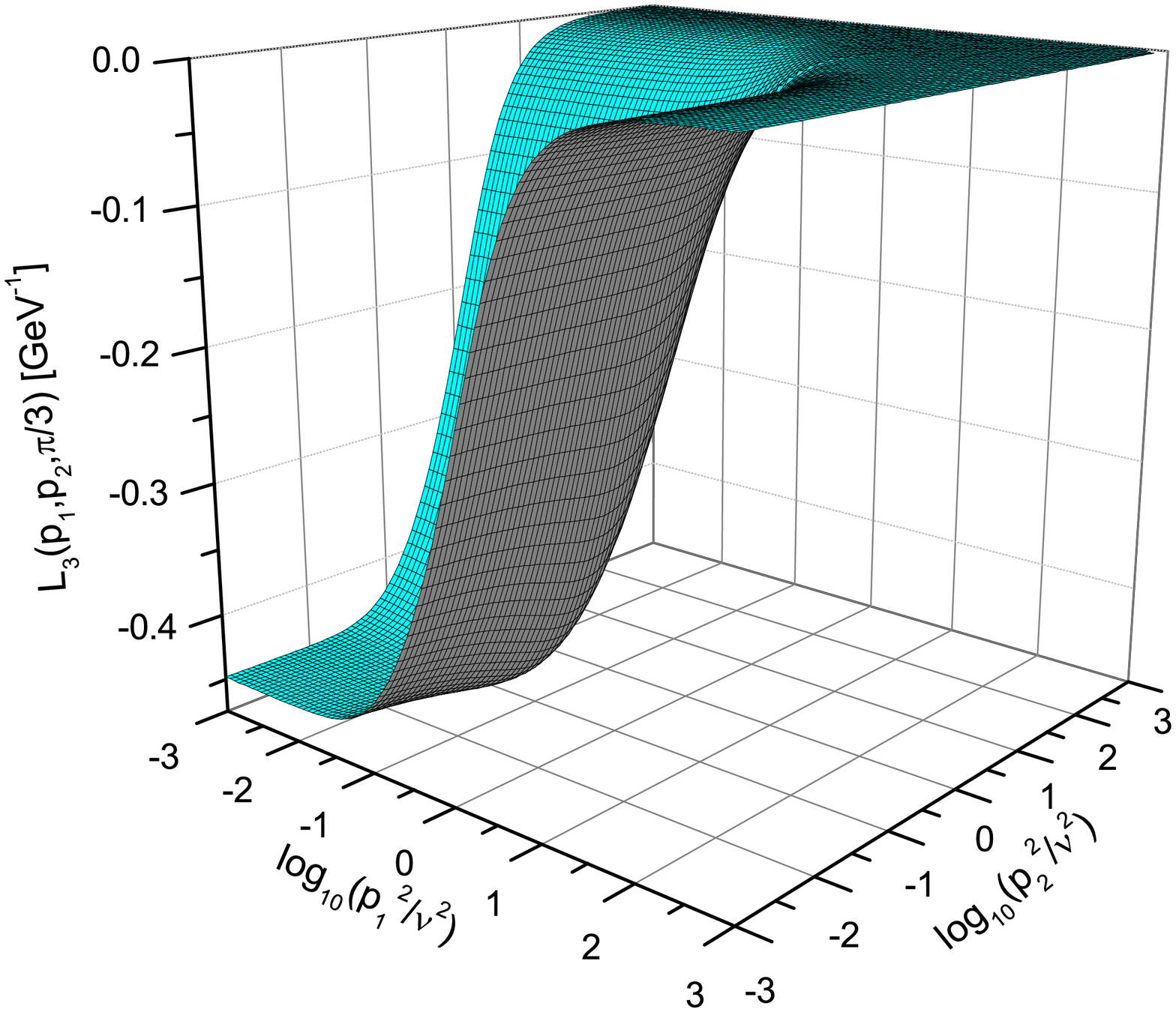}
\end{minipage}
\begin{minipage}[b]{0.45\linewidth}
\centering
\includegraphics[scale=0.35]{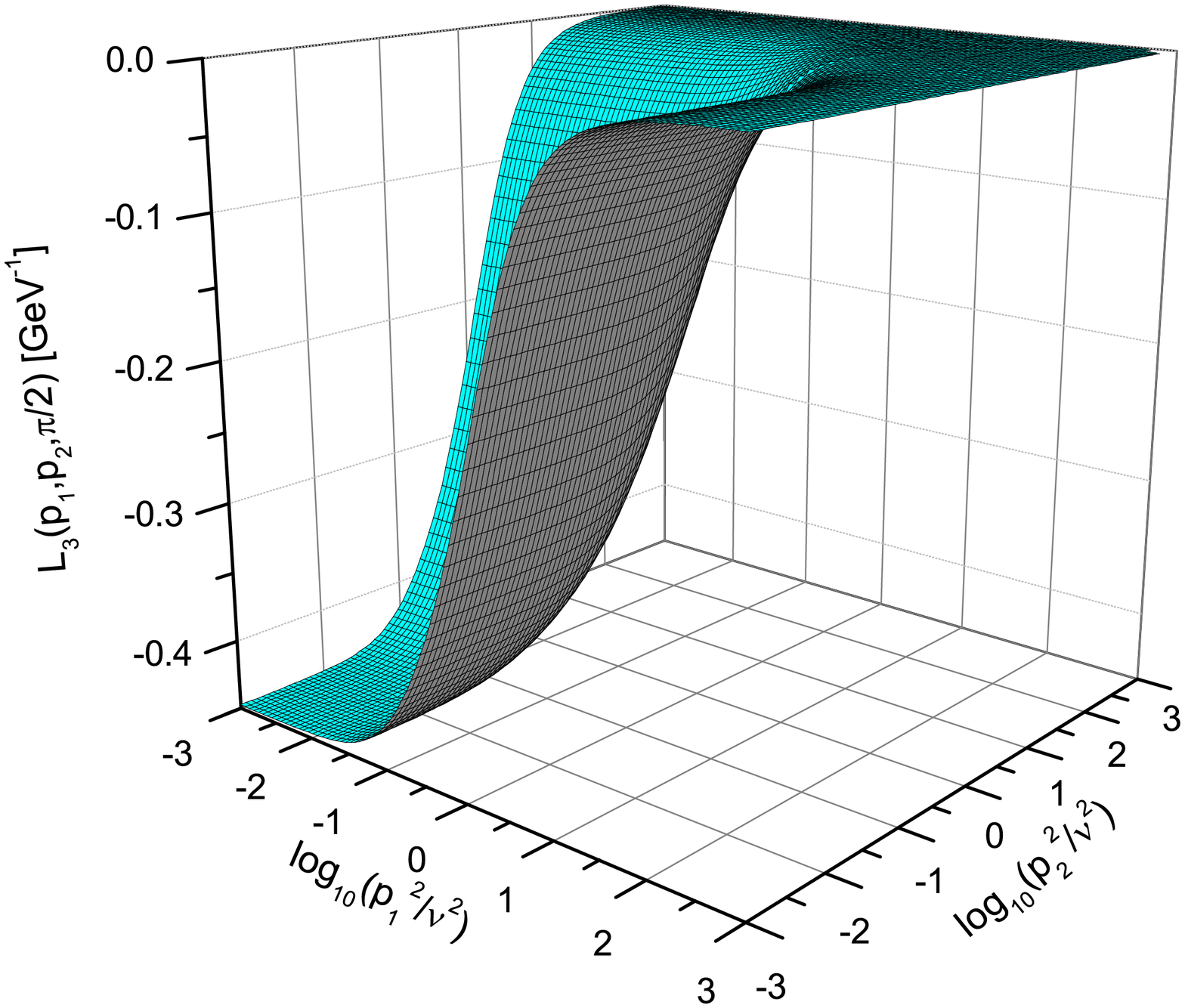}
\end{minipage}
\hspace{0.5cm}
\begin{minipage}[b]{0.50\linewidth}
\includegraphics[scale=0.35]{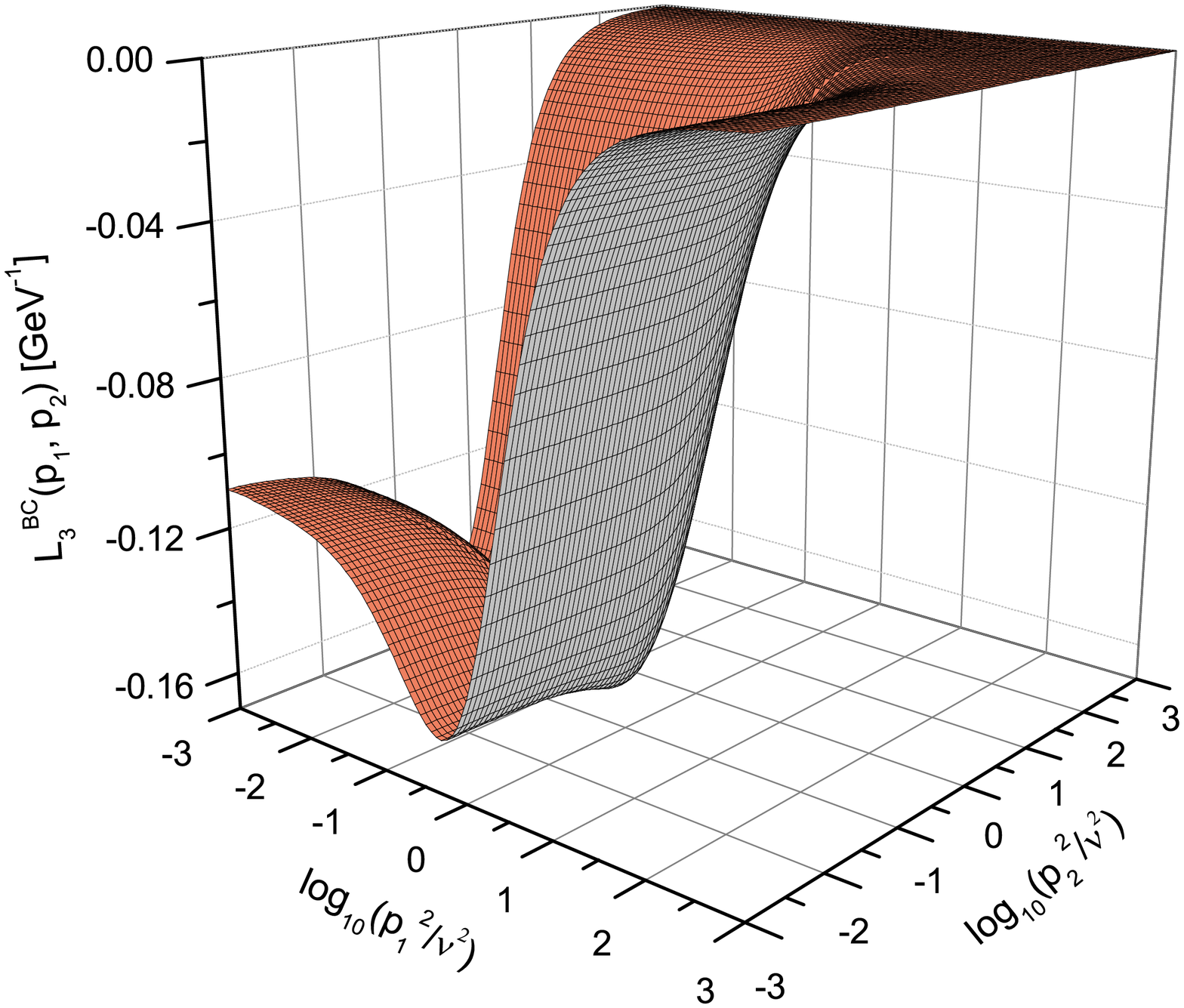}
\end{minipage}
\vspace{-1.25cm}
\caption{\label{fig:vertexL3} $L_3(p_1,p_2,\theta)$ for $\theta=\pi/6$, $\theta=\pi/3$, and $\theta=\pi/2$, together with $L_3^{\rm{BC}}$ (bottom right panel).}
\end{figure}

\begin{figure}[!ht]
\begin{minipage}[b]{0.45\linewidth}
\centering
\includegraphics[scale=0.35]{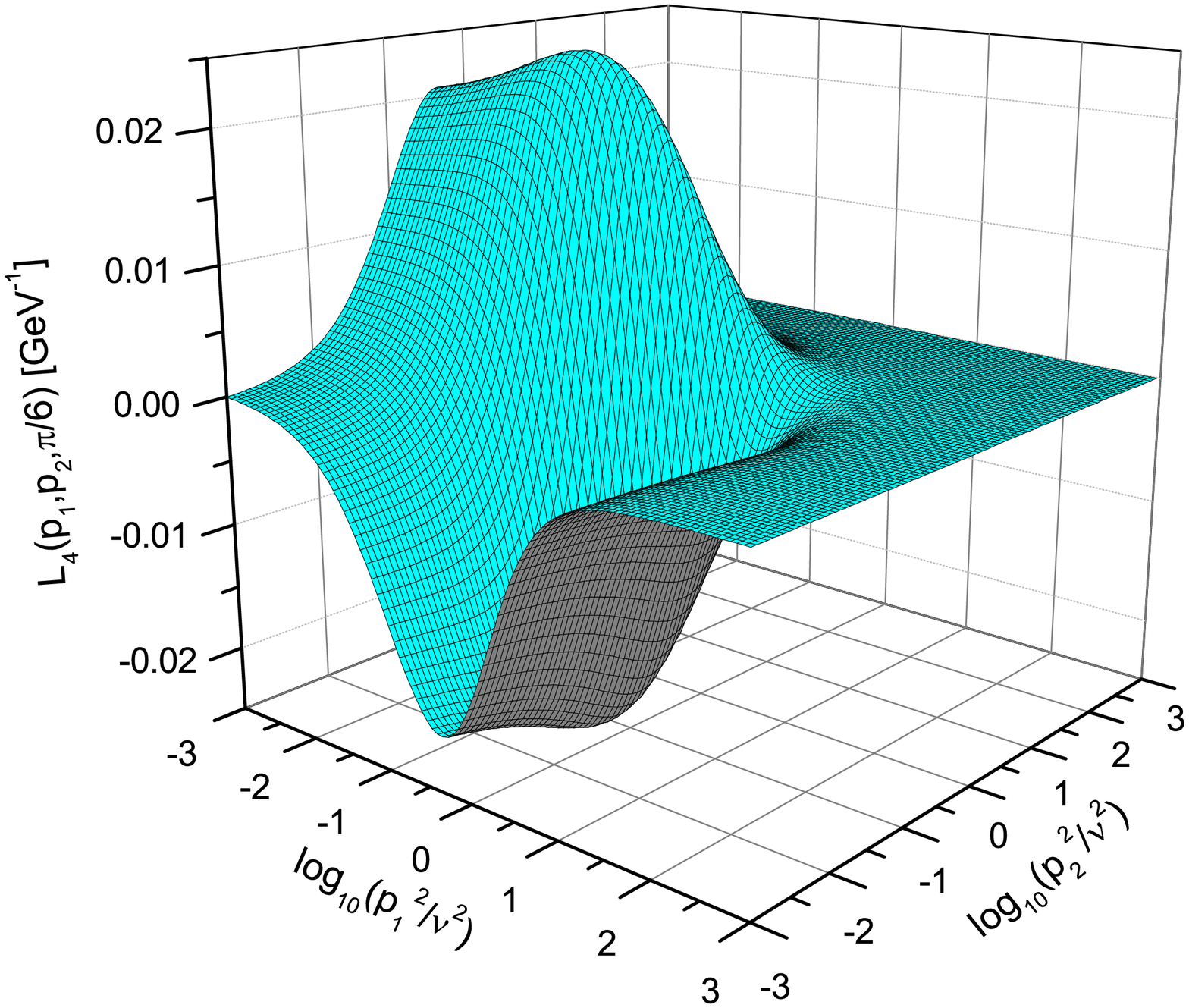}
\end{minipage}
\hspace{0.5cm}
\begin{minipage}[b]{0.50\linewidth}
\includegraphics[scale=0.35]{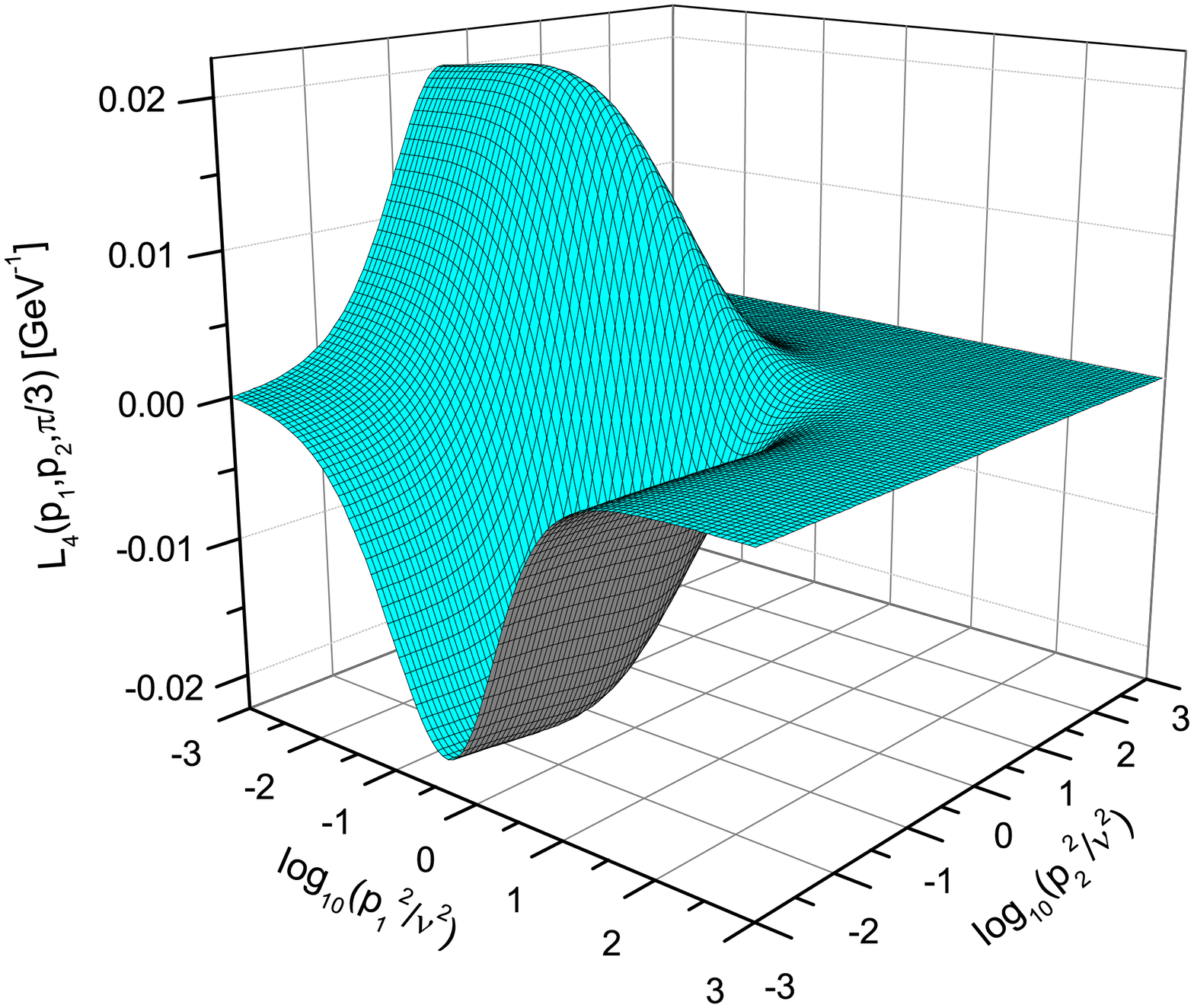}
\end{minipage}
\begin{minipage}[b]{0.45\linewidth}
\centering
\includegraphics[scale=0.35]{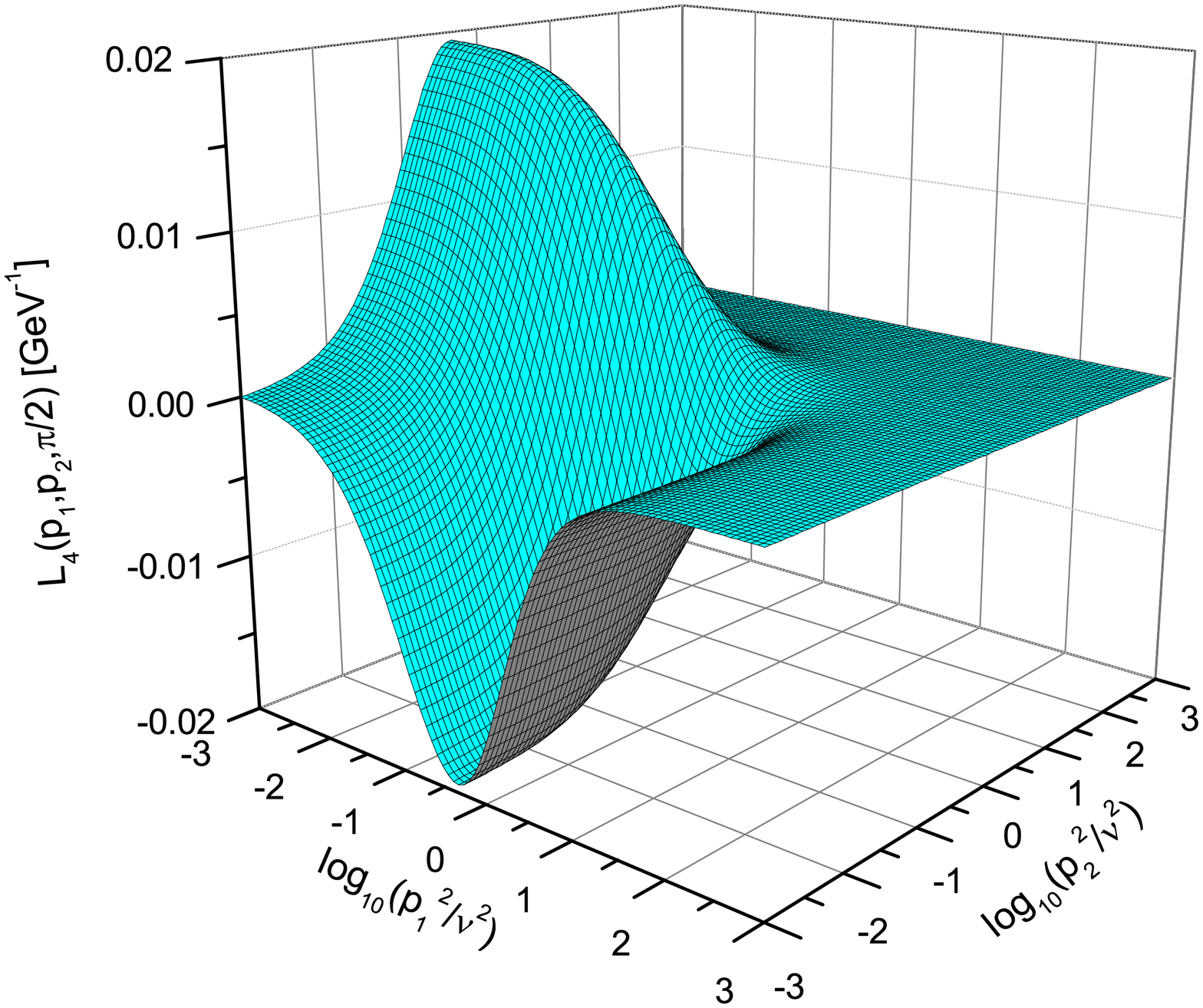}
\end{minipage}
\vspace{-0.75cm}
\caption{\label{fig:vertexL4} $L_4(p_1,p_2,\theta)$ for $\theta=\pi/6$, $\theta=\pi/3$, and $\theta=\pi/2$. Note that $L_4^{\rm{BC}}$ vanishes identically.}
\end{figure}

The results for $L_1$, $L_2$, $L_3$, and $L_4$ are shown in 
Figs.~\ref{fig:vertexL1},~\ref{fig:vertexL2},~\ref{fig:vertexL3}, and~\ref{fig:vertexL4}, respectively. 
In each of these figures, we present the corresponding form factor for three representative values of the angle  $\theta$, 
namely \mbox{$\theta=\pi/6$} (top left panels), $\theta=\pi/3$ (top right panels), and  $\theta=\pi/2$ (bottom left panels). In order to facilitate a direct 
visual comparison, in the bottom right panels of these figures we plot the corresponding Abelian form factors, $L_i^{\rm{BC}}$, given by Eq.~\eqref{BC_vertex},  
which, by construction, are independent of the angle $\theta$. The results for the individual form factors may be summarized as follows.

\begin{figure}[!ht]
\begin{minipage}[b]{0.45\linewidth}
\centering
\includegraphics[scale=0.35]{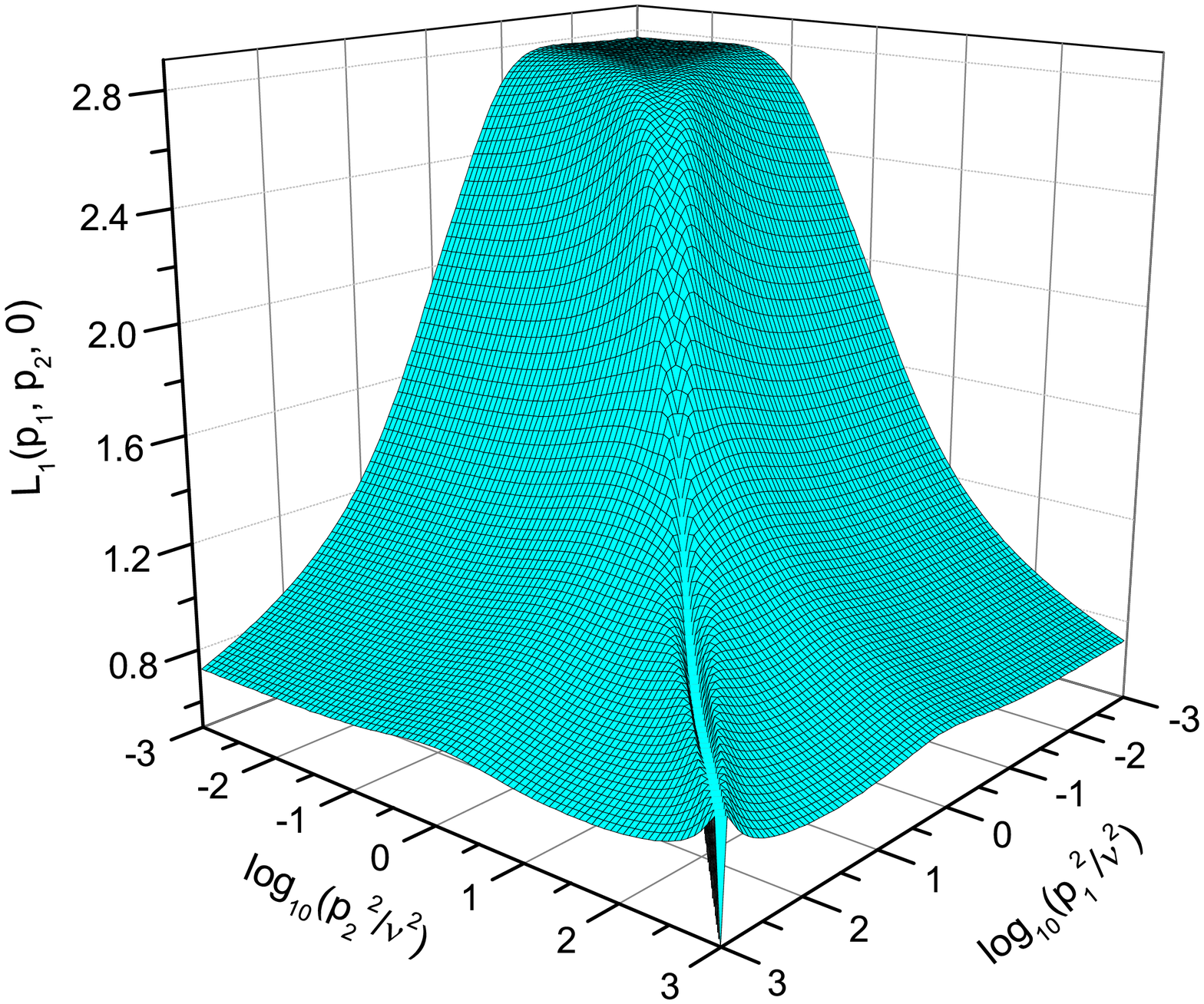}
\end{minipage}
\hspace{0.5cm}
\begin{minipage}[b]{0.50\linewidth}
\includegraphics[scale=0.35]{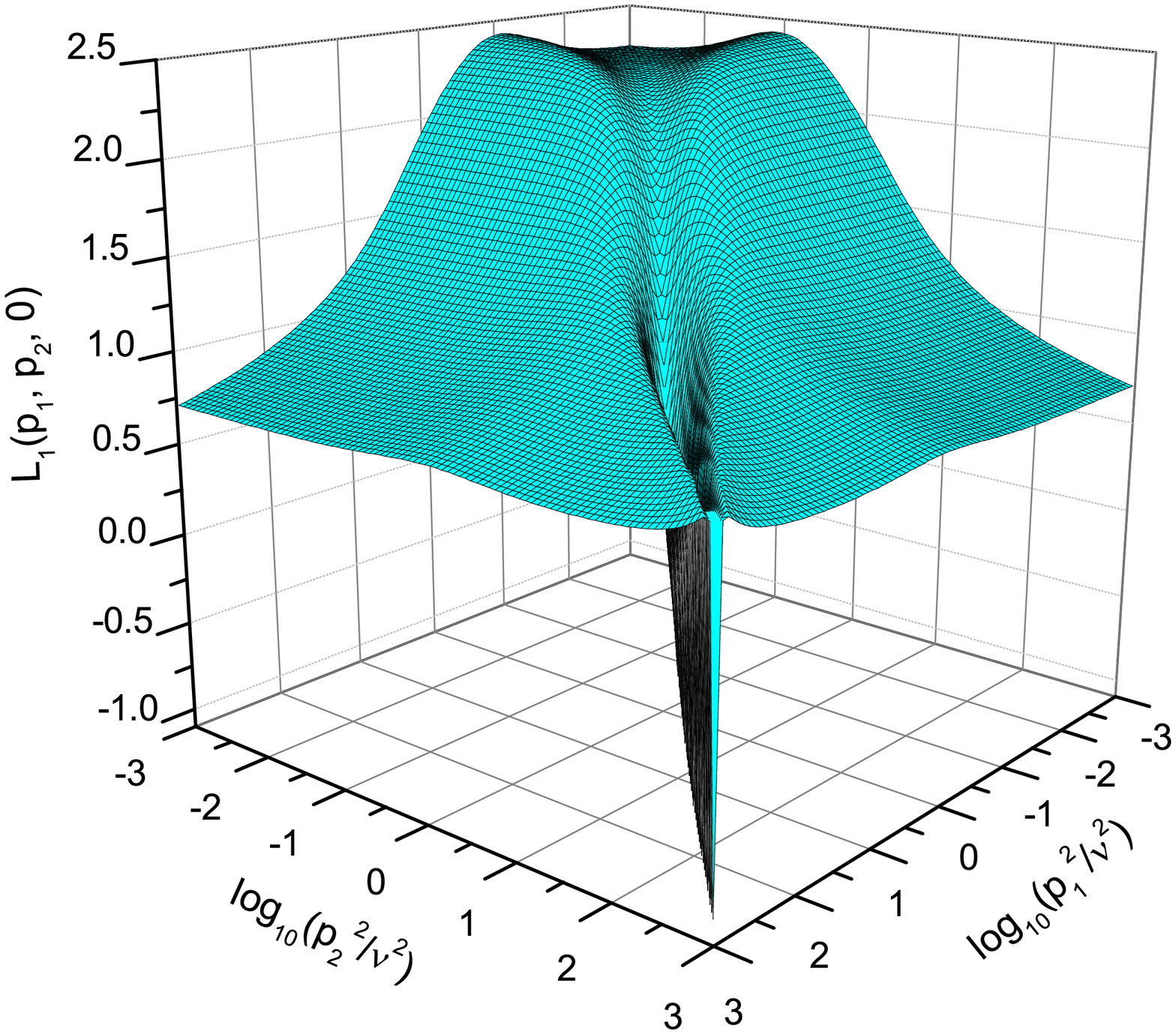}
\end{minipage}
\vspace{-0.75cm}
\caption{\label{fig:vertexL1angle0} The form factor $L_1(p_1,p_2,0)$  computed using either the Ansatz given by Eq.~\eqref{vertex1} (left panel) or
the one of Eq.~\eqref{vertex2} (right panel). The plane defined by \mbox{$p_1=p_2$} 
gives the result for  the soft gluon configuration. Note that, in order to make 
the narrow ``slit'' more visible, we have rotated the plot  with respect to Fig.~\ref{fig:vertexL1}, {\it i.e.}, the axes $p_1$ and $p_2$ have been interchanged.}
\end{figure}

%
\begin{figure}[!ht]
\begin{minipage}[b]{0.45\linewidth}
\centering
\includegraphics[scale=0.35]{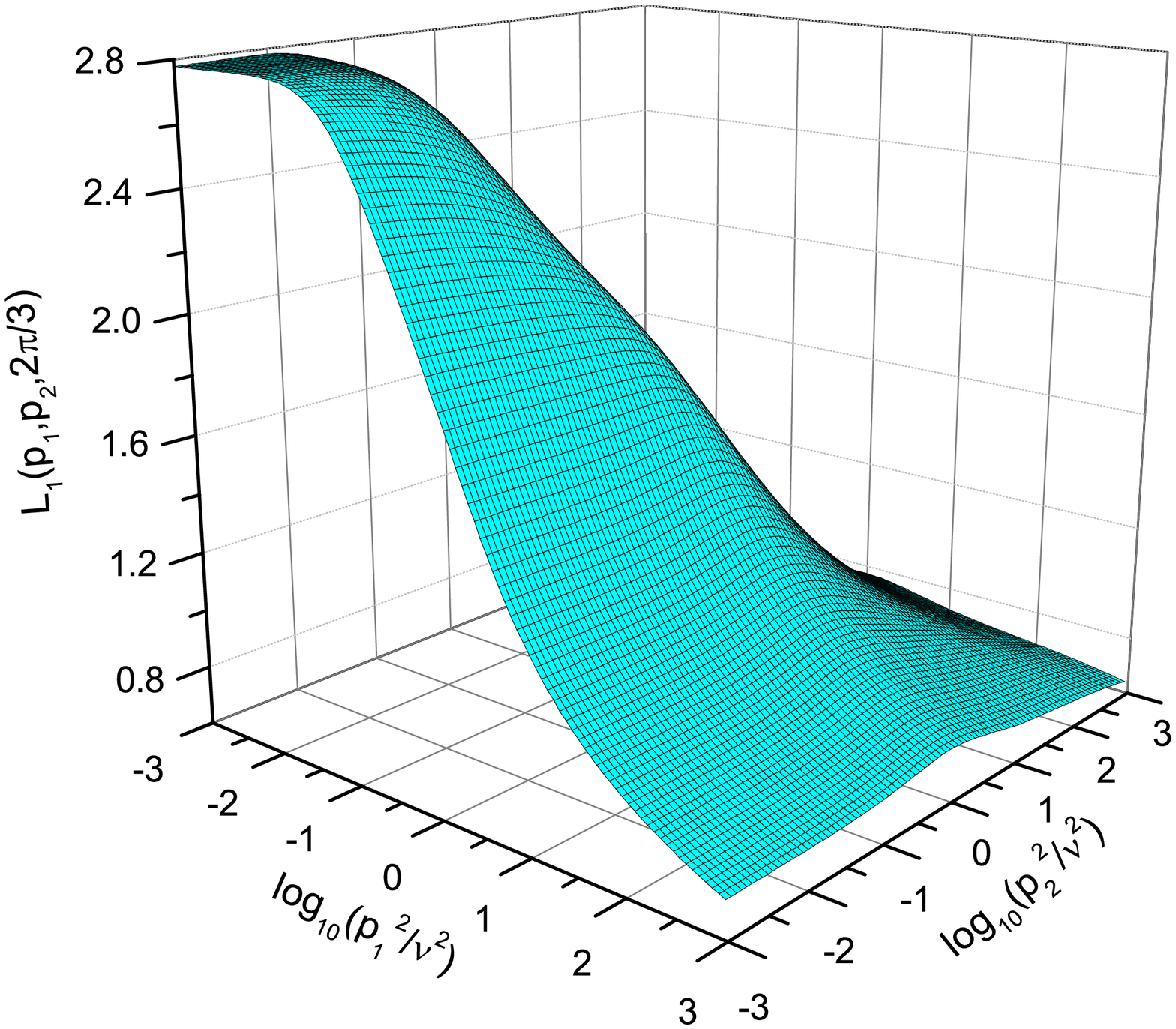}
\end{minipage}
\hspace{0.5cm}
\begin{minipage}[b]{0.50\linewidth}
\includegraphics[scale=0.35]{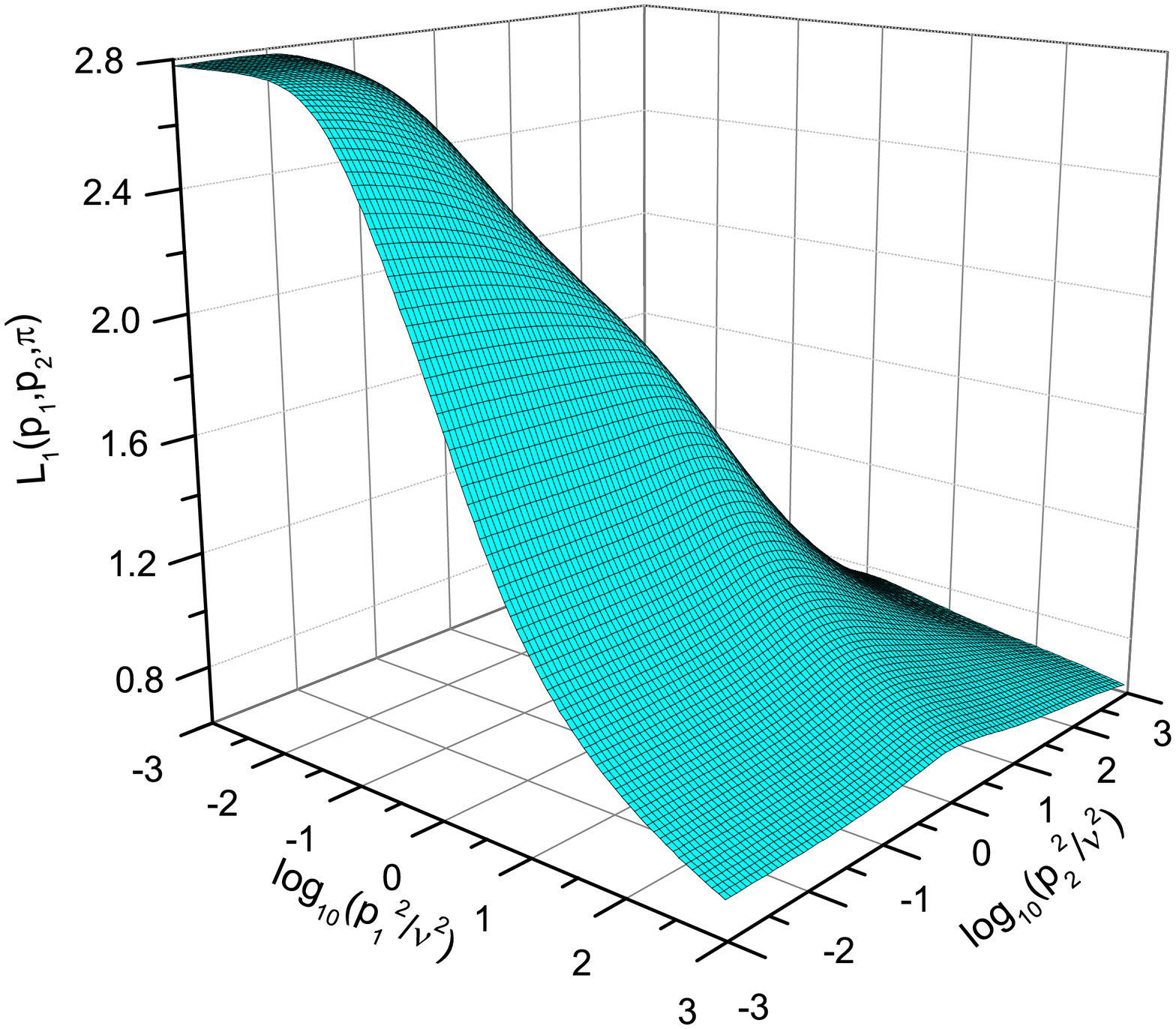}
\end{minipage}
\vspace{-0.75cm}
\caption{\label{fig:vertex_mix} The form factors $L_1(p_1,p_2,\theta)$ for $\theta =2\pi/3$ (left panel) and $\theta=\pi$ (right panel).} 
\end{figure}

\n{i} From Fig.~\ref{fig:vertexL1} it is clear that $L_1$ displays  a very mild dependence on $\theta$, except in the vicinity of $\theta=0$, 
which, due to its particularity, is shown separately in Fig.~\ref{fig:vertexL1angle0}. In this figure,  we clearly see that region
located in the proximity of the  slice defined  by $p_1 =p_2$ is drastically affected
by the type of the quark-gluon vertex Ansatz we employ in the calculation of the various $X_i$. More specifically, the  left panel shows $L_1(p_1,p_2,0)$ obtained with the Ansatz given by Eq.~\eqref{vertex1}, whereas in the right panel we show the result obtained with the vertex of Eq.~\eqref{vertex2}.
The origin of the ``slit''  in the right panel of Fig.~\ref{fig:vertexL1angle0} 
can be traced back to the presence of an extra $F(l-p_2)$ in the structure of the kernel of the Eq.~\eqref{kernelsH} which is introduced by the Ansatz of Eq.~\eqref{vertex2}. Notice  that, only when $p_1=p_2$ (soft gluon limit), the arguments of both $F$, appearing
in Eq.~\eqref{kernelsH}, become exactly the same. It is precisely the $F^2(l-p_1)$ that causes  steeper decrease observed in the right panel. Notice that, in the left panel, where the kernel of Eq.~\eqref{kernelsH} counts with a unique $F(l-p_1)$,   the ``slit''  is practically inexistent.  We emphasize that, with the exception of $L_1(p_1,p_2,0)$, all other $L_1(p_1,p_2,\theta)$  display only small quantitative changes (smaller than $18\%$) when both Ans\"atze are employed. Even though, evidently, further future analysis related to this point is required,
the sensitivity of $L_1(p_1,p_2,0)$ to the particular shape of the vertex employed, suggests that the tensorial structures omitted in 
both Ans\"atze given by Eq.~\eqref{vertex1} and Eq.~\eqref{vertex2} may play an important role
for the complete elimination of the ``slit'' appearing in Fig.~\ref{fig:vertexL1angle0}.

\n{ii} Turning to the $L_2$ shown in Fig.~\ref{fig:vertexL2}, we note that it displays a slightly stronger dependence on $\theta$ 
than $L_1$, which affects mainly the size of the peak located in the intermediate region of momenta. As we can see,  $L_2$ is one order of magnitude smaller compared to $L_1$.  
In addition, while $L_2^{\rm{BC}}$ is negative for all momenta, 
$L_2$ contains some small  positive regions (peaks). Moreover, they are clearly similar in the deep infrared region.

\n{iii} From Fig.~\ref{fig:vertexL3} we infer that the angular dependence of $L_3$  is very mild.
In addition, $L_3$ is always negative and tends to zero in the limit of large momenta (either $p_1$ or $p_2$, or both).
Moreover, we see that $L_3$ reaches sizable values (in modulo) for values of $p_1$ and $p_2$ smaller than $10^{-1}\,\mbox{GeV}^2$.
As a final remark, we notice that although  $L_3^{\rm{BC}}$ 
is  more suppressed than $L_3$, its shape is very similar to that of $L_3$; the region of momenta where the difference
is more pronounced is in deep infrared, where, unlike $L_3$, the $L_3^{\rm{BC}}$ displays a minimum.

\n{iv} As can be seen in Fig.~\ref{fig:vertexL4}, the  angular dependence of $L_4$ is  essentially negligible, and 
the most prominent characteristic is its suppressed structure within the entire range of momenta, reaching a maximum value of 
at most $0.027\,\mbox{GeV}^{-2}$. We recall here that $L_4^{\rm{BC}}$ vanishes identically 
[see Eq.~\eqref{BC_vertex}]. 

\n{v} For later convenience, we show in the Fig.~\ref{fig:vertex_mix}
the results for $L_1$ when $\theta=2\pi/3$ (left panel)  and  $\theta=\pi$ (right panel)  which will be used to determine some special kinematic configurations.

\begin{figure}[t]
\begin{minipage}[b]{0.45\linewidth}
\centering
\includegraphics[scale=0.35]{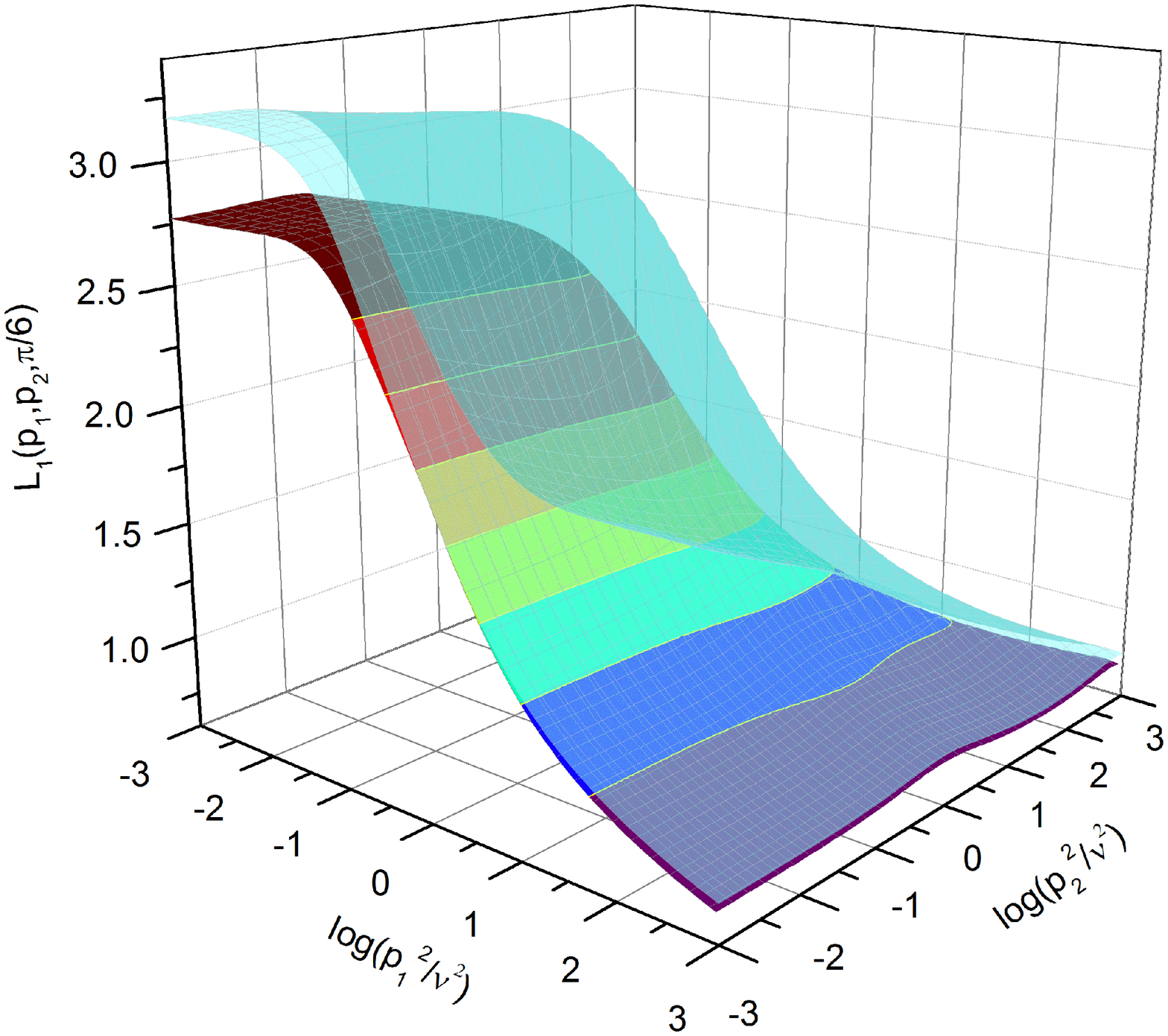}   
\end{minipage}
\hspace{0.5cm}
\begin{minipage}[b]{0.50\linewidth}
\includegraphics[scale=0.35]{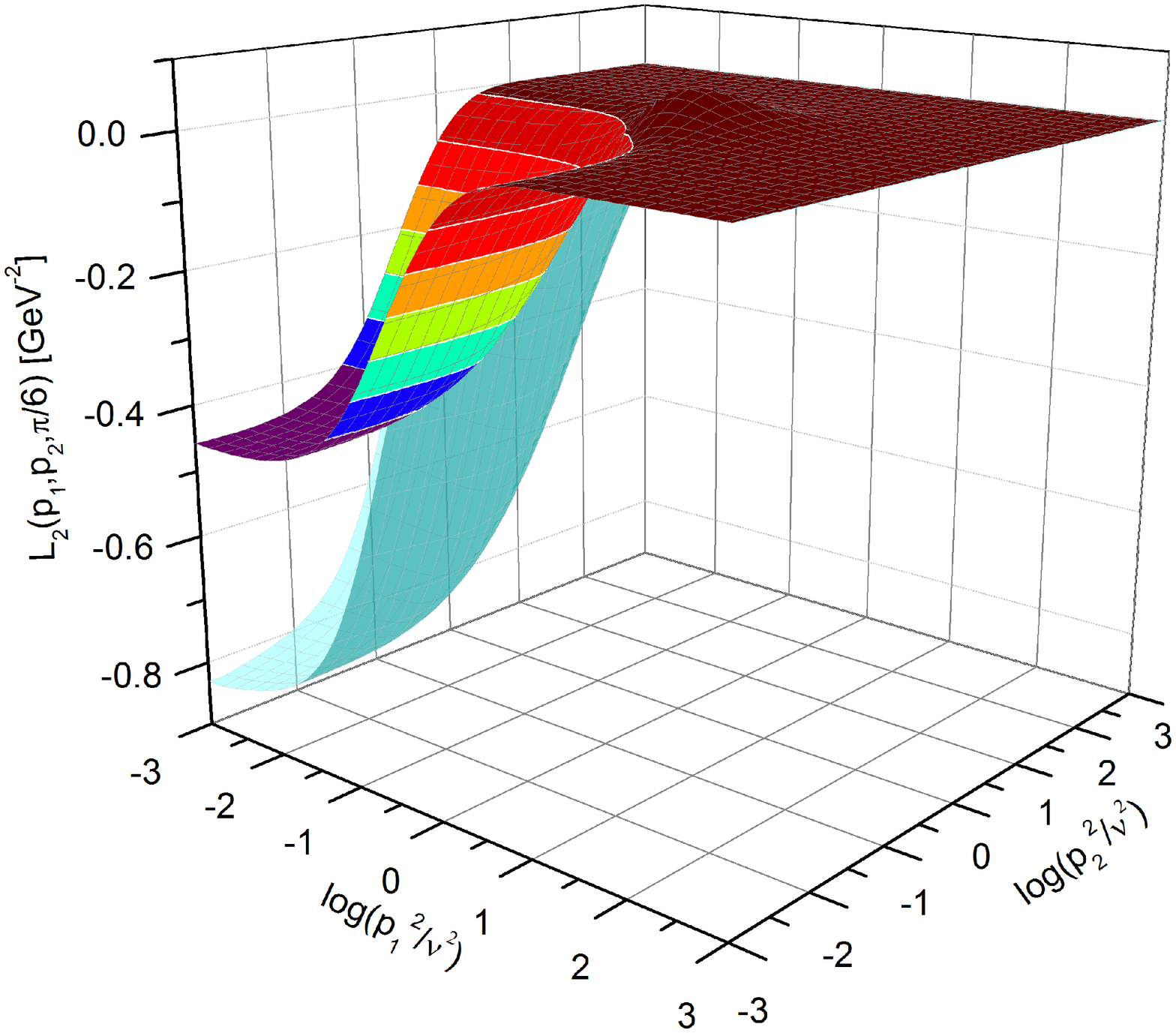}     
\end{minipage}
\vspace{0.5cm}\begin{minipage}[b]{0.45\linewidth}
\centering
\includegraphics[scale=0.35]{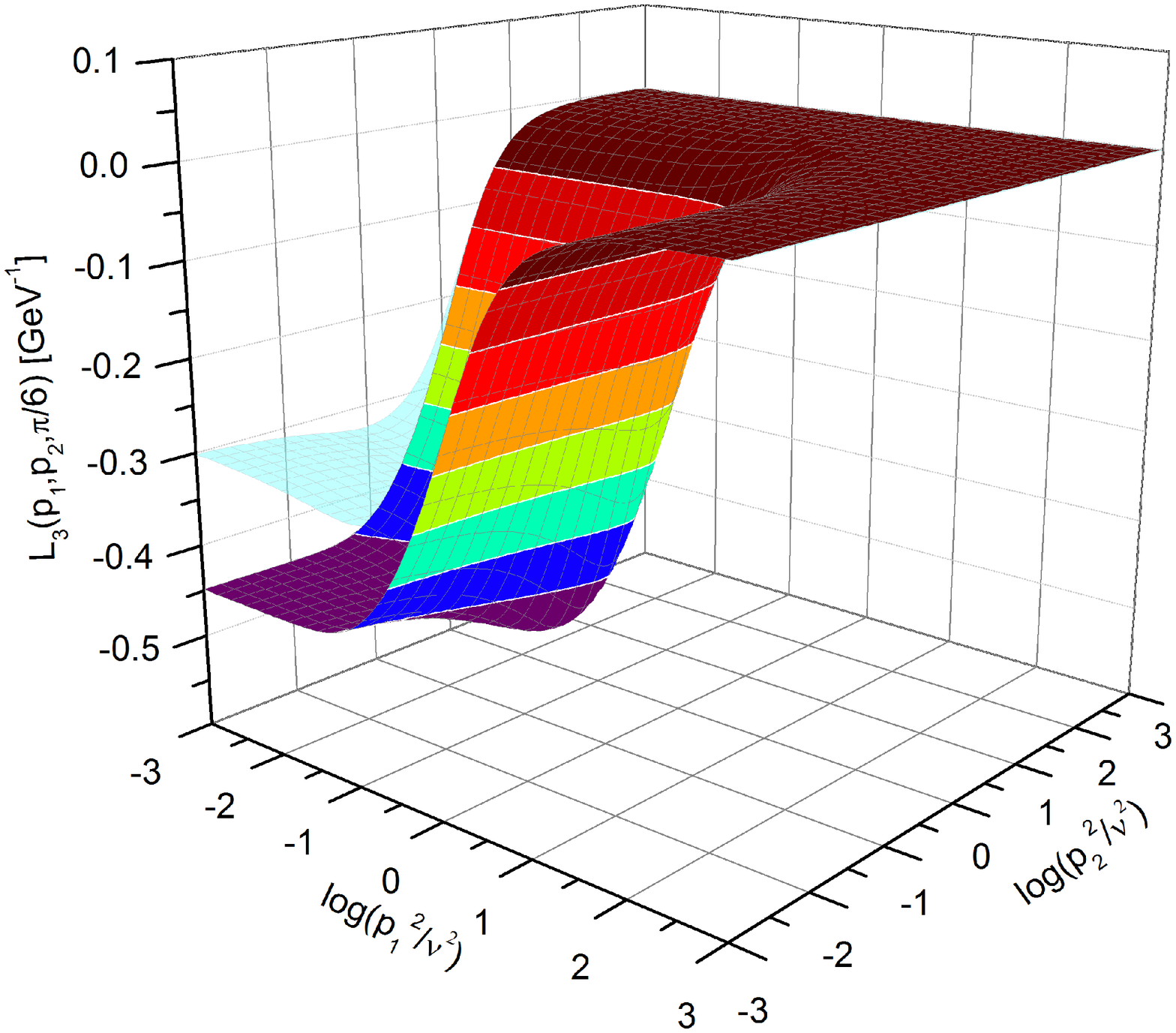}       
\end{minipage}
\caption{\label{fig:comp_fbc} Comparison of the form factors $L_i(p_1,p_2,\theta)$ (colored surface) with the``minimal'' non-Abelian Ansatz,  $F(q)L_i^{\rm{BC}}$ (cyan surface) for $\theta =\pi/6$.}
\end{figure}

\n{vi} Finally, in order to fully appreciate the numerical impact of the quark-ghost scattering kernel
on the form factors $L_i$, in Fig.~\ref{fig:comp_fbc} we compare our results (colored surface)  with those  
obtained when the quark-ghost scattering is fixed at its tree-level value (cyan surface) for a fixed angle ($\theta =\pi/6$). More specifically, setting in Eq.~\eqref{expLi} $X_0 = {\overline X}_0=1$ and $X_i = {\overline X}_i=0$, for $i \geq 1$, we obtain the ``minimal'' non-Abelian  Ansatz for the quark-gluon vertex, $F(q)L_i^{\rm{BC}}$~\cite{Fischer:2003rp,Fischer:2006ub,Aguilar:2010cn}.  In Fig.~\ref{fig:comp_fbc}, 
we clearly see that our results for  $L_1$ and $L_2$ (colored surface)
are  significantly more suppressed (in modulo)  compared with those obtained with the  ``minimal'' non-Abelian Ansatz (cyan surface). 
On the other hand, in the case of $L_3$ we observe the opposite effect.

\subsection{\label{CP_results} Form factors of the quark-gluon vertex for ${\mathcal M}(0) = 450$ MeV.}

We next analyze the differences that a higher value of the quark mass ${\mathcal M}(p^2)$ and a nearly monotonic $A(p^2)$ may produce in the overall shape of the $L_i$. 
To that end, we have recomputed the  $X_i$ and  $L_i$, using as ingredients the $A(p^2)$ and $B(p^2)$ 
that give rise to \mbox{$\mathcal{M}(0) =450$ MeV}  (see the blue-dashed line in Fig.~\ref{AA:plot}).

It turns out that the qualitative behavior for all $X_i$ are very similar to those already presented in the Fig.~\ref{Xstheta0:plot}.  More specifically, we notice that
for $X_0$, $X_1$ and $X_2$  the differences in their extrema
are at most of the order of $10\%$. In the case of $X_0$ the maximum of the curve changes
from  $1.13$ to $1.11$, whereas for $X_1$ and $X_2$ their extrema increase from 
\mbox{$\pm 0.20\,\mbox{GeV}^{-1}$} to  \mbox{$\pm 0.22\,\mbox{GeV}^{-1}$} (see~Fig.~\ref{Xstheta0:plot}). On the other hand, $X_3$ is the one which suffers the biggest suppression ($\approx 27.5\%$ in modulo),  saturating now at \mbox{$-0.40\, \mbox{GeV}^{-2}$} instead of \mbox{$-0.51\, \mbox{GeV}^{-2}$}. Since the new set of $X_i$ is qualitatively very similar to the previous one, we will  omit the corresponding plots.

\begin{figure}[h]
\begin{minipage}[b]{0.45\linewidth}
\centering
\includegraphics[scale=0.35]{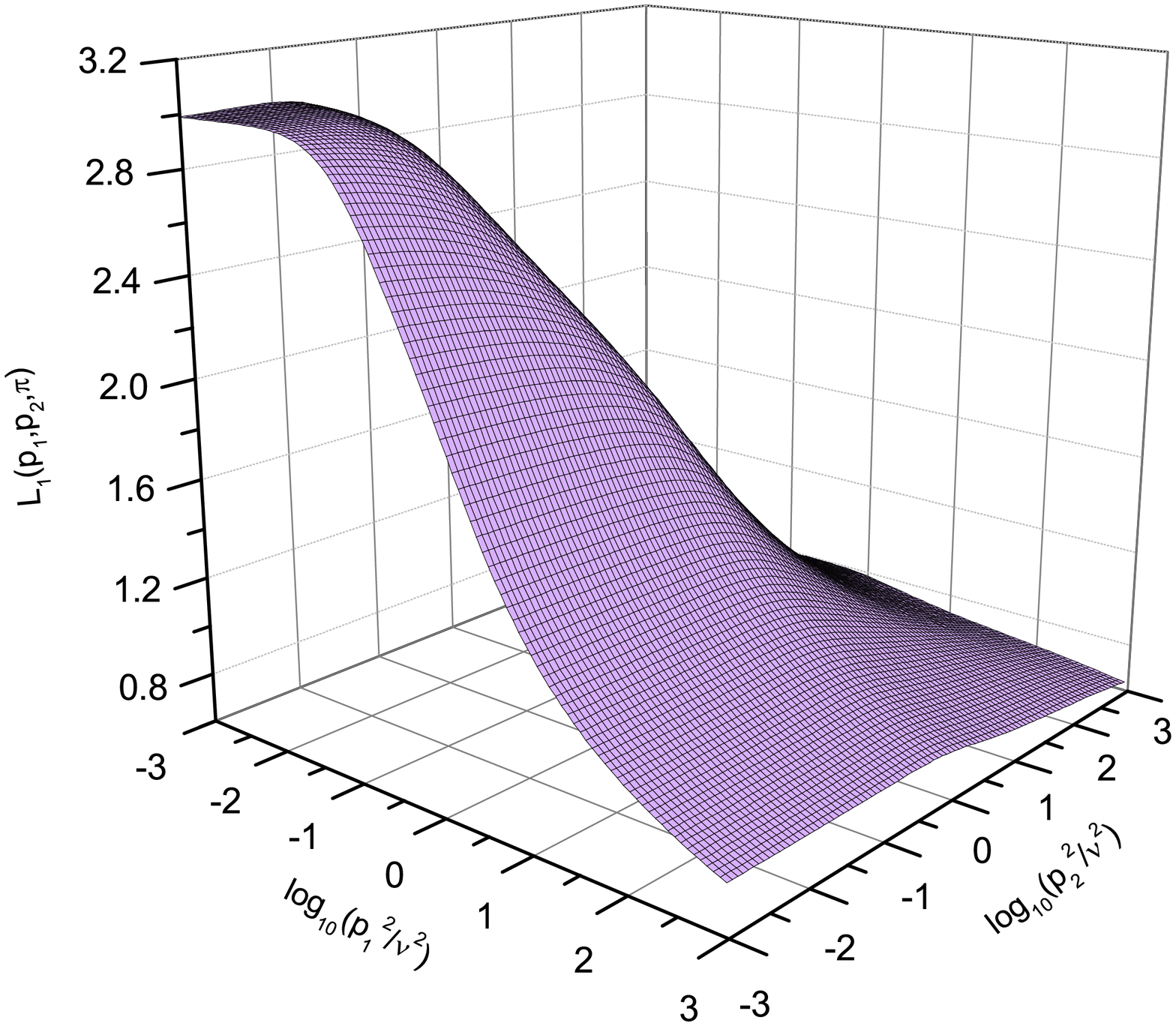}
\end{minipage}
\hspace{0.5cm}
\begin{minipage}[b]{0.50\linewidth}
\includegraphics[scale=0.35]{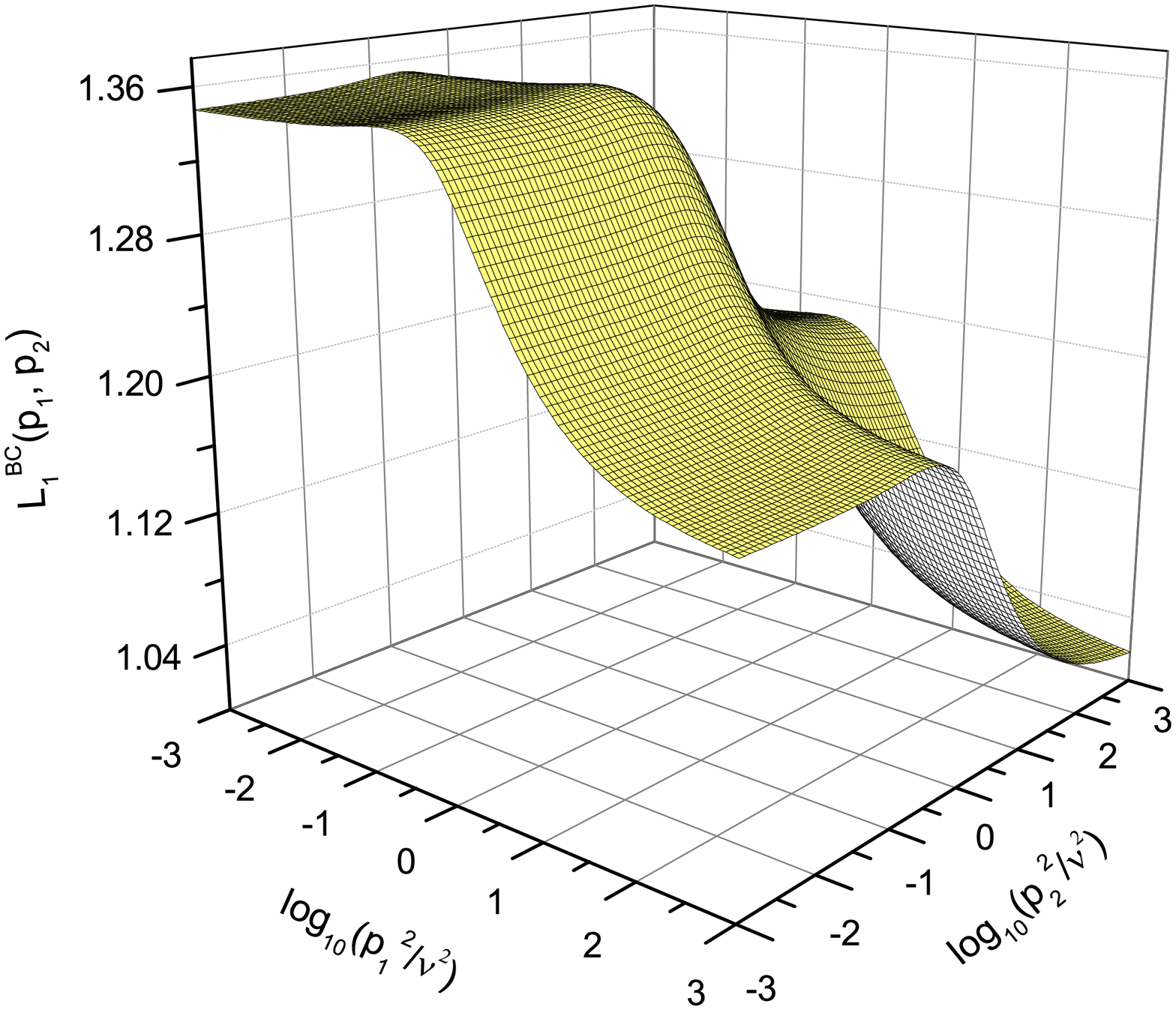}
\end{minipage}
\begin{minipage}[b]{0.45\linewidth}
\centering
\includegraphics[scale=0.35]{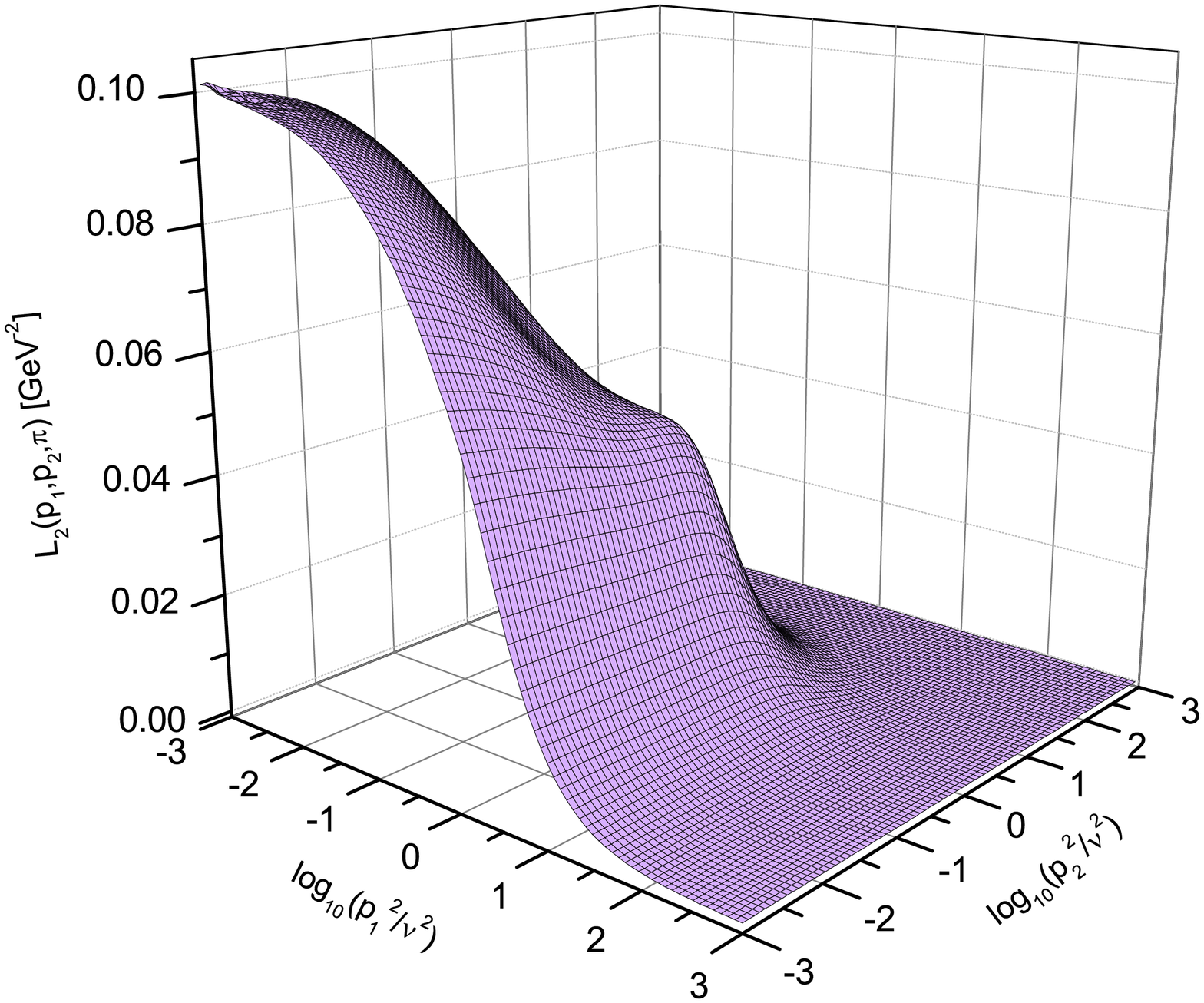}
\end{minipage}
\hspace{0.5cm}
\begin{minipage}[b]{0.50\linewidth}
\includegraphics[scale=0.35]{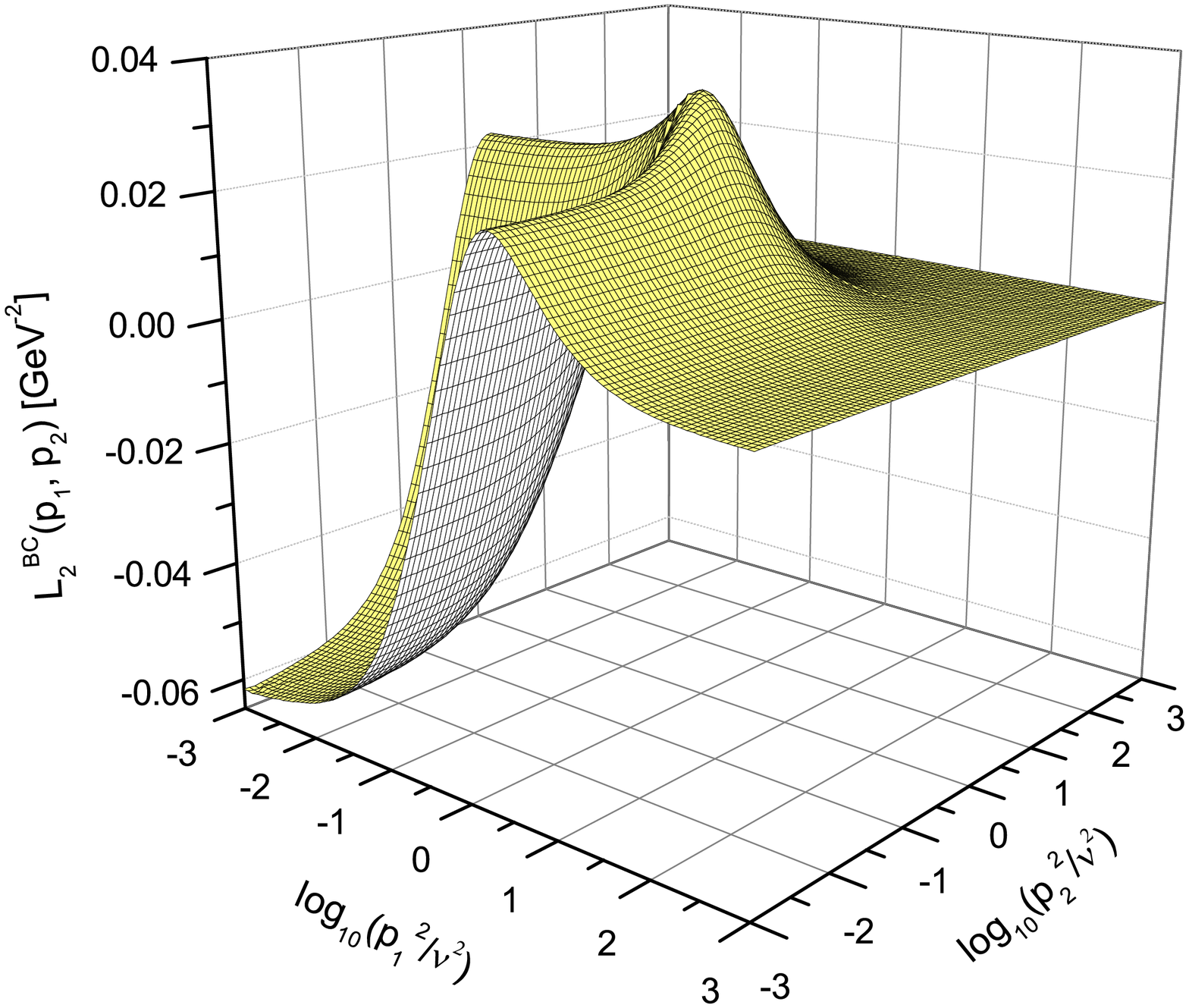}
\end{minipage}
\caption{\label{fig:vertexL12CP} The form factors $L_1$ and $L_2$ when $\theta=\pi$ (left panels) and the corresponding $L_1^{\rm{BC}}$ and $L_2^{\rm{BC}}$(right panels) for a higher value of the dynamical quark mass.}
\end{figure}

\begin{figure}[h]
\begin{minipage}[b]{0.45\linewidth}
\centering
\includegraphics[scale=0.35]{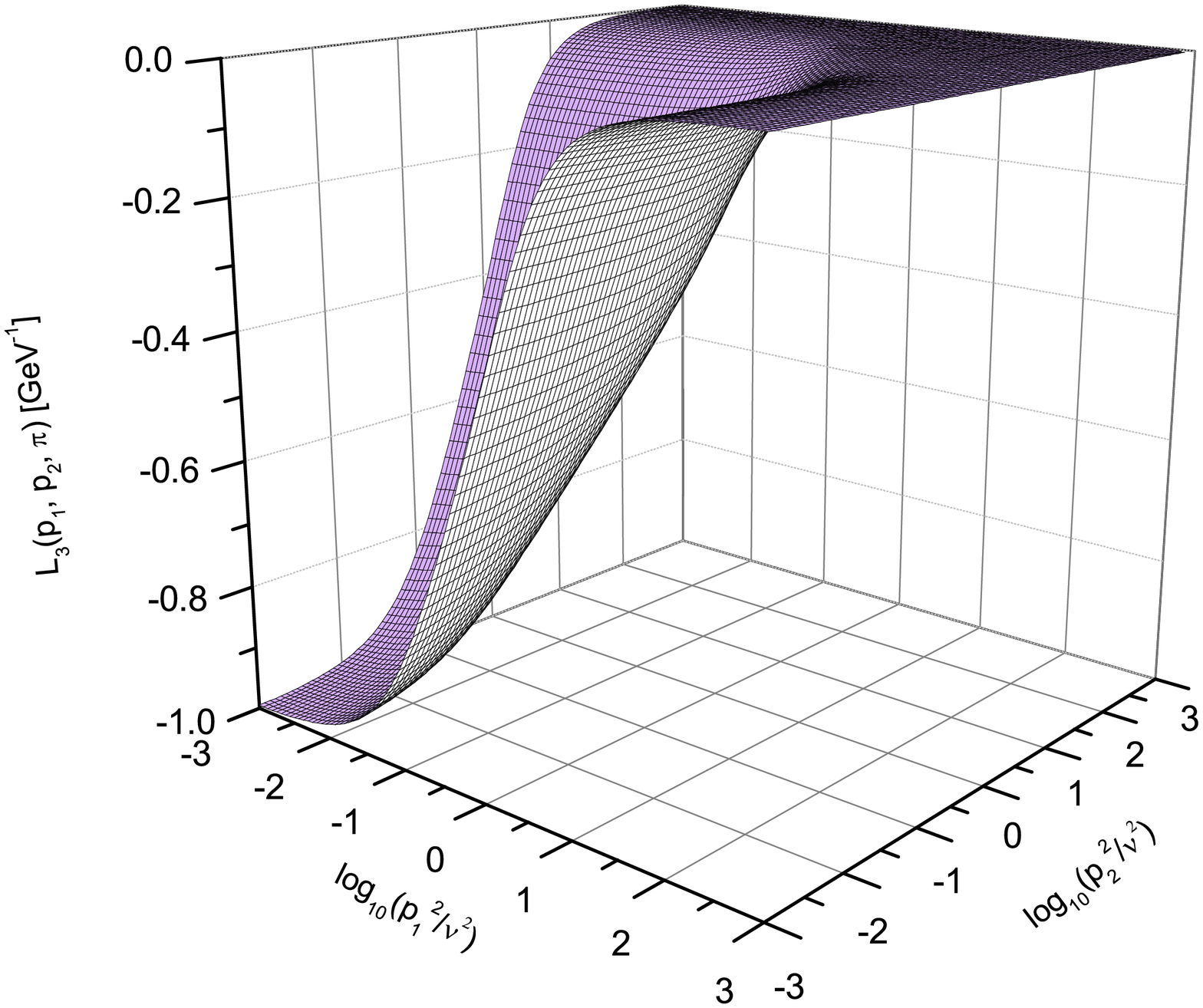}
\end{minipage}
\hspace{0.5cm}
\begin{minipage}[b]{0.50\linewidth}
\includegraphics[scale=0.35]{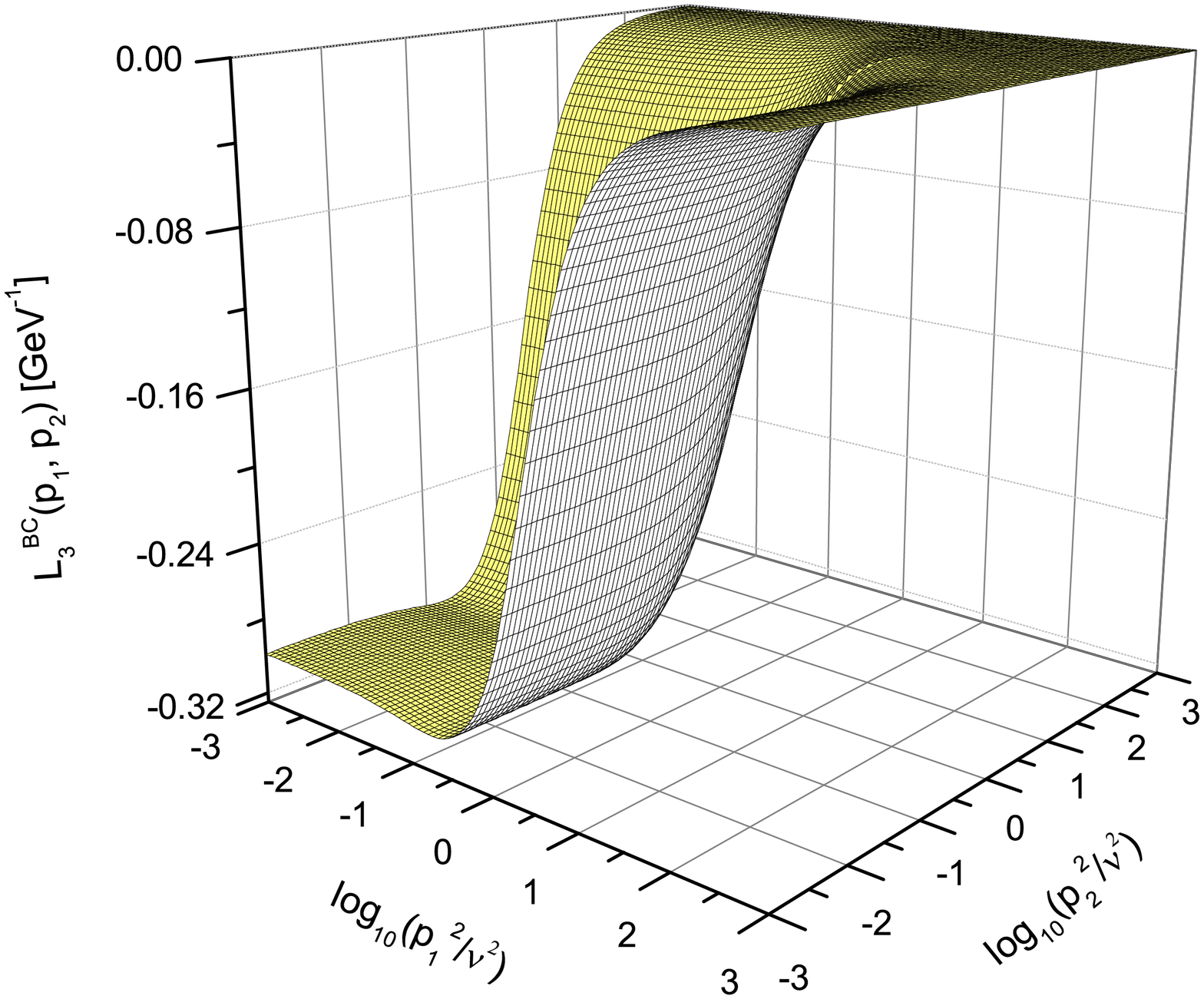}
\end{minipage}
\vspace{-1.25cm}
\caption{\label{fig:vertexL3CP} The form factors $L_3$ for $\theta=\pi$ (left panel) and the corresponding $L_3^{\rm{BC}}$ (right panel) for a higher value of the dynamical quark mass.}
\end{figure}

\begin{figure}[h]
\begin{minipage}[b]{0.45\linewidth}
\centering
\includegraphics[scale=0.4]{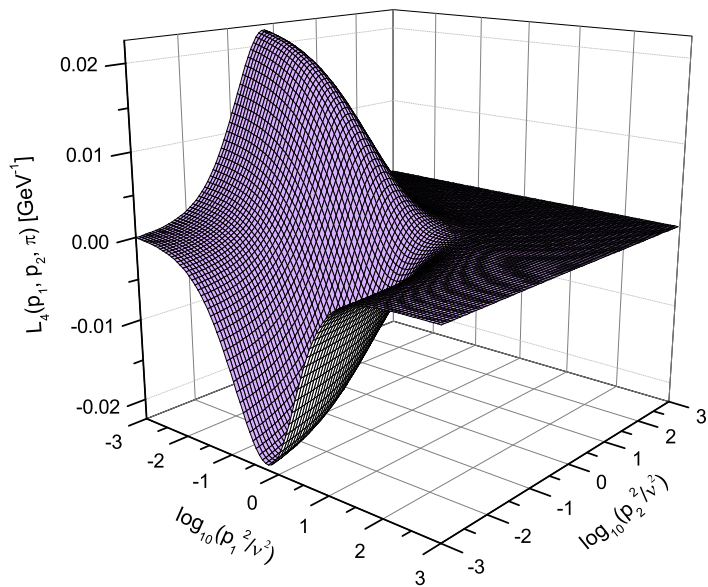}
\end{minipage}
\vspace{-0.6cm}
\caption{\label{fig:vertexL4CP} The form factor $L_4$ 
when $\theta=\pi$ for a higher value of the dynamical quark mass.}
\end{figure}

The next step is the determination of the $L_i$ from Eq.~\eqref{expLi}; the results for $\theta=\pi$ are shown 
in the left panels of Fig.~\ref{fig:vertexL12CP} and~\ref{fig:vertexL3CP}. As before, 
in order to expose the non-Abelian content of these form factors, on the right panels 
we plot their Abelian counterparts, $L_i^{BC}$. Moreover, in  Fig.~\ref{fig:vertexL4CP} 
we show the result for $L_4$  alone, since $L_4^{BC}=0$. 
The comparison of the results in Figs.~\ref{fig:vertexL12CP},~\ref{fig:vertexL3CP} and~\ref{fig:vertexL4CP}  with those presented  previously 
in the sequence of Figs.~\ref{fig:vertexL1},~\ref{fig:vertexL2},~\ref{fig:vertexL3}, and~\ref{fig:vertexL4}, allows us to make the following remarks.

\n{i} The results for $L_1$, $L_3$ and $L_4$ are qualitatively rather similar, and do not seem 
especially sensitive to  the particular shape of $A(p^2)$ and $B(p^2)$, nor to the amount of dynamical quark mass generated. 

\n{ii} Instead, $L_2$ changes completely its shape in the infrared,  displaying  a  structure which is not so smoother compared  to that of Fig.~\ref{fig:vertexL2}. We notice that, when $A(p^2)$ is nearly monotonic in the infrared, $L_2$ reverses its sign and becomes positive in the entire range of momenta,  saturating in the infrared  around  \mbox{$0.1\,\mbox{GeV}^{-2}$} [see left bottom panel of Figs.~\ref{fig:vertexL12CP}]. This rather abrupt change may be traced back to   
the fact that,  $A(p^2)$ enters in the expression for $L_2$  multiplied by a particular
combination of $X_0$ and $X_3$ [see Eq.~\eqref{expLi}]; 
in the limit of $p_1\to p_2$, this combination reduces to a derivative-like term. Since the corrections of both $X_0$ and $X_3$ are of the same order, any change in their infrared values may lead to an oscillation in the sign of $L_2$. It is interesting to notice that $L_3$ also contains a similar term [see Eq.~\eqref{expLi}]; however, in this case, the $A(p^2)$ is  multiplied by a combination of $X_1$ and $X_2$, which tends to be very small in the infrared region, furnishing a subleading contribution to the overall shape of $L_3$.

\n{iii} Finally, as one might have intuitively expected, the Abelian form factors $L_i^{BC}$ are significantly more sensitive  
to the precise functional forms of $A(p^2)$ and $B(p^2)$. In particular, 
we notice that  $L_1^{BC}$ of Fig.~\ref{fig:vertexL12CP} displays a much smoother 
behavior when compared with the one plotted in Fig.~\ref{fig:vertexL1}. Evidently, this is a direct consequence of 
having switched to a nearly monotonic $A(p^2)$, since the 
planes where either $p_1=0$ or $p_2=0$ in the $L_1^{BC}$ of Fig.~\ref{fig:vertexL12CP}  should
reproduce (by construction) the same functional form of $A(p^2)$ but shifted and multiplied by a constant, {\it i.e.}  $A(p^2)/2 + A(0)/2$.

\section{\label{special} Some special kinematic limits: 2-D plots}

In this section  we concentrate  on the determination of the form factors $L_i$ in five  different kinematic limits. 
The cases discussed will be the following: 

{\it(i)} The  soft gluon limit,  which is defined when both momenta $p_1$ and $p_2$ 
have the same magnitude {\it i.e.} \mbox{$|p_1|=|p_2|=|p|$} and 
the angle between them is $\theta=0$;

{\it(ii)}  The quark symmetric configuration, where again the two momenta have the same magnitude, \mbox{$|p_1|=|p_2|=|p|$}, 
but now they are anti-parallel {\it i.e.}, $\theta=\pi$; 

{\it(iii)} The soft anti-quark limit  
obtained when the anti-quark momentum vanishes, {\it i.e.} $p_1\to 0$ and $p_2\to p$; 

{\it(iv)} The soft quark limit where $p_2\to 0$ and $p_1\to p$ ; 

{\it(v)} The totally symmetric limit defined when the square of the three momenta of vertex are all equal, {\it i.e.} \mbox{$p_1^2=p_2^2=q^2=p^2$} and the angle $\theta=2\pi/3$. 

Evidently, in all configurations listed above, 
the $L_i$s  become  functions of a single momentum variable, namely $p$. We will refer to the $L_i$s in each configuration as $L_i^{g}(p)$, $L_i^{sym}(p)$, $L_i^{\overline{q}}(p)$, $L_i^{q}(p)$ and $L_i^{\s{\rm TS}}(p)$, respectively.

\subsection{Special kinematic limits}

The determination of $L_i$ in any of the above kinematic configurations may be performed in two different ways.  
The first is to consider the limit of interest directly at the level of Eq.~\eqref{generalx}, 
and then use the results obtained in  Eq.~\eqref{expLi}. This particular procedure requires certain care,   
due to the presence of the function $h$ in the various denominators. Specifically, in Euclidean space 
\be
h = p_1^2p_2^2 \sin^2 \theta\,,
\label{hfunctionEuc}
\ee
which vanishes when implementing the limits defining the cases {\it(i)}-{\it(iv)}. 
Therefore, the numerators of the corresponding expressions 
in Eq.~\eqref{generalx}  must be appropriately expanded, and the potentially divergent terms  
explicitly canceled, by virtue of the exact vanishing of certain angular integrals. 
A detailed description of this procedure for the cases {\it(i)} and {\it(ii)}  will  be outlined in the Appendix~\ref{app:taylor}.
The second way is to exploit directly the numerical results 
obtained  for general configurations, since all special kinematic limits  constitute particular projections (``slices'') 
of the 3-D results. Evidently, the results obtained with both methods ought to coincide.

\begin{figure}[t]
\begin{minipage}[b]{0.45\linewidth}
\centering
\includegraphics[scale=0.32]{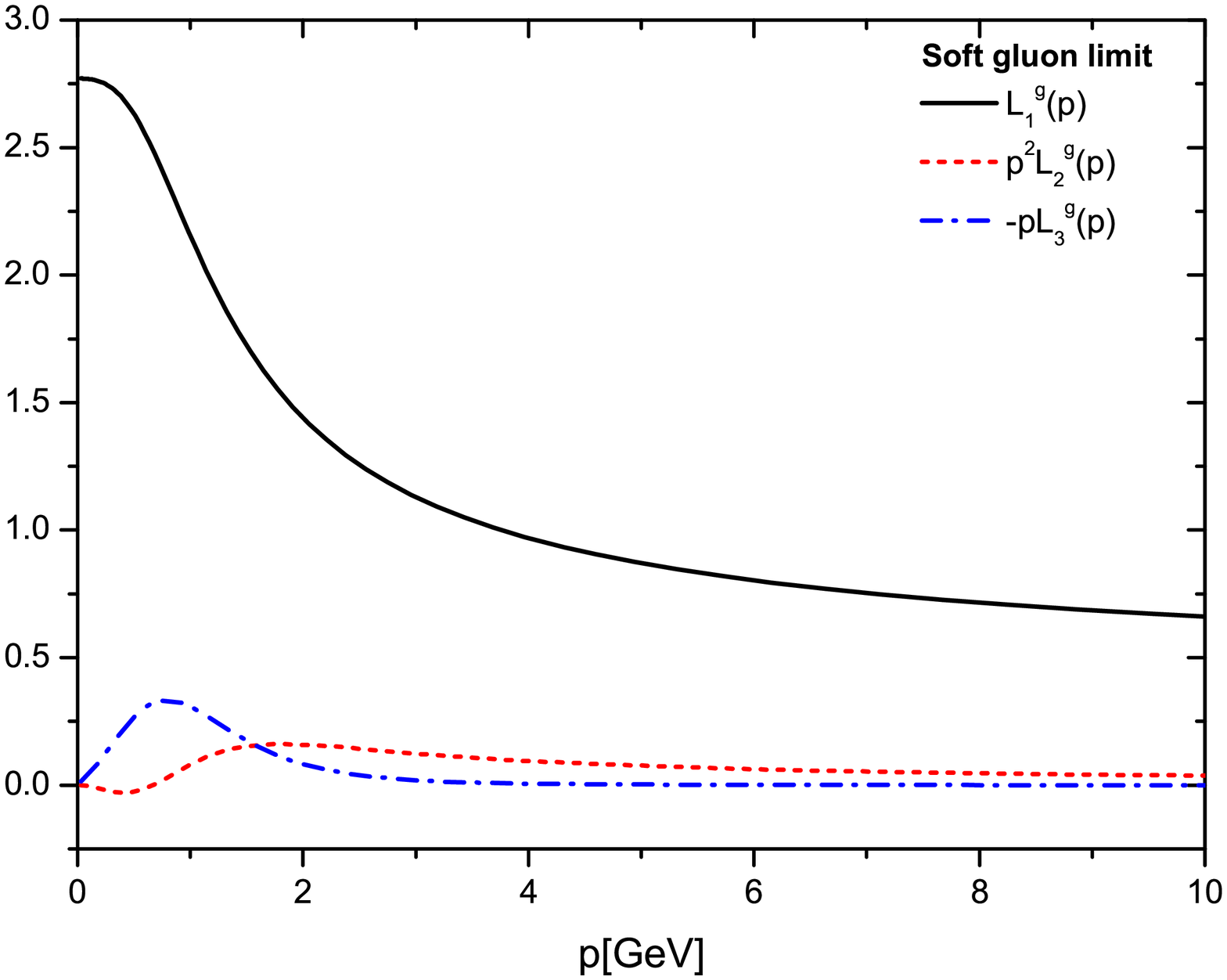}
\end{minipage}
\hspace{0.5cm}
\begin{minipage}[b]{0.50\linewidth}
\includegraphics[scale=0.32]{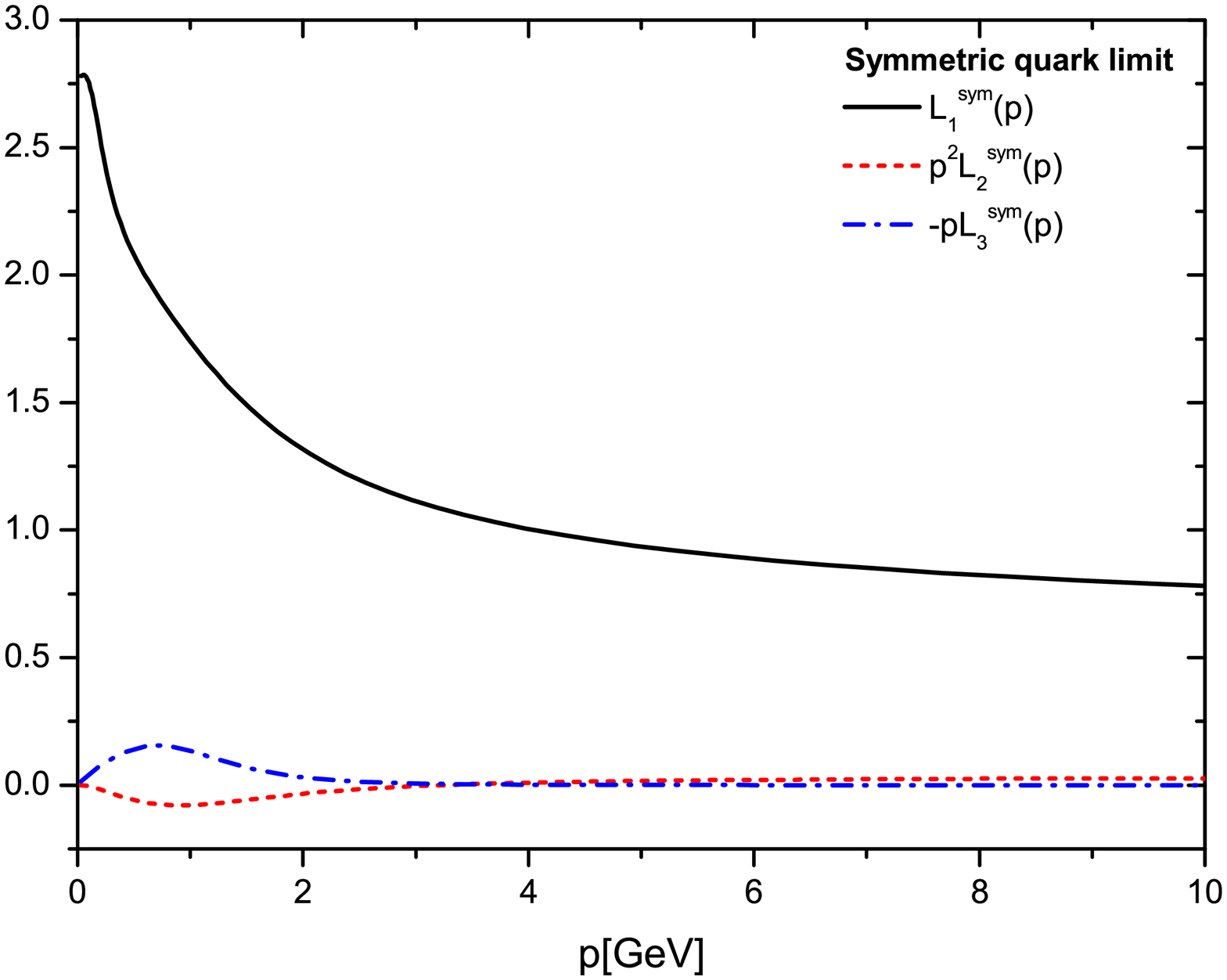}
\end{minipage}
\vspace{0.5cm}\begin{minipage}[b]{0.45\linewidth}
\centering
\vspace{0.2cm}
\includegraphics[scale=0.32]{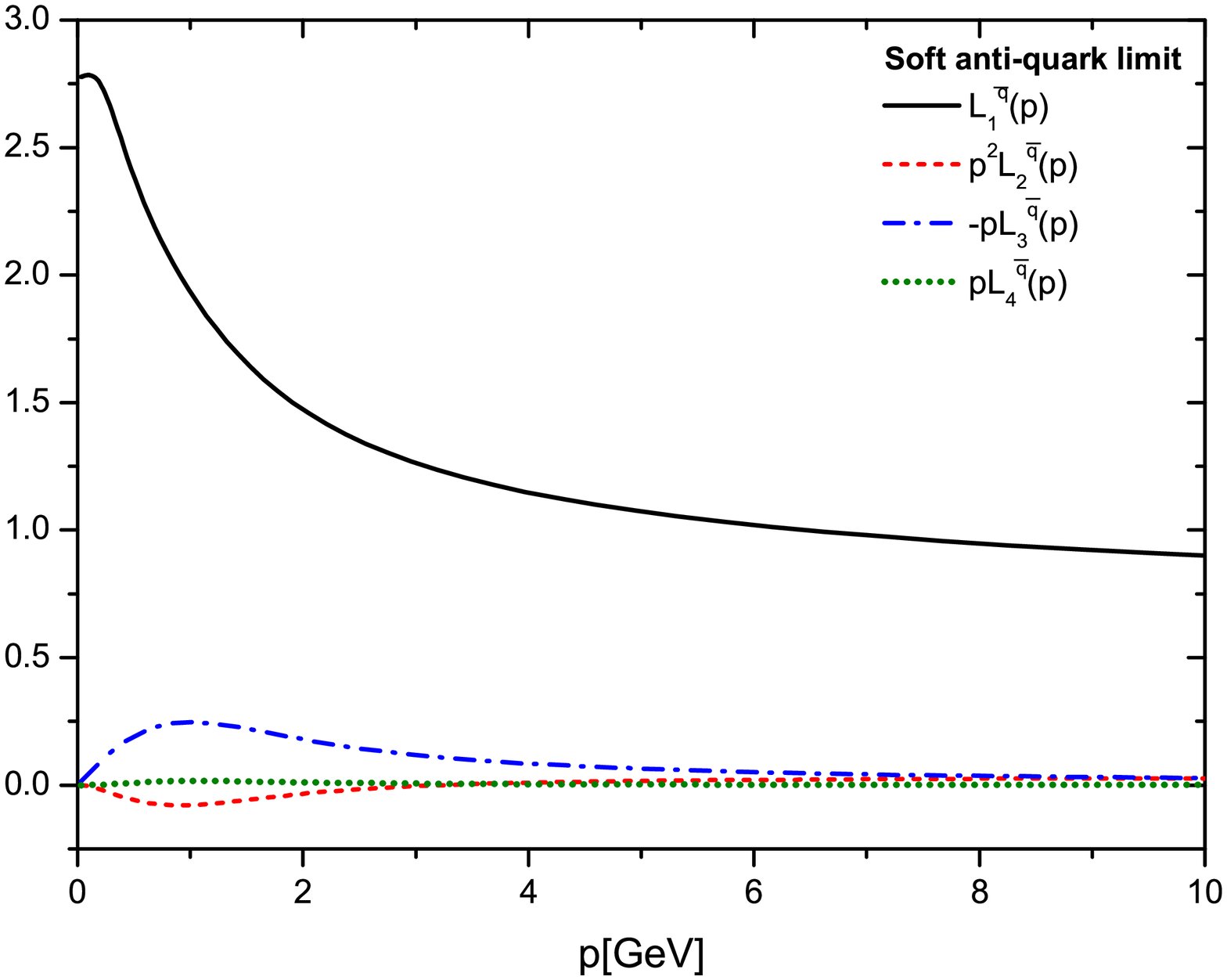}
\end{minipage}
\hspace{0.5cm}
\begin{minipage}[b]{0.50\linewidth}
\includegraphics[scale=0.32]{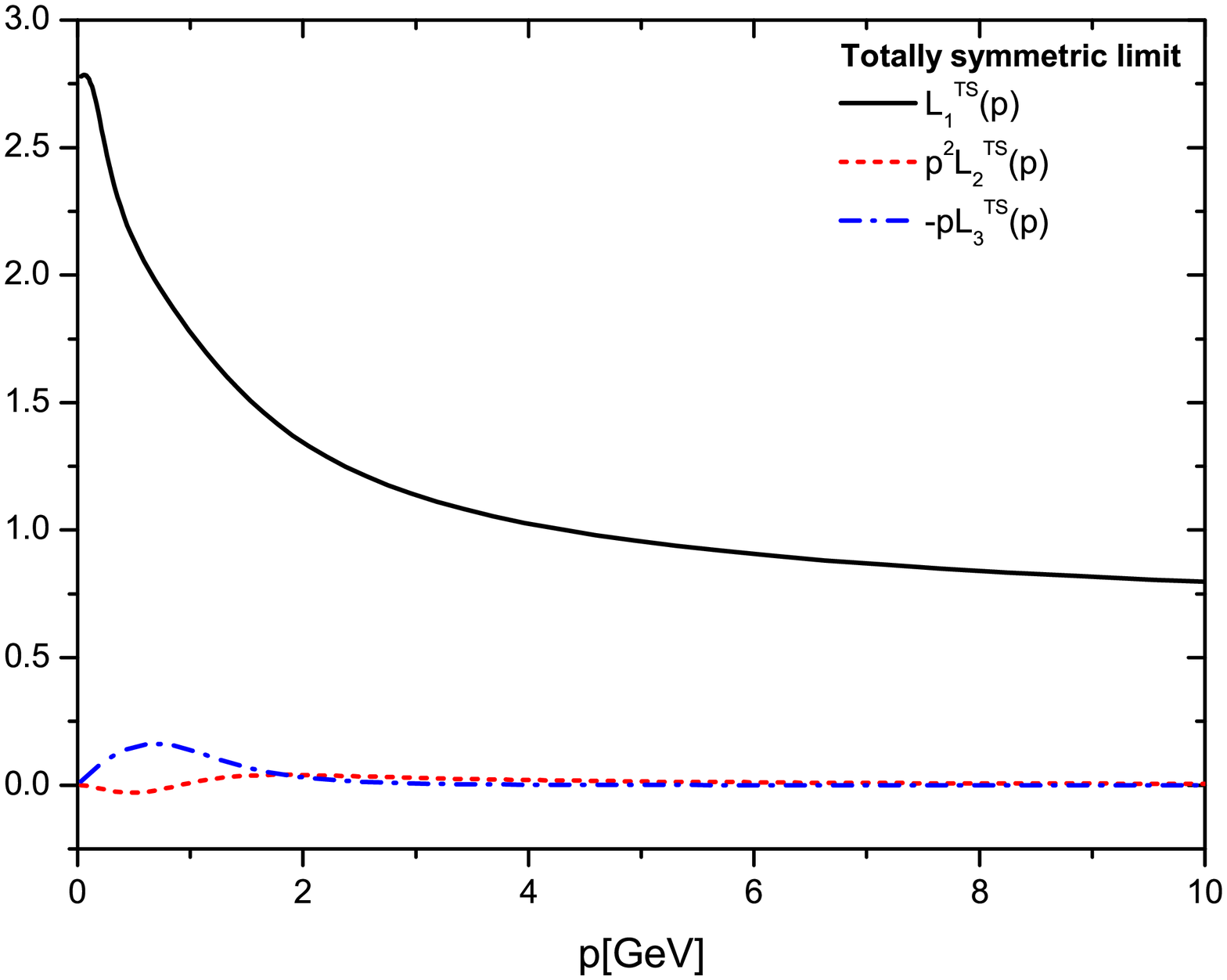}
\end{minipage}
\vspace{-1.5cm}
\caption{\label{fig:Lsoft}  The form factors $L_i$s for different kinematic configurations. The $L_i^{g}$ (top left panel) are the form factors in the soft gluon configuration. The  $L_i^{sym}$ (top right panel) 
represents the quark symmetric case. The 
$L_i^{\overline{q}}$  (or $L_i^{q}$) (bottom left panel) is
 in the soft anti-quark (or quark) limit, whereas  $L_i^{\s{\rm TS}}$(bottom right  panel) is the totally symmetric configuration.}
\end{figure}

For example, the plane defined by the equation  $p_1=p_2$ in the left panel of Fig.~\ref{fig:vertexL1angle0}  
corresponds to the slice that defines $L_1^{g}(p)$, since  the angle in this figure is fixed at $\theta=0$.  
This particular slice  was isolated in the top left panel of Fig.~\ref{fig:Lsoft}  and it is represented by the black continuous line.

  As we can see, $L_1^{g}(p)$ displays a smooth behavior, decreasing monotonically towards the ultraviolet region.  It is important to stress
  that the fact that  \mbox{$L_1^{g}(\mu)\neq 1$}  is not in contradiction with the renormalization condition employed in the calculation [see  
 Eq.~\eqref{cond_mom}],  which  ensures that the $L_1$ will be equal to the unity in the totally symmetric point, defined when  \mbox{$p_1^2=p_2^2=q^2=\mu^2$}.

Now, following the same procedure outlined before, we can extract the other non-vanishing form factors,
 namely $L_2^{g}(p)$ and $L_3^{g}(p)$. For the purpose of comparison, we also 
plot in the top left panel of Fig.~\ref{fig:Lsoft} the dimensionless combinations
$p^2L_2^{g}(p)$ (red dashed line) and  $-pL_3^{g}(p)$ (blue dash-dotted line).
We clearly see that both  vanish in the infrared limit, and they are evidently much more suppressed than $L_1^{g}(p)$. 

Next, we turn to the quark symmetric configuration. In the top right panel of the Fig.~\ref{fig:Lsoft}, we show the various  $L_i^{sym}(p)$. In particular, the projection of $L_1^{sym}(p)$ (black continuous line)  corresponds to the slice defined by the plane $p_1=p_2$ of the right panel of Fig.~\ref{fig:vertex_mix} ($\theta=\pi$). Even though 
the tensorial structures $\lambda_2^{\mu}$, $\lambda_3^{\mu}$ and $\lambda_4^{\mu}$ defined in Eq.~\eqref{tens_long} vanish in the 
quark symmetric limit, the form factors $L_2^{sym}(p)$ and $L_3^{sym}(p)$ are nonvanishing.
For this reason, we show in the same plot $p^2L_2^{sym}(p)$ (red dashed) and  $-pL_3^{sym}(p)$ (blue dash-dotted); again,
both  quantities are rather suppressed when compared to  $L_1^{sym}(p)$.

The next quantities of interest are the $L_i^{\overline{q}}(p)$, shown in the bottom left panel of Fig.~\ref{fig:Lsoft}. 
Note that, as can be easily confirmed by means of an analytic  derivation, $L_i^{\overline{q}}(p)$  is independent of the angle $\theta$. 
For this reason, when we select the plane  where $p_1=0$ (for practical purposes $p_1\approx 30$ MeV),  in  any of the plots shown in the Figs.~\ref{fig:vertexL1},~\ref{fig:vertexL2},~\ref{fig:vertexL3} and~\ref{fig:vertexL4}, we obtain the same result for all $L_i^{\overline{q}}(p)$, respectively.

Turning to $L_i^{q}(p)$, note that when we combine Eq.~\eqref{sym}  with Eqs.~\eqref{sum_long} and \eqref{expLi}, one concludes that \mbox{$L_i^{\overline{q}}(p)=L_i^{q}(p)$} with $i=1,2,3$, 
while \mbox{$L_4^{\overline{q}}(p)=-L_4^{q}(p)$}; this happens because the first three tensorial structures 
in Eq.~\eqref{sum_long} are symmetric under $p_1 \leftrightarrow p_2$, while the fourth is antisymmetric.
Therefore, the numerical results for $L_1^{q}(p)$, $p^2L_2^{q}(p)$ and $-pL_3^{q}(p)$  coincide with those shown in  the bottom left panel of the Fig.~\ref{fig:Lsoft}, except for  $pL_4^{q}(p)$ (green dotted line), which reverses its sign.

Finally, $L_i^{\s{\rm TS}}(p)$  is obtained by selecting the plane where \mbox{$p_1^2=p_2^2$} in the 3-D plots where $\theta=2\pi/3$, such as the 
one for $L_1$ shown on the left panel of Fig.~\ref{fig:vertex_mix}.  
The results of these projections  are shown in the bottom right panel of Fig.~\ref{fig:Lsoft}. 

It is interesting to notice that all $L_i$ display a very similar pattern. More specifically, all $L_1$  have narrow peaks of similar size 
located in the region of a few MeV, and then decrease logarithmically in the ultraviolet, whereas $p^2L_2$ and $-pL_3$ are rather suppressed quantities,
vanishing in both the infrared and ultraviolet limits.

\subsection{\label{closer}A closer look at the soft gluon limit}

\begin{figure}[t]
\includegraphics[scale=0.7]{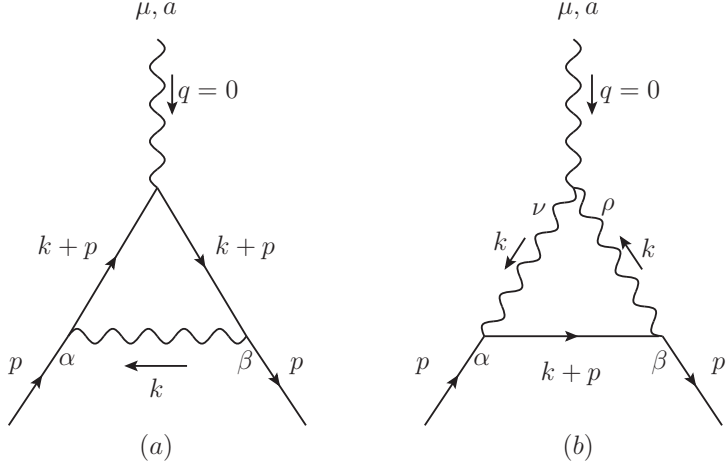}
\vspace{-0.5cm}
\caption{\label{fig:vertper} The one-loop diagrams of the quark-gluon vertex
in the soft gluon limit.}
\end{figure}

We will next consider the one-loop result for  $L_1^{g}(p)$, which will furnish
some additional insights on the asymptotic (ultraviolet) behavior of this form factor, 
shown in the top left panel of Fig.~\ref{fig:Lsoft}.  The  
derivation of the one-loop expression for $L_1^{g}(p)$
may proceed in two different ways:
the first is based on the direct calculation of the one-loop  diagrams shown in Fig.~\ref{fig:vertper}, for this particular kinematics;
the second consists of substituting 
one-loop results for the ingredients entering in the all-order relation captured by 
Eq.~\eqref{expLi}. Evidently, the answers obtained with either method ought to coincide.

Following~\cite{Davydychev:2000rt}, we employ  dimensional regularization
for the direct one loop calculation; specifically, the measure of Eq.~\eqref{dqd} becomes
\be
\int_{l} \rightarrow \frac{\mu^{\epsilon}}{(2\pi)^{d}}\!\int\!\mathrm{d}^d l,
\label{dimreg}
\ee
where $d=4-\epsilon$ and $\mu$ is the 't Hooft mass scale.

It is then relatively straightforward to demonstrate that 
in the Landau gauge the ``abelian''  diagram (a) vanishes exactly in the soft gluon limit
(before renormalization)~\cite{Davydychev:2000rt}; note that the derivation of this result requires
the use of dimensional regularization 
formulas such as $\gamma_{\mu}\gamma_{\nu} \gamma^{\mu} = (2-d) \gamma_{\nu}$ for the corresponding 
``Diracology'', instead of the usual $\gamma_{\mu}\gamma_{\nu} \gamma^{\mu} = -2 \gamma_{\nu}$. 
The evaluation of diagram (b) yields (in Euclidean space)  
\begin{eqnarray}
L_{1\,{\rm pert}}^{g}(p) =1 +  \frac{C_{\rm A}\alpha_s}{16\pi}\left[ -3\ln\left(\frac{p^2+m_0^2}{\mu^2+m_0^2}\right)
 +\frac{m_0^4}{p^4}\ln\left(\frac{p^2+ m_0^2}{m_0^2}\right) \right. \\ \nonumber
 \left. -\frac{m_0^4}{\mu^4}\ln\left(\frac{\mu^2+ m_0^2}{m_0^2}\right) 
  -m_0^2 \left(\frac{1}{p^2} -\frac{1}{\mu^2}\right)\right] \,,
\label{pertvg}
\end{eqnarray}
where $m_0$ is the current quark mass, which guarantees the infrared finiteness of the 
result. In addition, note that the above expression was renormalized using the MOM scheme, imposing the condition $L_{1\,{\rm pert}}^{g}(\mu^2)=1$. 
Turning to the second way, the 
substitution of \mbox{$p_1=p_2=p$} ($q=0$)
into Eq.~\eqref{expLi}, and the use of the fact that $X_i=\overline{X}_i$, $X_0=1$ and $X_1=-X_2$ in this limit, 
yields (in Euclidean space)
\begin{align}
L_1^{g}(p) &= F(0)\left[A(p)(1+ 2p^2X_3(p)) - 2B(p)X_1(p)\right] \,.
\label{sglimit}
\end{align}

The determination of  $L_{1\,{\rm pert}}^{g}(p)$  from Eq.~\eqref{sglimit} is rather 
subtle, and involves the exact cancellation of two infrared divergent contributions
stemming from two of the ingredients appearing in it.   
We start by recalling that $F$ at one-loop level is given by
\begin{align}
F^{\rm pert}(q^2) =  1 - \frac{3C_{\rm A}\alpha_s}{16\pi}\ln\left(\frac{q^2}{\mu^2}\right) \,,
\label{dressing}
\end{align}  
which clearly displays an infrared divergence of the type ``$\ln 0$'' when $q^2\to 0$, due to the presence of the ``unprotected'' 
logarithm; of course, nonperturbatively the infrared divergence of this logarithm is known to be tamed, and a finite
value for $F(0)$ is obtained (see, e.g.,~\cite{Bogolubsky:2007ud}).
Since $L_{1\,{\rm pert}}^{g}(p)$  displays no such divergence,
an exact cancellation against a similar contribution must take place. To establish the precise mechanism that makes this happen,
we set in Eq.~\eqref{soft_gluon1} $p_1=p_2=p$, $\theta=0$,
and the tree-level expressions $A(p)=1$, $B(p)=m_0$, $F(q)=1$ and $\Delta(q)=1/q^2$, and after restoring the measure we obtain   
 \begin{align}
 X_{1}^{\rm pert}(p) &= i\frac{C_{\rm A}g^2}{6}
\int_k \left[ 2+ \frac{(k\cdot p)^2}{k^2p^2}\right]\frac{m_0}{k^4[(k+p)^2-m^2_0]} 
  \,, \nonumber\\
p^2X_{3}^{\rm pert}(p) &= -i\frac{C_{\rm A}g^2}{6}
\int_k \left[ 2p^2+ 3(k\cdot p) + \frac{(k\cdot p)^2}{k^2} \right]\frac{1}{k^4[(k+p)^2-m^2_0]}\,.
\label{x3pert}
\end{align}  

 A systematic inspection of the above terms reveals that the only source of such a divergent contribution  is the third term
in $p^2X_{3}^{\rm pert}(p)$. Focusing on this term and applying Feynman parametrization, 
one may eventually identify a contribution of the form
$\frac{3C_{\rm A}\alpha_s}{16\pi} \int_0^1 dx/x$, which allows for the necessary cancellation to go through.
As it should, the final result coincides with that of Eq.~\eqref{pertvg}.

\begin{figure}[t]
\includegraphics[scale=0.32]{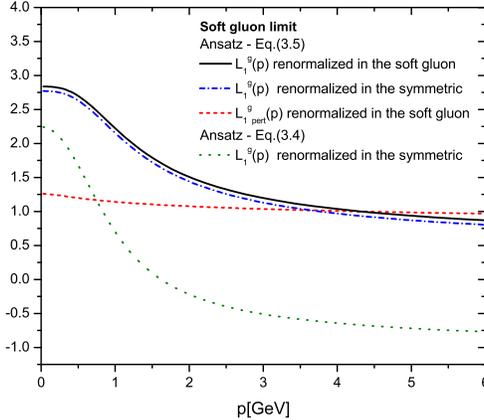}
\vspace{-0.75cm}
\caption{\label{fig:L1pert}
Comparison of the one-loop result $L_{1\,\mbox{pert}}^{g}(p)$ (red dashed), given in Eq.~\eqref{pertvg} with  $L_1^{g}(p)$  obtained with the Ansatz given by Eq.~\eqref{vertex1} and renormalized  in the two different configurations: (i) soft gluon configuration, {\it i.e.}  $L_1^{g}(\mu)=1$ (black continuous) and (ii)  totally symmetric configuration given by Eq.~\eqref{cond_mom} (blue dash-dotted line). In addition, we show the $L_1^{g}(p)$ renormalized in the totally symmetric configuration  obtained with the Ansatz given by Eq.~\eqref{vertex2} (green dotted line).}    
\end{figure}

In Fig.~\ref{fig:L1pert} we compare our nonperturbative result for $L_1^{g}(p)$ (black continuous line) with the one-loop expression of Eq.~\eqref{pertvg} (red dashed line). Notice that,  in order to perform a meaningful comparison, we renormalize both curves imposing  the same renormalization condition, {\it i.e.}, \mbox{$L_1^{g}(\mu)=L_{1\,{\rm pert}}^{g}(\mu)=1$}.  Clearly, we see a  qualitative agreement
between both curves for large values of $p$. Evidently, the small deviations between these curves in the ultraviolet is expected, and
can be attributed to the higher order loop corrections that $L_1^{g}(p)$ contains in it.       
   
We also show in the Fig.~\ref{fig:L1pert}  $L_1^{g}(p)$ renormalized
in the totally symmetric configuration (blue dash-dotted line), which satisfies
the renormalization condition imposed by Eq.~\eqref{cond_mom}. This curve is the same one (black continuous) shown in the top left panel of Fig.~\ref{fig:Lsoft}. Therefore, the small quantitative difference between the black continuous
and the blue dash-dotted  curves shown in the  Fig.~\ref{fig:L1pert} is merely the effect of the  imposition of different renormalization conditions.

  In order to expose how the Ansatz of Eq.~\eqref{vertex2}  affects dramatically the perturbative behavior of the soft gluon configuration, in Fig.~\ref{fig:L1pert} we show the result for $L_1^{g}(p)$  computed with Eq.~\eqref{vertex2} (green dotted line).  Notice that this curve corresponds to the ``slit'' (slice where $p_1=p_2$) shown in the right panel of Fig.~\ref{fig:vertexL1angle0}. Since the result
  in Fig.~\ref{fig:vertexL1angle0}  is renormalized in the totally symmetric configuration, the 
  (green) dotted curve should be compared with the (blue) dash-dotted line, which was obtained employing the  Ansatz of the Eq.~\eqref{vertex1}. The sizable deviation of the  green dotted line from the expected perturbative region, where all the others curves are located, clearly indicates that the Ansatz of Eq.~\eqref{vertex2} is not an  appropriate choice. Evidently, the use of 
  Eq.~\eqref{vertex2} provides excessive strength in the single  component of the vertex considered;  
  in fact, as we notice 
  in Fig.~\ref{fig:comp_fbc}, the combination 
  $F(q)L_1^{\rm{BC}}(p_1,p_2)$ is indeed more enhanced than the  solution
  for the complete $L_1(p_1,p_2,\theta)$.

\begin{figure}[t]
\includegraphics[scale=0.32]{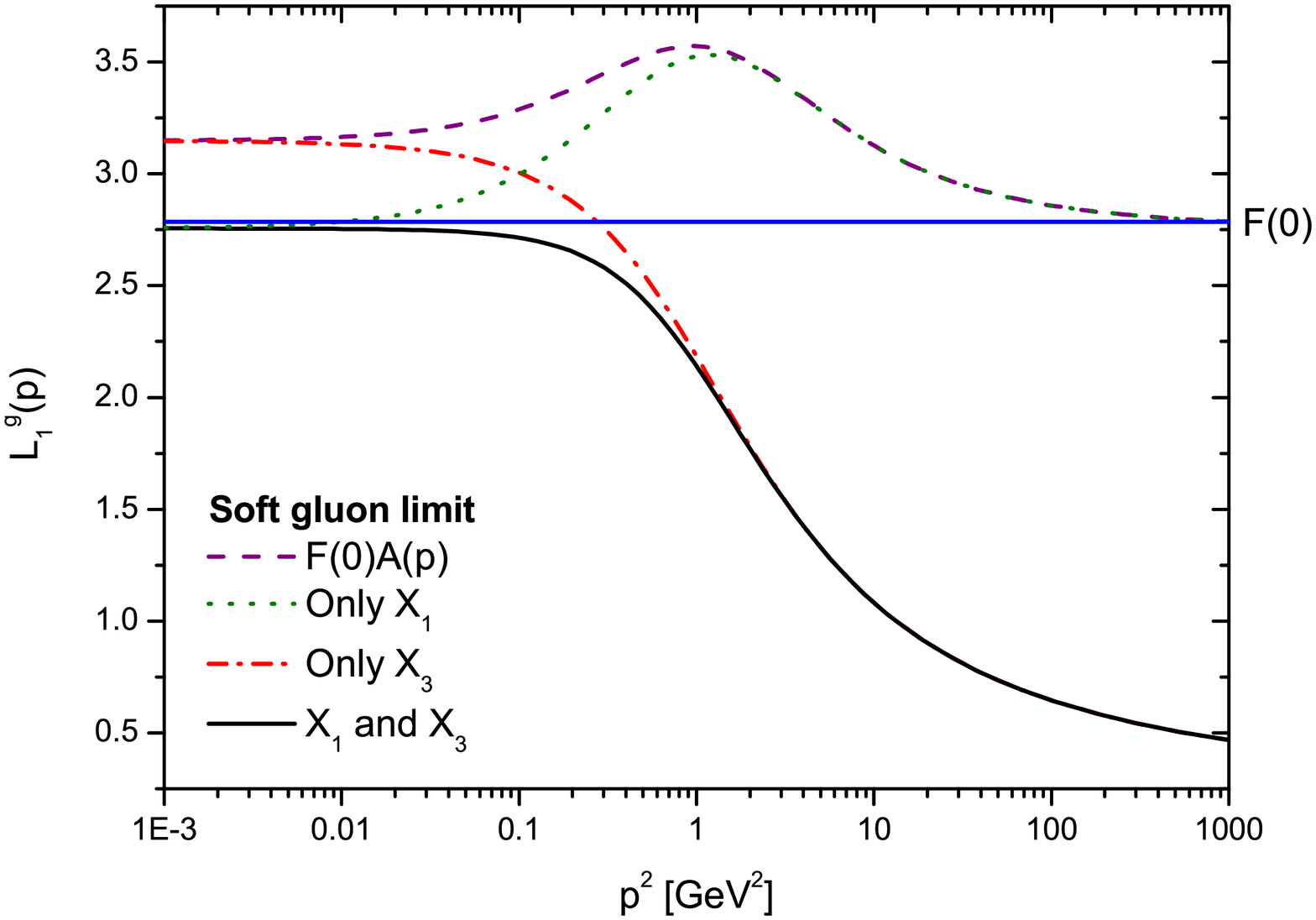}
\vspace{-0.75cm}
\caption{\label{fig:Lsoft1}  The individual contributions entering
into the definition of $L_1^{g}(p)$, given in Eq.~\eqref{sglimit}.}
\end{figure}
 
For completeness, in Fig.~\ref{fig:Lsoft1},  we show the contribution of each of the individual  terms
appearing in the Eq.~\eqref{sglimit}. We start showing the contribution of the first term which corresponds to the ``minimal'' non-Abelian BC vertex (purple dashed curve), $F(0)A(p)$. In the plot, we clearly see that when we neglect all  the $X_i$ contributions, $L_1^{g}(p)\to F(0)$ when ${p\to \infty}$.  Now, let us turn on the 
contribution of the  $X_1$ (green dotted line); given that $X_1$ is multiplied by $B(p)$,  it is clear that
it  can only modify the infrared and intermediate regions of $L_1^{g}(p)$, leaving the ultraviolet behavior intact.
Focusing on the most interesting term,  namely $X_3$, we first note that, since it is multiplied by $p^2$,  
it does not change the infrared limit (red long-dashed dotted line). However, since $X_3$
is negative and quite sizable (see bottom right panel of Fig.~\ref{Xstheta0:plot}), it has a considerable
influence on the intermediate and ultraviolet regions of  $L_1^{g}(p)$, producing a notable decrease in its behavior,
as shown by the case where all contributions of  Eq.~\eqref{sglimit} are taken into account (black continuous curve).

As an additional remark in  this subsection,  we emphasize that  the numerical
calculation of the full $X_3$ when $\theta\approx 0$ and $p_1^2\approx
p_2^2$ expressed  by Eq.~\eqref{soft_gluon1} is rather  delicate. More
specifically, when one fixes the values of the  external momenta $p_1^2$
and $p_2^2$ at infrared or intermediate scales (order of $10^{-3} - 1 \,
\mbox{GeV}^2$ ),  the resulting integrand  of Eq.~\eqref{soft_gluon1}
is relatively smooth. However,  as one increases the momenta towards
the ultraviolet  region, the integrand  develops sharp  peaks, whose
width decreases  as the  momenta increase or/and  as $p_1$ gets  close to
$p_2$. A precise numerical treatment of these peaks requires a refined
grid, especially because minor errors in the integration 
may be subsequently amplified due to the multiplication by $p^2$,
as    happens   in    the   case of $L_1^{g}(p)$    given   by
Eq.~\eqref{sglimit}. In fact, an earlier analysis~\cite{Aguilar:2016lbe} using a less sophisticated numerical  treatment
of these peaks gave rise to an artificial increase   of   $L_1^{g}(p)$   in  the   ultraviolet
region.  Interestingly  enough,  the recent  lattice  simulations  for
$L_1^{g}(p)$  of Ref.~\cite{Oliveira:2016muq,Sternbeck:2017ntv}  found
a similar increase around the same region,   which  the  authors  seem to attribute to lattice artifacts.

  We end this section by suggesting to the reader that, in view 
of the above observations, the results presented in this work  for the soft-gluon limit
ought to be considered as provisional. Indeed, even though an appropriate choice of the quark-gluon
Ansatz used in evaluating $H$, \ie the transition from \1eq{vertex2} to \1eq{vertex1}, 
appears to alleviate considerably the problem of the unnatural ultraviolet suppression,
and despite a vast array of checks implemented on our integration routines,  
the possibility that an unresolved
numerical issue may still be lurking has not been conclusively discarded.

\section{\label{comp}Comparison with previous works}

  In this section we compare our results for the quark-gluon vertex in the soft gluon configuration
with those obtained in a variety of earlier works appearing in the literature.
The reason for choosing this particular configuration is because it is the most widely explored 
in the literature, and because is one of the few that can be individually
isolated in lattice simulations without being ``contaminated'' by transverse contributions.

\begin{figure}[t]
\begin{minipage}[b]{0.45\linewidth}
\centering
\includegraphics[scale=0.32]{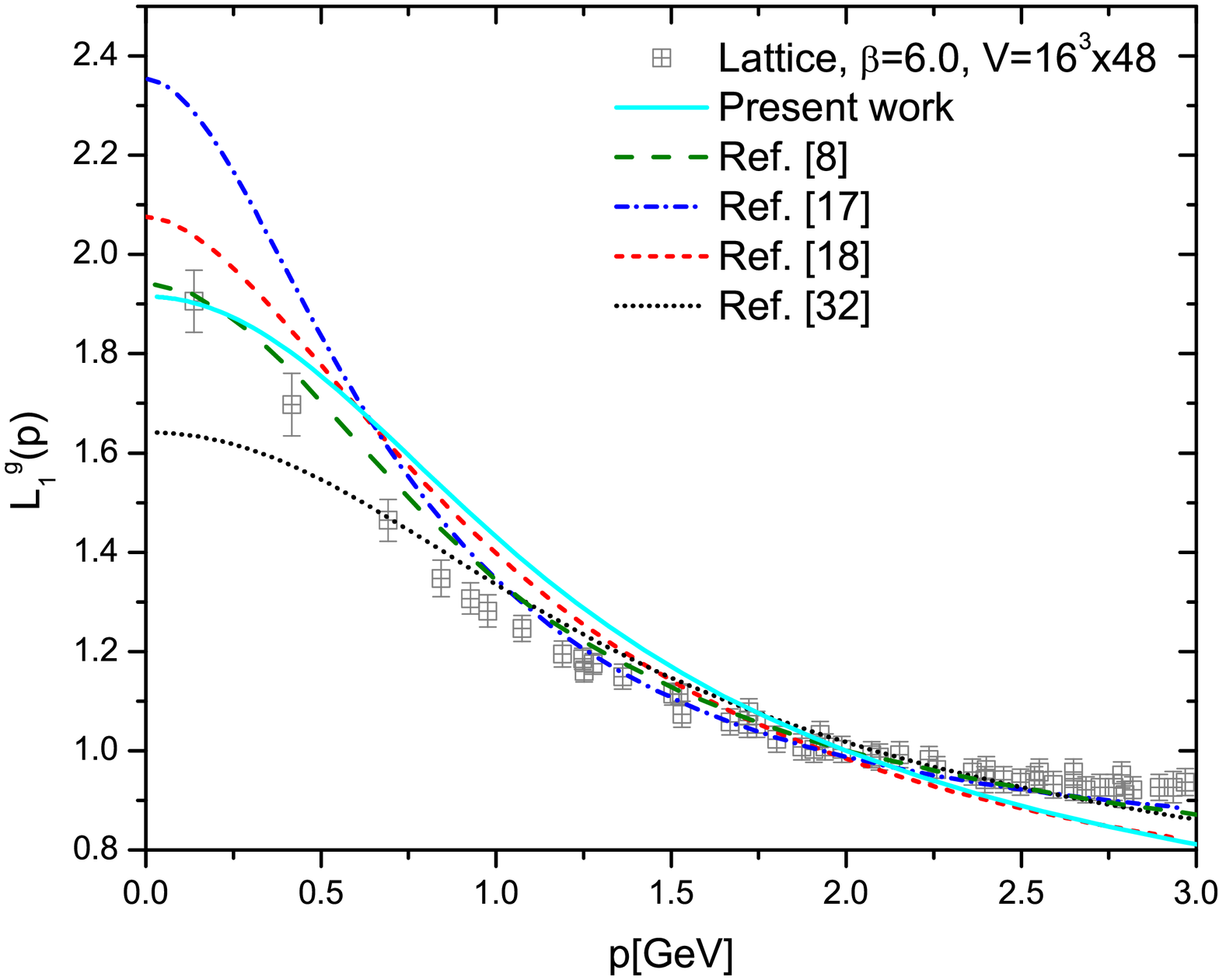}
\end{minipage}
\hspace{0.5cm}
\begin{minipage}[b]{0.50\linewidth}
\includegraphics[scale=0.32]{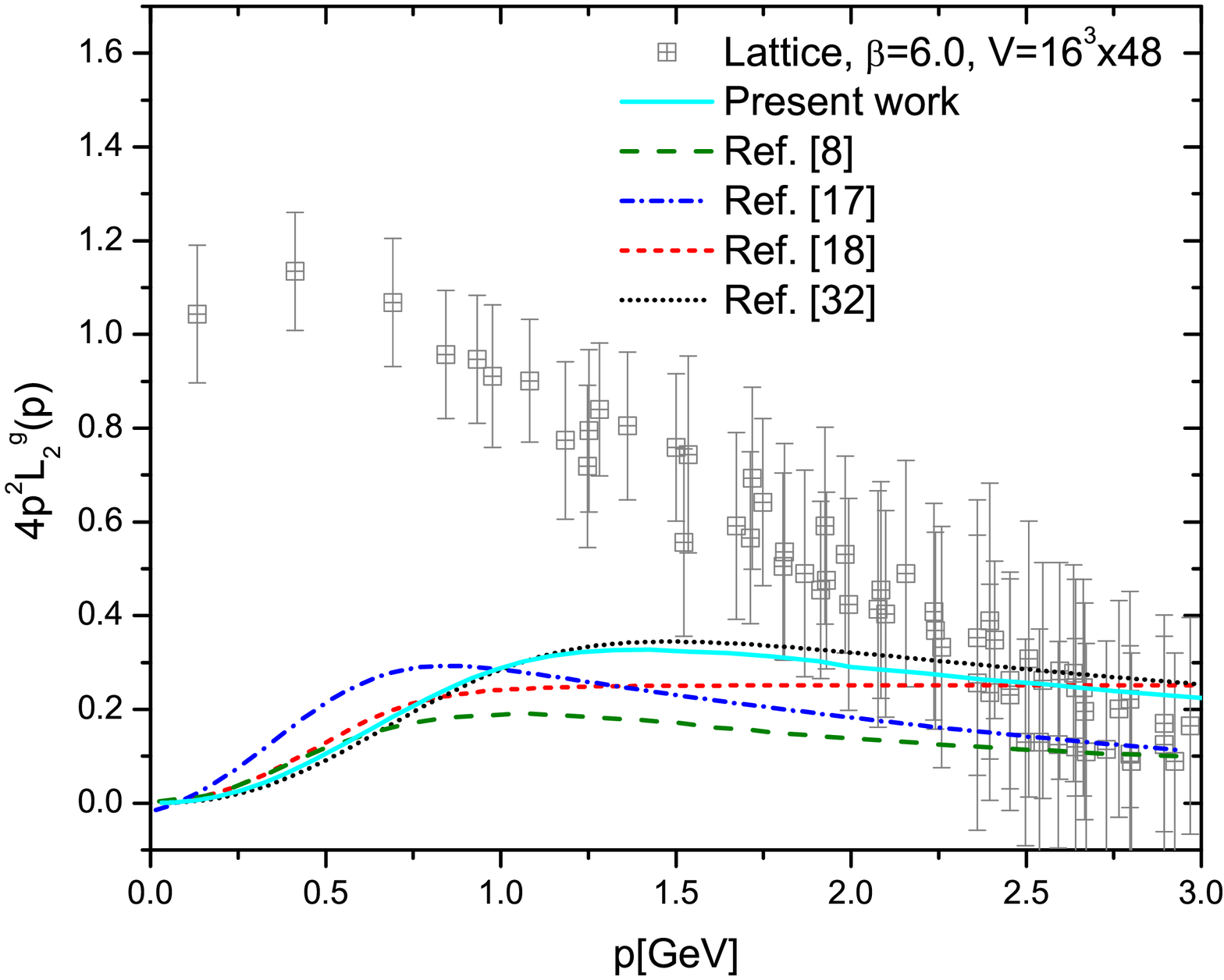}
\end{minipage}
\vspace{0.5cm}\begin{minipage}[b]{0.45\linewidth}
\centering
\vspace{-0.2cm}
\includegraphics[scale=0.32]{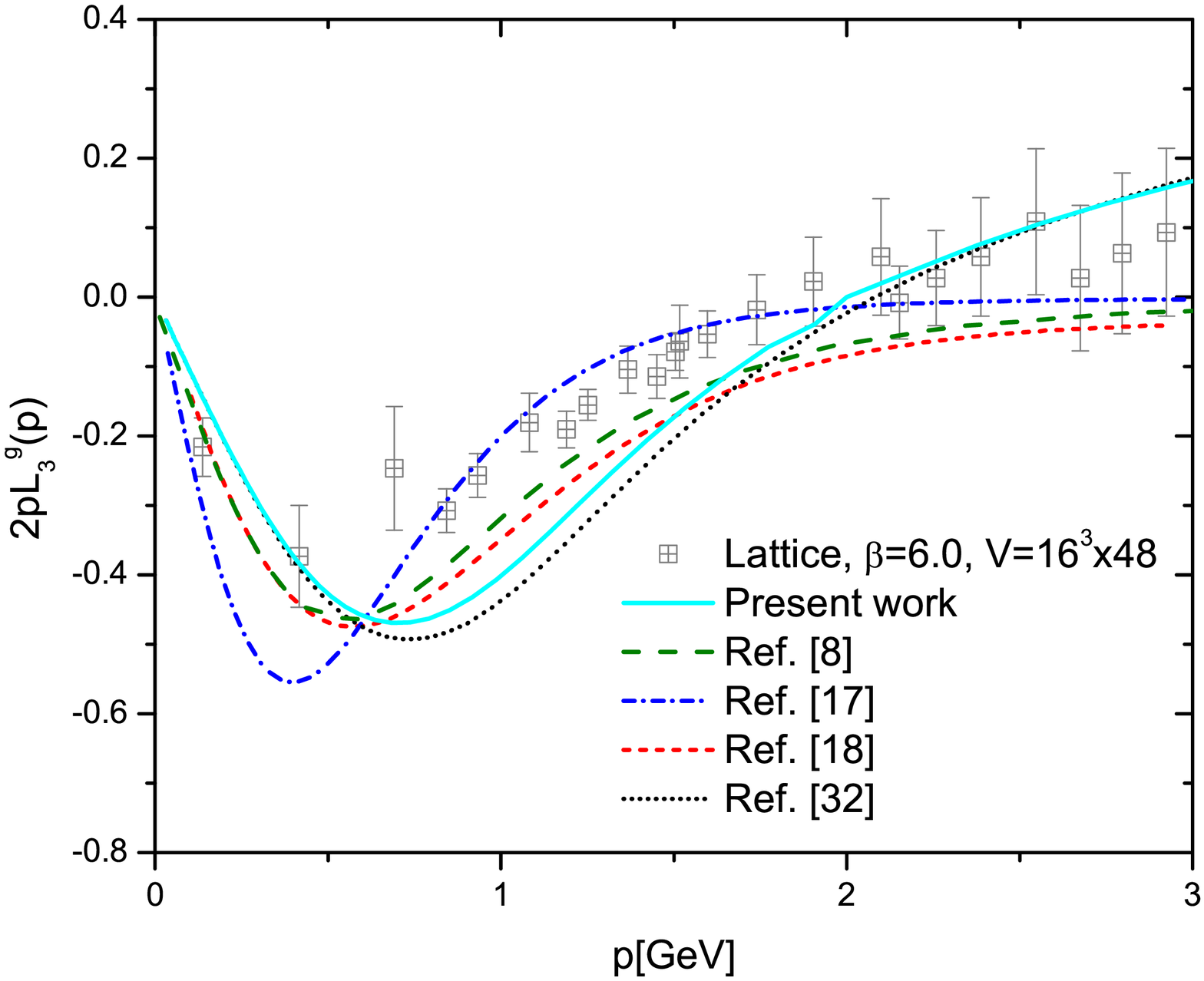}
\end{minipage}
\vspace{-1.0cm}
\caption{\label{fig:comp_latt} {Comparison of our results (cyan continuous line) in the soft gluon configuration with the  previous  analytical results obtained in the Refs.~\cite{Williams:2015cvx} (green dashed), \cite{Bhagwat:2004kj} (blue dashed dot), \cite{LlanesEstrada:2004jz} (red short dashed) and \cite{Aguilar:2014lha} (black dotted), and the 
available lattice data (squares) of Ref.~\cite{Skullerud:2003qu,Skullerud:2004gp}}.}
\end{figure}

Here we will concentrate on the results presented in the  Refs~\cite{Williams:2015cvx, Bhagwat:2004kj,LlanesEstrada:2004jz,Aguilar:2014lha}.
Let us start by recalling that in  Refs.~\cite{Bhagwat:2004kj,LlanesEstrada:2004jz} 
an approximate version of the SDE governing the quark-gluon vertex was considered, retaining the one-loop dressed diagrams that compose the  skeleton expansion
of $\Gamma_{\mu}$.
The main difference between the works of Ref.~\cite{Bhagwat:2004kj} and Ref.~\cite{LlanesEstrada:2004jz} originates from 
the functional form employed for the gluon propagator, $\Delta(q^2)$. In the case of  Ref.~\cite{Bhagwat:2004kj},
the ``rainbow-ladder'' approximation was used, and the product   $g^2\Delta(q^2)$ was replaced by a phenomenologically motivated {\it Ansatz}~\cite{Maris:1999nt}.
Instead, in  Ref.~\cite{LlanesEstrada:2004jz}, the one-loop dressed diagrams  were computed using as nonperturbative ingredients
the quark and gluon propagators calculated within the ``ghost dominance'' formalism~\cite{Fischer:2003rp,Fischer:2006ub}. In Ref.~\cite{Williams:2015cvx}, the authors instead of using the Schwinger-Dyson approach, they construct the  three-particle irreducible (3PI) effective action to three loops to determine the quark-gluon vertex structure. To do that, they use as input the gluon and the ghost propagators which are in 
agreement with  lattice results of Ref.~\cite{Sternbeck:2005tk}. Finally, in Ref.~\cite{Aguilar:2014lha} an improved version of the gauge technique was employed, where the transverse form factors were estimated
by resorting to the so-called transverse Ward identities~\mbox{\cite{Takahashi:1985yz,Kondo:1996xn,He:2000we,Pennington:2005mw,He:2006my,Qin:2013mta}}, and the nonperturbative ingredients such as  $\Delta(q^2)$ and $F(q^2)$ were
taken from the lattice~\cite{Bogolubsky:2007ud}. 
Of particular interest for the present work are the numerical results for the soft gluon configuration reported in~\cite{Aguilar:2014lha}.

In order to make a direct contact with the 
lattice data of Refs.~\cite{Skullerud:2002sk,Skullerud:2003qu}, it is important to mention that
their results were obtained using a current quark mass of \mbox{$m_0=115$\, MeV}. Moreover, the relevant
form factors were renormalized at the scale \mbox{$\mu^{\prime}=2$ GeV}. Therefore, for the 
sake of comparison, we will  also employ, exclusively in this section,  a new set of inputs for $A(p)$, $B(p)$, $\Delta(q)$, $F(q)$
with the aforementioned characteristics.
To obtain these new inputs,  we follow the same procedure outlined in the Ref.~\cite{Aguilar:2014lha}, fixing $\alpha(\mu^{\prime})=0.45$,
which permits a better agreement  with the lattice data.

In Fig.~\ref{fig:comp_latt} we    
compare our results for the soft gluon form factors $L_1^{g}$, $4p^2L_2^{g}(p)$ and $2pL_3^{g}(p)$ (cyan curves)  with those obtained  in the  analysis presented in  Refs.~\cite{Williams:2015cvx} (green dashed),~\cite{Bhagwat:2004kj} (blue dashed-dot) and~\cite{LlanesEstrada:2004jz} (red short dashed). In addition, on the same plot, we show the results of Ref.~\cite{Aguilar:2014lha} (black dotted), and the lattice data obtained in  Ref.~\cite{Skullerud:2003qu,Skullerud:2004gp} (squares). 
In the case of the form factors  $L_1^{g}$ and  $2pL_3^{g}$, 
our results show rather good agreement
with  both the lattice simulations and the previous analytical studies.
In the case of $4p^2L_2^{g}$ [top right panel of  Fig.~\ref{fig:comp_latt}], our result 
agrees with the general pattern found by all previous analytic determinations; in particular, all curves share the characteristic
feature of vanishing at the origin. However, as may be plainly established from Fig.~\ref{fig:comp_latt}, 
our result, and all others, are vastly different from the curve found on the lattice.


\section{\label{RGIQ} RGI interaction kernels}

\begin{figure}[!ht]
\begin{center}
\includegraphics[scale=0.6]{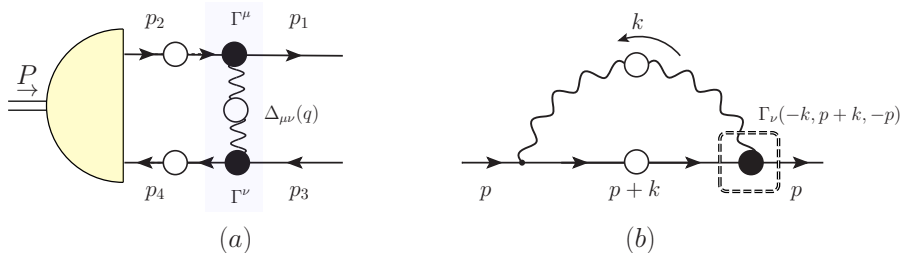}
\end{center}
\vspace{-1.0cm}
\caption{Panel (a): The one-loop dressed gluon exchange that typically appears in the
Bethe-Salpeter equation. Panel (b): The complete quark self-energy entering in the definition of the  gap equation.}
\label{fig_rgi}
\end{figure}

 As a direct application of some of the results obtained in the previous sections, 
we turn to the construction of RGI ($\mu$-independent)
combinations that, depending on the truncation schemes employed, 
quantify the strength of {\it a particular piece} of 
the effective  (momentum-dependent) interaction between quarks. 
Specifically, we will consider the dressed ``one-gluon exchange'' approximation of 
the kernel appearing in a typical Bethe-Salpeter equation [panel (a) of Fig.~\ref{fig_rgi}],
and the corresponding kernel entering in the
gap equation that determines the dynamically 
generated constituent quark mass  [panel (b) of Fig.~\ref{fig_rgi}].

\subsection{\label{assump} Underlying assumptions}

The main objective of this section is to consider
the part of the interaction kernel 
that is of the form $L_1 \gamma_\mu P^{\mu\nu}(q) \gamma_\nu L_1$, neglecting all other
tensorial structures. 
Motivated by the discussion presented in section~\ref{sec:scatter} regarding the way that the choice of the
basis for $\Gamma_\mu^{\rm{(ST)}}$ may affect amplitudes containing the combination
$\Gamma_{\mu}^{\rm{(ST)}} P^{\mu\nu}(q)\Gamma_{\nu}^{\rm{(ST)}}$,
it is important to emphasize at this point some of underlying assumptions of the ensuing analysis.  

\begin{enumerate}

\item
  As has been explained following \1eq{newLprime}, if one considers the combination
  $P^{\mu\nu}(q)\Gamma_{\nu}^{\rm{(ST)}}$, with no reference to the transverse parts,
  the result depends crucially on the choice of the basis for $\Gamma_\mu^{\rm{(ST)}}$.
  In particular, as has been exemplified with the case of $L_1$, the transversely projected
  $\Gamma_{\nu}^{\rm{(ST)}}$ may acquire any value whatsoever, through appropriate choice
  of the basis elements.

\item
  The class of bases that we consider contain the classical tensor 
   $\gamma_\mu$ as a genuine element of the
  tensors that span $\Gamma_\mu^{\rm{(ST)}}$, {\it without} any admixture of transverse components. 
  Put in the language of  section~\ref{sec:scatter}, in \1eq{barL} we have $c=0$, or, equivalently, 
  in \1eq{treelincom} we have $\Gamma_\mu^{[0]} = \lambda_{1,\mu}$. Evidently, the BC basis employed throughout 
  our analysis is precisely of this particular type.
  
\item
  One may rephrase the previous point in the following way. Let us imagine for a moment that
  all quantum corrections are switched off; then, unambiguously, 
  \mbox{$\Gamma_{\mu} P^{\mu\nu}(q)\Gamma_{\nu} \to \Gamma_{\mu}^{[0]} P^{\mu\nu}(q)\Gamma_{\nu}^{[0]} = d-1$}.
  Therefore, one may fix the amount of $L_1$ by requiring that, when one sets $L_1=1$, the
  above result is reproduced (that forces $c=0$).

\item  Of course, the above ``normalization'' does not fix
  the values of the ``$c_i$'' that control   
  the amount of $L_2$, $L_3$ and $L_4$ entering in the answer. In what follows 
  we will simply set $L_2=L_3=L_4=0$ by hand, even though the basis used
  (BC) does {\it not} furnish  $c_i=0$ for them.

\end{enumerate}

\subsection{\label{BSK} Bethe-Salpeter kernel}

As  mentioned above,  one can see in the panel (a) of Fig.~\ref{fig_rgi}, that the kernel of the Bethe-Salpeter equation    
receives  contributions not only from the full gluon propagator but also from the fully-dressed quark-gluon 
vertices. To simplify the analysis, 
let us consider only the part of $\Gamma_{\mu}$ that is proportional to $\gamma_{\mu}$, \ie  $\Gamma_{\mu}=L_1\gamma_{\mu}$.  Then, the strength of this effective interaction may be described 
by means of a scalar quantity, to be denoted by ${\mathcal Q}(q,-p_1,p_3)$, given by  
\begin{align}
{\mathcal Q}(q,-p_1,p_3)= \alpha_s \Delta(q)\left[\frac{L_1(q,p_2,-p_1)L_1(-q,p_3,-p_4)}{A(p_2)A(p_4)}\right] \,. 
\label{qprime}
\end{align}
    This particular quantity is $\mu$-independent, as one may verify by employing the standard renormalization relations 
\bea
S_R(p ; \mu)&=& Z_F^{-1}(\mu)S(p)\,,
\nonumber\\
\Delta_R(q; \mu) &=& Z_{A}^{-1}(\mu) \Delta(q)\,,
\nonumber\\
\Gamma^{\nu}_R(p,k,q;\mu)  &=& Z_{1}(\mu)\Gamma^{\nu}(p,k,q)\,,
\nonumber\\
g_R(\mu) &=& Z_g^{-1}(\mu) g = Z_{1} Z_F^{-1} Z_A^{-1/2} g\,, 
\label{renconst}
\eea
where $Z_F$, $Z_{A}$, $Z_{1}$, and $Z_g$ are the corresponding
renormalization constants\footnote{In the Landau gauge, $Z_F=1$ at {\it one loop}, and, therefore, one may omit the factor
  $A^{-1}(p_2)A^{-1}(p_4)$ 
  in the definition of ${\mathcal Q}(q,-p_1,p_3)$, which would then be $\mu$-independent at that order. Note, however, that
  higher loops make $Z_F$ non-trivial~\cite{Nachtmann:1981zg}, and thus, the inclusion of this factor becomes necessary.}.

It is interesting to compare ${\mathcal Q}$ 
to a closely related quantity, defined in the recent literature~\cite{Binosi:2014aea}. 
In particular, a field-theoretic construction based on the pinch technique 
allows the definition of a process- and $\mu$-independent combination, denoted by 
\begin{align}
\widehat{d}(q^2) = \frac{\alpha_s(\mu)\Delta(q^2)}{[1+G(q^2)]^2}\,,
\label{rgihat1}
\end{align}
where $G(q^2)$ is the transverse component of a special Green's function~\cite{Grassi:1999tp,Binosi:2002ez}, akin to  
a ghost-gluon ``vacuum polarization'', which arises in contemporary applications 
of the aforementioned technique~\cite{Binosi:2009qm,Aguilar:2009nf,Aguilar:2010gm,Binosi:2014aea}. 
From $\widehat{d}(q^2)$ one may define the dimensionless quantity 
\begin{align}
{\mathcal I}_{\widehat{d}}(q^2)= q^2 \widehat{d}(q^2) \,,
\label{coupd}
\end{align}
which, as explained in detail in~\cite{Binosi:2014aea}, makes direct contact with the 
interaction strength obtained from a 
systematic ``bottom-up" treatment, where bound-state data are  fitted within a well-defined truncation
scheme~\cite{Qin:2011dd}.

Given that  $\widehat{d}(q^2)$ is a function of a single kinematic variable, a meaningful 
comparison with ${\mathcal Q}(q,-p_1,p_3)$ 
may be accomplished by computing the latter in a special kinematic limit. Specifically, we choose 
to evaluate both  $L_1(q,p_2,-p_1)$ and $L_1(-q,p_3,-p_4)$ at their corresponding totally symmetric points, namely 
$p_1^2=p_2^2=q^2$ and $p_3^2=p_4^2=q^2$, thus converting ${\mathcal Q}$ to a function of the  single variable,   
\begin{align}
{\mathcal Q}(q^2)= \alpha_s(\mu)\Delta(q^2)\left[\frac{L_1^{\s{\rm TS}}(q^2)}{A(q^2)}\right]^2 \,,
\label{rgi_sym}
\end{align}
where the behavior of $L_1^{\s{\rm TS}}(q)$ is given by the bottom right panel of Fig.~\ref{fig:Lsoft}.
Evidently, the dimensionless quantity analogous to ${\mathcal I}_{\widehat{d}}(q^2)$ is given by 
\begin{align}
{\mathcal I}_{\mathcal Q}(q^2)= q^2 {\mathcal Q}(q^2)\,.
\label{coupQ}
\end{align}
Note that, for asymptotically large $q^2$, both ${\mathcal I}_{\widehat{d}}(q^2)$ and ${\mathcal I}_{\mathcal Q}(q^2)$
capture the one-loop running coupling of QCD~\cite{Aguilar:2009nf}.

\begin{figure}[!t]
\begin{minipage}[b]{0.45\linewidth}
\centering
\includegraphics[scale=0.37]{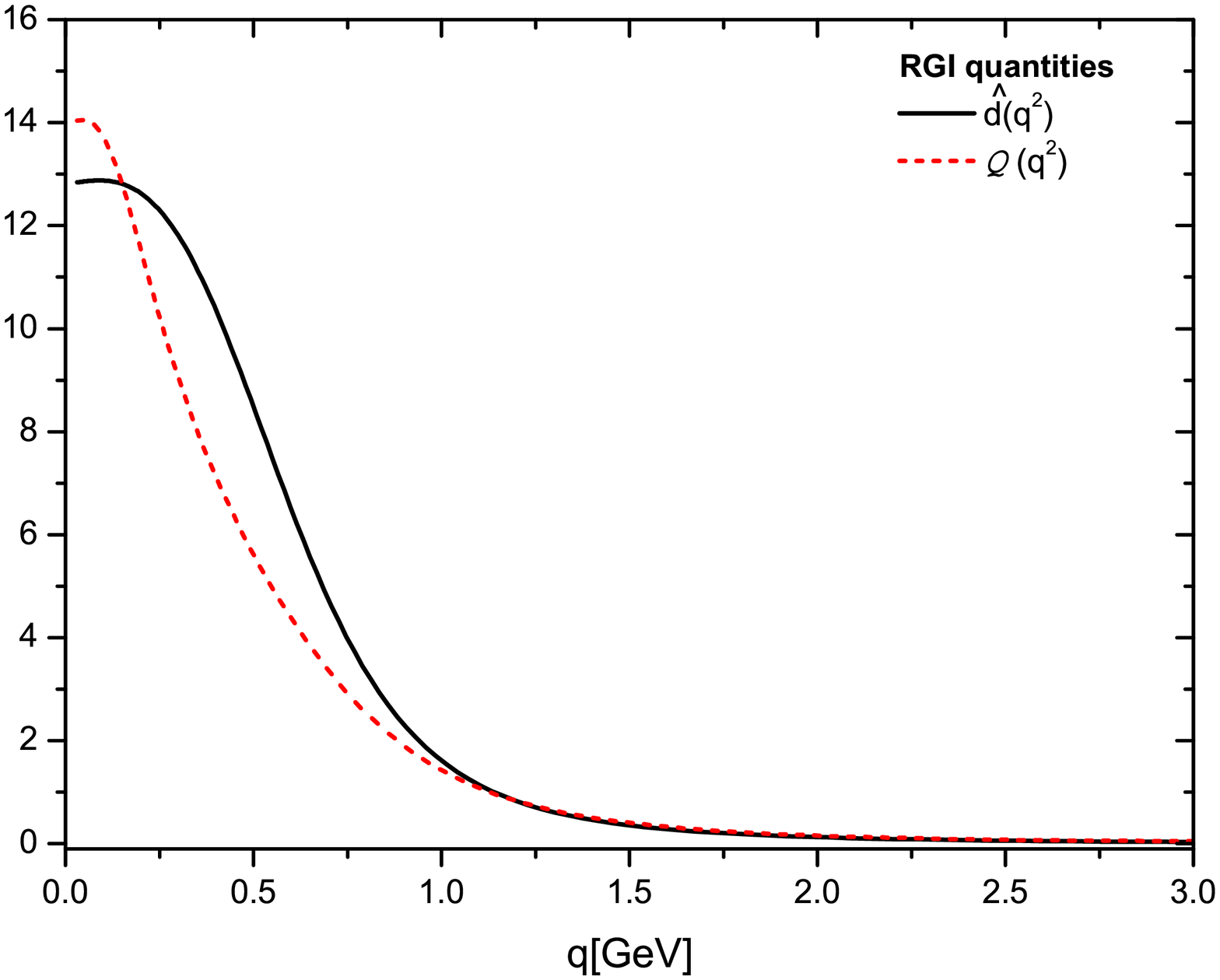}
\end{minipage}
\hspace{0.5cm}
\begin{minipage}[b]{0.50\linewidth}
\includegraphics[scale=0.37]{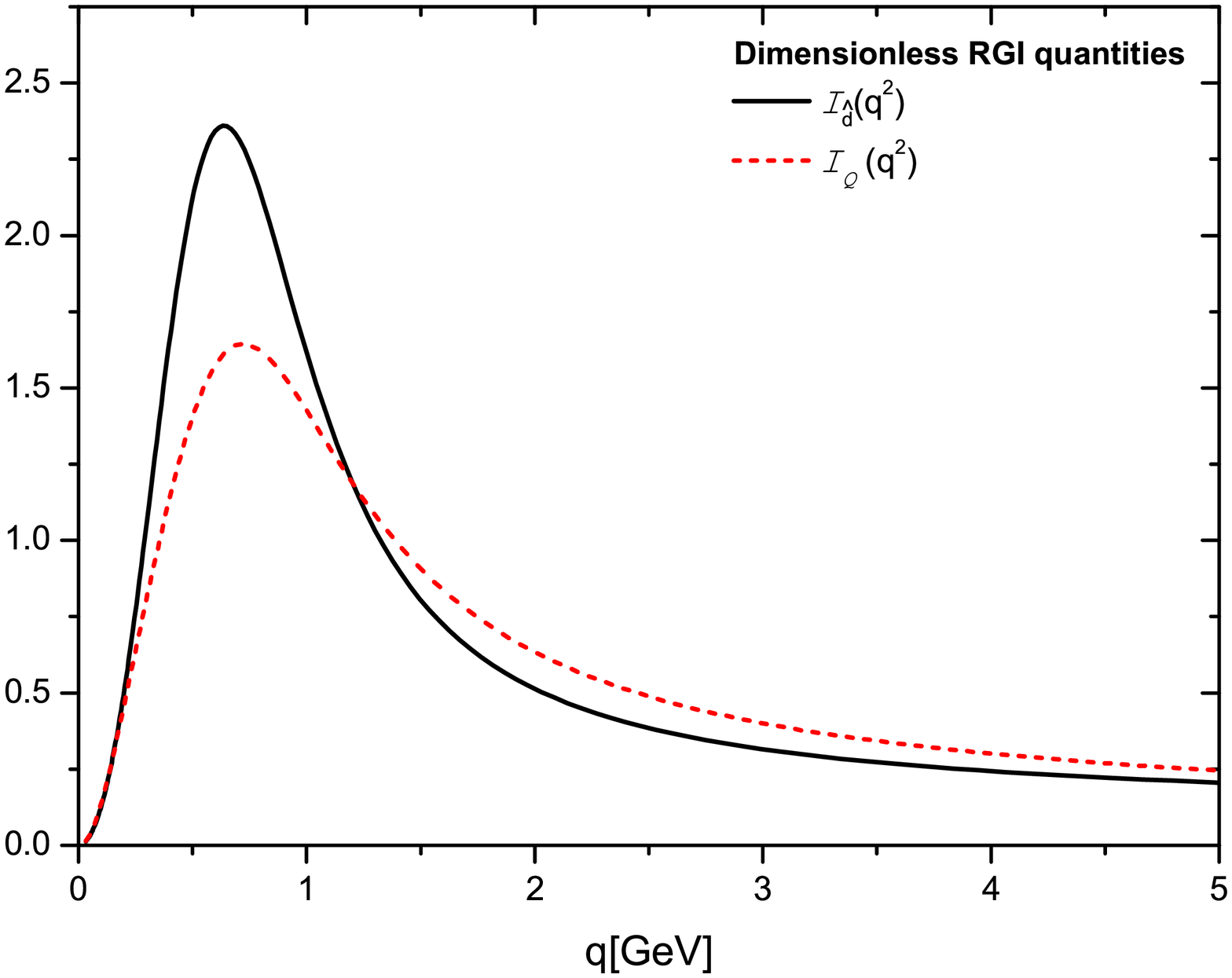}
\end{minipage}
\vspace{-1.25cm}
\caption{\label{fig:rgi} Comparison of the dimensionful RGI quantities $\widehat{d}(q^2)$  and ${\mathcal Q}(q^2)$ defined in the Eqs.~\eqref{rgihat1} and \eqref{rgi_sym}, respectively (left panel) and their 
 dimensionless counterparts ${\mathcal I}_{\widehat{d}}(q^2)$ and ${\mathcal I}_{\mathcal Q}(q^2)$ given by Eqs.~\eqref{coupd} and~\eqref{coupQ} (right panel).}
\end{figure}

On the left panel of Fig.~\ref{fig:rgi} we compare ${\mathcal Q}(q^2)$ and $\widehat{d}(q^2)$.   For obtaining $\widehat{d}(q^2)$ (black continuous line), we use the same value 
for  $\alpha_s(\mu)$ as in~\cite{Binosi:2014aea}, namely $\alpha_s(\mu)=0.22$ for  \mbox{$\mu=4.3$\,GeV}; the  determination of this value entails a subtle combination of 
4-loop perturbative results, nonperturbative information included in the vacuum condensate of dimension two, 
and the extraction of $\Lambda_{\rm QCD}$ from lattice results of the ghost-gluon vertex 
in the Taylor kinematics~\cite{Boucaud:2008gn}.  Instead,
for computing ${\mathcal Q}(q^2)$ (red dashed line)  
self-consistency dictates that one must set in Eq.~\eqref{rgi_sym}
$\alpha_s(\mu)=0.28$, because this is precisely the value used for obtaining
$A(q^2)$ and $L_1^{\s{\rm TS}}(q^2)$ in Sections~\ref{ingred} 
and~\ref{special}, respectively.

On the right panel of the same figure,  we show 
${\mathcal I}_{\mathcal Q}(q^2)$ and ${\mathcal I}_{\widehat{d}}(q^2)$.
We clearly see that, ${\mathcal I}_{\widehat{d}}(q^2)$ displays a higher peak in the region of about \mbox{$650$ MeV},  while ${\mathcal I}_{\mathcal Q}(q^2)$ has its peak around \mbox{$730$ MeV}. Notice that ${\mathcal I}_{\mathcal Q}(q^2)$ is consistently higher in the interval between \mbox{$1.25-5$ GeV}.

In order to obtain an indication of the {\it integrated strength} furnished by these curves,  following~\cite{Binosi:2014aea} we may introduce the integral    
\begin{align}
W_{\mathcal I} = \int_0^{q^2_0}\!\!\! dq^2 \,{\mathcal I}(q^2) \,,
\label{intstr}
\end{align}
where $q_0\approx 5$ GeV is the value of the momentum 
where the two ${\mathcal I}(q)$ practically merge into each other in Fig.~\ref{fig:rgi}.
The results of the integration are 
~$W_{\mathcal I} = 10.4$~$\mbox{GeV}^2$ when ${\mathcal I}={\mathcal I}_{\widehat{d}}$ and ~$W_{\mathcal I} = 11.3$~$\mbox{GeV}^2$  when ${\mathcal I}={\mathcal I}_{\mathcal Q}$. 

Interestingly, while ${\mathcal Q}(q^2)$
appears quite suppressed relative  to ${\widehat{d}}(q^2)$ in the range of momenta  between \mbox{$0.15-1.1$ GeV}, the corresponding 
integrated strengths turn out to be rather close to each other; in fact,
${\mathcal Q}(q^2)$ is $8.6\%$  larger than ${\widehat{d}}(q^2)$.

Even though the amount of  physical information contained in ${\mathcal
  Q}(q^2)$ and ${\widehat{d}}(q^2)$ is  a-priori different, given that
the two  quantities originate from distinct truncation schemes, the
simple analysis presented above seems  to suggest a certain similarity
in  the structures obtained  using either  of them.
One  should  keep  in  mind, of course, that  all  remaining  tensorial
structures  of the quark-gluon vertex (transverse and non-transverse), which are certainly non-negligible, must be properly taken into account (within both frameworks), before any robust conclusion on this matter may be drawn.

\subsection{\label{GEK} Gap equation kernel}

We end with a preliminary look at the kernel that enters in the   
standard quark gap equation formulated in the Landau gauge. 
In particular, the fully-dressed quark-gluon vertex $\Gamma_{\mu}$ 
constitutes a central ingredient of the quark self-energy, shown in the panel (b) of Fig.~\ref{fig_rgi}; 
evidently, $(q, p_2, -p_1) \to (p-k, k, -p)$, and, eventually, in order to treat the 
full question of chiral symmetry breaking, 
two out of the three momenta of this particular vertex 
are to be integrated over, since the virtual momentum $k$ circulating in the loop enters in them.

For the purposes of this introductory discussion, we consider a ``frozen'' kinematic configuration, 
\ie no integration over  $k$ will be implemented,  
and simplify the analysis by using the approximation  $\Gamma_{\mu}=L_1\gamma_{\mu}$.
Then, the gap equation assumes the form 
\begin{align}
S^{-1}(p) = \slashed{p} - 4\pi C_F\int_{k}\gamma_{\mu}
\frac{1}{(\slashed{p}+\slashed{k}) -{\mathcal M}(p+k)}
\gamma_{\nu}P^{\mu\nu}(k)K_{\rm gap}(-k,k+p,-p)\,, 
\label{gap}
\end{align}
where  $C_F=4/3$ is the Casimir eigenvalue in the fundamental representation, and 
\begin{align}
K_{\rm gap}(-k,k+p,-p) = \alpha_s(\mu)\Delta(k^2)F(k^2)\left[\frac{L_1(-k,k+p,-p)}{A(p+k)}\right]\,.
\label{kcsb}
\end{align}

As has been discussed in detail in~\cite{Aguilar:2010cn},   
the appearance of the factor  $F(k^2)$ in $K_{\rm gap}$ accounts in an effective way 
for contributions originating from the transverse part of the quark-gluon vertex, which, if properly treated, 
would combine with the renormalization constant $Z_1$ that multiplies the quark self-energy, 
furnishing the correct value for the anomalous dimension of the quark mass obtained  
(for an earlier treatment along the same lines, see~\cite{Fischer:2003rp}). 
The main upshot of these arguments for our present purposes is that, just as ${\mathcal Q}(q,-p_1,p_3)$ and $\widehat{d}(q^2)$, 
the quantity $K_{\rm gap}(-k,k+p,-p)$ defined in Eq.~\eqref{kcsb} is also $\mu$-independent. In fact, $K_{\rm gap}(-k,k+p,-p)$ is composed of two individually $\mu$-independent factors, namely
\begin{align}
  K_{\rm gap} = \left\{\alpha^{1/2}_s(\mu)\Delta^{1/2}(k^2) \left[\frac{L_1(-k,k+p,-p)}{A(p+k)}\right]\right\}
 \bigg[\alpha^{1/2}_s(\mu)\Delta^{1/2}(k^2)F(k^2)\bigg] \,.
\label{kcsbprod}
\end{align}    

Evidently, the combination in curly brackets is essentially ${\mathcal Q}^{1/2}$, while the one 
in square brackets is RGI due to the nonperturbative
relation that holds between $F$ and $1+G$, as discussed in detail in~\cite{Aguilar:2009nf}.

\begin{figure}[!ht]
\begin{minipage}[b]{0.45\linewidth}
\centering
\includegraphics[scale=0.35]{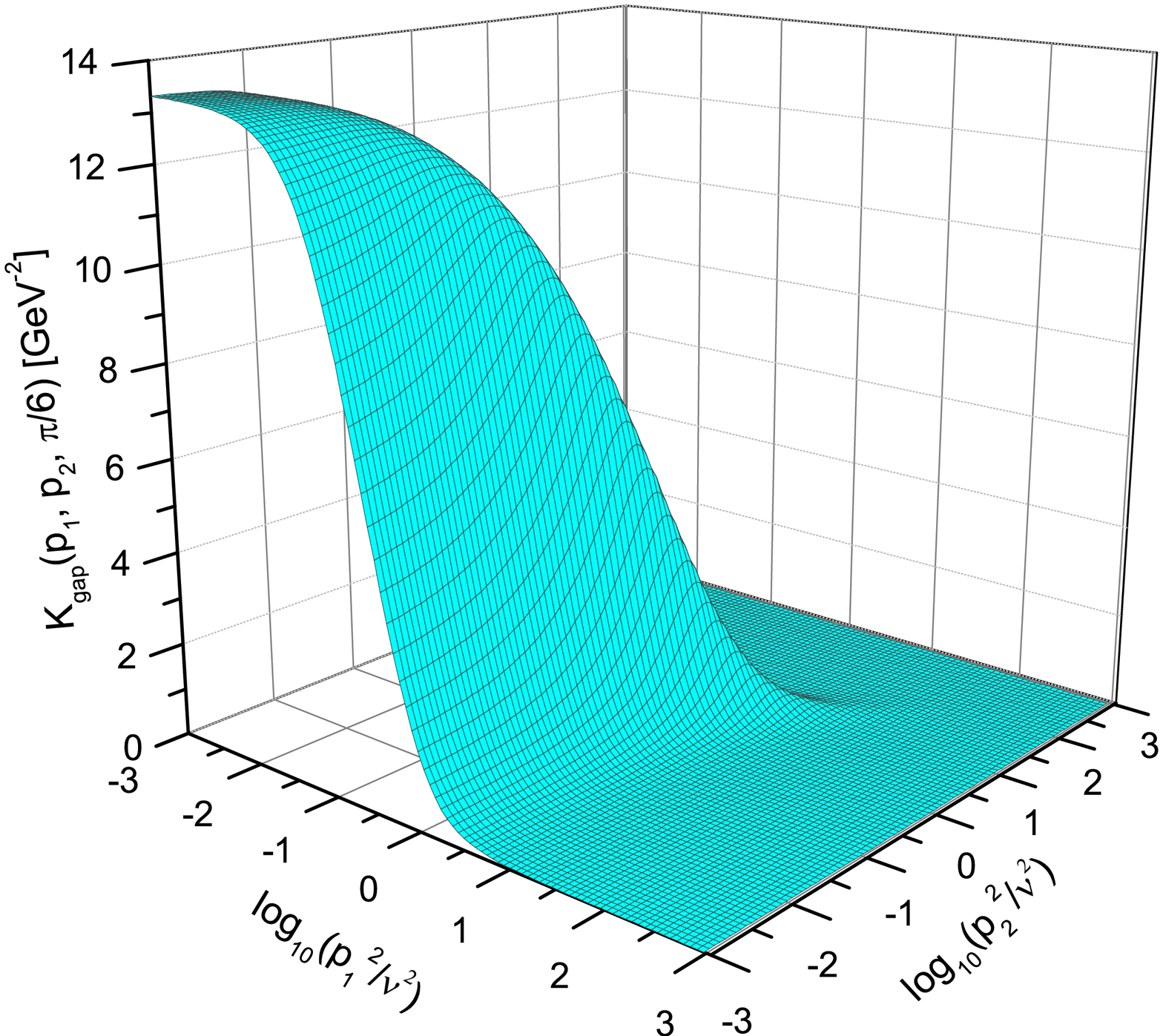}
\end{minipage}
\hspace{0.5cm}
\begin{minipage}[b]{0.50\linewidth}
\includegraphics[scale=0.35]{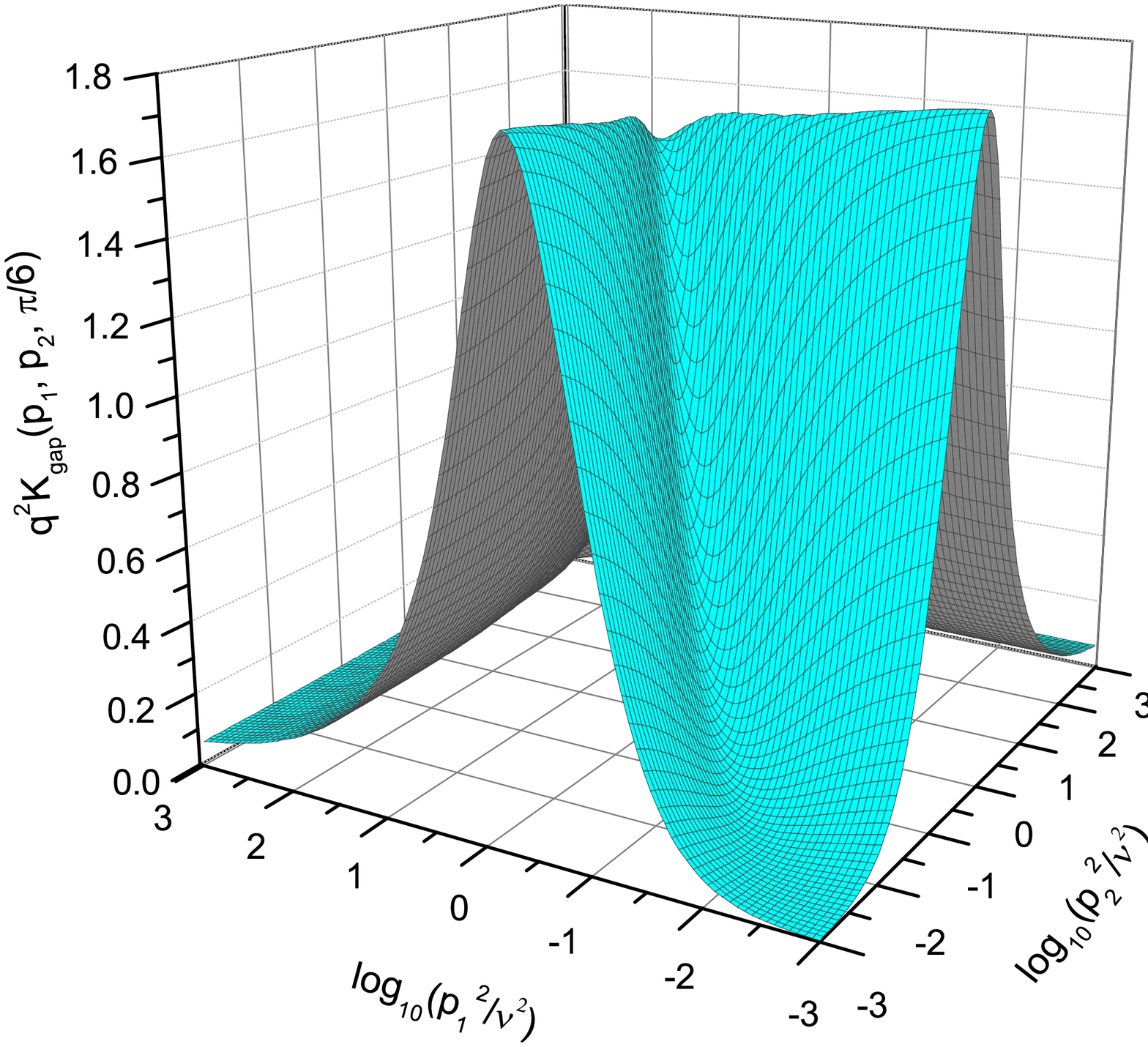}
\end{minipage}
\vspace{-0.75cm}
\caption{\label{fig:kernelcsb} The kernel $K_{\rm gap}(p_1,p_2,\pi/6)$  defined in Eq.~\eqref{kcsb} (left panel), and its dimensionless counterpart $q^2K_{\rm gap}(p_1,p_2,\pi/6)$ (right panel). } 
\end{figure}

A representative case of  $K_{\rm gap}(p_1,p_2,\theta)$ when $\theta=\pi/6$ is shown on the left panel of Fig.~\ref{fig:kernelcsb}, while on the right panel we show the dimensionless quantity $q^2 K_{\rm gap}(p_1,p_2,\pi/6)$, 
which is the 3-D analogue of  ${\mathcal I}_{\mathcal Q}(q^2)$ and ${\mathcal I}_{\widehat{d}}(q^2)$.  

We clearly see that $q^2 K_{\rm gap}(p_1,p_2,\pi/6)$  displays two symmetric peaks, which appear  when one of the momenta(either $p_1$ or $p_2$) vanishes and the  other is of the order of  \mbox{$730$ MeV}. For other values of $\theta$ we found a similar pattern.  Therefore, it is reasonable to expect that the gap equation will receive more support around this region.

We end this section by emphasizing that a complete  treatment of both BSE
and gap equations requires the inclusion of  all vertex form factors
(the four non-transverse and the  eight transverse ones). Moreover,  both
equations  involve an angular  integration over $\theta$; 
consequently, variations  in the
angular structure of  the twelve form factors may  have a significant
impact on the  phenomenological parameters produced. Therefore, the 
analysis presented in the section should  serve as a 
simple rough estimate of  the possible impact of a unique form-factor, namely $L_1$.

\section{\label{concl}Discussion and Conclusions}

We have presented a novel nonperturbative computation of the ``non-transverse'' components  of the quark-gluon vertex, for general values of the Euclidean momenta 
entering in them. The starting point of this analysis is the 
STI that the quark-gluon vertex satisfies, which determines completely its 
four  form factors in terms of the quark propagator, the ghost dressing function, and 
the quark-ghost scattering kernel. 
The inclusion of these last two ghost-related 
quantities implements the  non-Abelian conversion of the usual QED-inspired 
BC Ansatz employed for the quark-gluon vertex. 
Even though the procedure we have followed is in principle applicable for any value of the 
gauge fixing parameter, in practice all ingredients 
relevant to the calculation have been computed in the Landau gauge.  
The form factors of the quark-ghost scattering kernel have been computed within the 
``one-loop dressed'' approximation, which involves a single diagram, 
where all propagators are fully dressed, while certain simplifying approximations have been employed 
for the corresponding vertices.
The results obtained, displayed in various 3-D plots, indicate considerable deviations from the  
Abelian BC expressions. In addition, several typical kinematic configurations considered in the 
related literature, such as the ``soft gluon'' 
or ``symmetric'' limits,  have been extracted from the general 3-D results 
through appropriate 2-D ``projections''.

The most natural context where the results of the present analysis may be applied is 
the study of chiral symmetry breaking and dynamical quark mass generation by means of the 
standard Landau gauge gap equation, 
along the lines presented in subsection \ref{GEK},
where, however, only the effect of a special kinematic configuration of $L_1$ was considered.
Of course, the precise 3-D form of all form factors 
is bound to affect the overall strength of the kernel of the 
gap equation, and the ensuing integration over virtual momenta will 
``peak'' the strength of the full kernel around a particular mass-scale, 
whose value, as is well-known, is crucial for the final amount of quark mass that can be 
generated dynamically. Note, in particular, that in an earlier approach~\cite{Aguilar:2010cn} a rudimentary 
version of the non-Abelian BC vertex was constructed by setting $X_1=X_2=X_3=0$, and using only 
a particular 2-D ``slice'' of  $X_0$. The gap equation with 
this particular vertex gave rise to a running quark mass, ${\cal M}(q^2)$, 
whose value at the origin was ${\cal M}(0)\approx 300$ MeV, when the value of the strong coupling used was  
$\alpha_s\approx 0.29$ . 
 It would be therefore important to study the same issue  
using instead the more complete version of the non-Abelian BC vertex obtained in the present work. 
In fact, a detailed analysis of the gap equation combining the 
non-Abelian BC part derived here and the purely transverse part employed in~\cite{Binosi:2016wcx}, imposing the 
physical constraints applied in this latter work, may single out a rather limited set of 
the possible vertex Ans\"atze that would be compatible with contemporary QCD phenomenology. 

An additional issue, related to the present work as well as the prospect of applying the results 
to the study of the gap equation, has to do with the treatment of the set of dynamical equations 
that enter into the problem under study.
In particular, one of the main technical shortcomings of the present work is the 
treatment of the Dirac components of the quark propagator [$A(p^2)$ and ${\cal M}(q^2)$], shown in Fig.~(\ref{AA:plot}) 
as {\it external quantities}, in the 
sense that they were obtained from a gap equation that was solved in isolation, using an Ansatz for the 
quark-gluon vertex corresponding precisely to the simplified non-Abelian BC vertex mentioned above~\cite{Aguilar:2010cn}.
The amelioration of this drawback requires the treatment of all dynamical equations involved  as a system of   
coupled integral equations, whose simultaneous solution must be determined through numerical iteration.
We hope to be able to report considerable progress in this direction in the near future.

\acknowledgments 

The research of J.~P. is supported by the Spanish  Ministerio de Econom\'ia y Competitividad (MEYC) under grants FPA2014-53631-C2-1-P and SEV-2014-0398, and Generalitat Valenciana  
under grant Prometeo~II/2014/066.
The work of  A.~C.~A, J. C. C.  and M. N. F. are supported by the Brazilian National Council for Scientific and Technological Development (CNPq) under the grants 305815/2015, 141981/2013-0
and  147433/2014-2, respectively. The authors thank D.~Binosi for valuable communications.
This research was performed using the Feynman Cluster of the
John David Rogers Computation Center (CCJDR) in Institute of Physics ``Gleb
Wataghin", University of Campinas.

\appendix

\section{\label{app:taylor} Taylor expansions of $\mathcal{K}$}

In this Appendix we outline the Taylor expansions of Eq.~\eqref{euclidean_xi}  
needed for the derivation of certain special kinematic limits. For concreteness we will work in some detail  
the derivation of the $X_i$ in the soft gluon and  the quark symmetric cases, 
which, as discussed in the section~\ref{special}, are obtained 
by taking the limit  $\sin\theta \to 0$ in Eq.~\eqref{euclidean_xi}. 
The corresponding expansions of  $\mathcal{K}$
around $p_1=0$ ($p_2= 0$), relevant for the soft anti-quark (quark) configuration, 
proceed following completely analogous steps.
 
Consider the Eq.~\eqref{euclidean_xi}, and expand  the kernel $\mathcal{K}(p_1,p_2,l)$ around $\sin\theta=0$, 
\begin{align}
\mathcal{K}(p_1,p_2,l)=\mathcal{K}_0 +\sin\theta \, \mathcal{K}_0^{\prime}+\mathcal{O}(\sin^2\theta)\,,
\label{taylor_sin}
\end{align}
where we have introduced the compact notation 
\begin{align}
\mathcal{K}_0 &= \mathcal{K}(p_1,p_2,l)\Big\vert_{\sin\theta=0} \,, \qquad
\mathcal{K}_0^{\prime} = \left.\frac{\partial\mathcal{K}(p_1,p_2,l)}{\partial\sin\theta}\right\vert_{\sin\theta=0} \,, \nonumber 
\end{align}
Concentrating on the contribution of $\mathcal{K}_0$ in Eqs.~\eqref{generalx}, we clearly see that the 
only dependence of $\mathcal{K}(p_1,p_2,l)$ on $\varphi_2$ stems from $D(l-p_1)$, namely
\begin{align}
D(l-p_1)=D(l^2+p_1^2-2lp_1[\cos\theta\cos\varphi_1+\sin\theta\sin\varphi_1\cos\varphi_2]) \,.
\label{ghost_arg}
\end{align}
Thus, in the limit $\sin\theta =0$, $\mathcal{K}_0$
is completely independent of  $\varphi_2$, and the integration over this variable becomes trivial. 
Then  we notice that, in the expressions for $X_i$, the terms with a $\sin\theta$ in the denominator are always proportional to $\cos\varphi_2$, which leads to the following vanishing angular integration
\begin{align}
\int_0^\pi \!\!\!d\varphi_2\sin\varphi_2\cos\varphi_2=0 \,.
\label{intvan}
\end{align}
Therefore, all contributions of the  Eq.~\eqref{generalx} containing $\mathcal{K}_0$ are finite. 
Evidently, the contribution of the 
second term in the expansion of Eq.~\eqref{taylor_sin} is automatically finite, 
given that is it explicitly multiplied by a $\sin\theta$ 
that cancels directly the corresponding term in the denominator of $\mathcal{K}$.

Implementing the above procedure in Eq.~\eqref{euclidean_xi}, we find that the form factors in the limit $\sin\theta=0$ reduce to 
\begin{align}
X_0(p_1,p_2,\theta=0,\pi)&=1+\frac{C_\mathrm{A}g^2}{4}\int_{l_E} \frac{A(l^2)}{s^2}\mathcal{K}_0\left\lbrace p_2^2l^2\sin^2\varphi_1-(l-p_2)^2p_1p_2\cos\theta\right.\nonumber\\
&\left.+[p_1l\cos\theta\cos\varphi_1-p_1p_2\cos\theta](p_2 l\cos\varphi_1-p_2^2)\right\rbrace\,, \nonumber\\
X_1(p_1,p_2,\theta=0,\pi)&=\frac{C_\mathrm{A}g^2}{4}\int_{l_E}\frac{B(l^2)}{s^2}\left\lbrace\mathcal{K}_0\left[s^2-l^2\sin^2\varphi_1\cos^2\varphi_2\right]\right.\nonumber\\
&\left.+l(l\cos\varphi_1-p_2)\left(\frac{p_2}{p_1}-\cos\theta\right) \mathcal{K}_0^{\prime}\sin\varphi_1\cos\varphi_2 \right\rbrace\,, \nonumber \\
X_2(p_1,p_2,\theta=0,\pi)&=\frac{C_\mathrm{A}g^2}{4}\int_{l_E}\frac{B(l^2)}{s^2}\left\lbrace\mathcal{K}_0\left[ (l\cos\varphi_1-p_2)^2\left(1-\frac{p_1}{p_2}\cos\theta\right) -s^2
\right.\right.\nonumber\\
&\hspace{-2.5cm}\left.\left.+\frac{p_1l^2}{p_2}\cos\theta\sin^2\varphi_1\cos^2\varphi_2\right]  -l(l\cos\varphi_1-p_2)\cos\theta\left(1-\frac{p_1}{p_2}\cos\theta\right)\mathcal{K}_0^{\prime} \sin\varphi_1\cos\varphi_2 \right\rbrace\,, \nonumber\\
X_3(p_1,p_2,\theta=0,\pi)&=\frac{C_\mathrm{A}g^2}{4}\int_{l_E}\frac{A(l^2)}{s^2}\left\lbrace\mathcal{K}_0\left[ l^2\sin^2\varphi_1\cos^2\varphi_2 -s^2\frac{l}{p_2}\cos\varphi_1\right] \right. \nonumber \\ 
&\left.
-l^2(l-p_2\cos\varphi_1)\left(\frac{1}{p_1}-\frac{\cos\theta}{p_2}\right)\mathcal{K}_0^{\prime} \sin\varphi_1\cos\varphi_2 \right\rbrace\,.
\label{soft_gluon1}
\end{align}
where the variable $s^2$ was defined below Eq.~\eqref{euclidean_xi}.
For the actual determination of $\mathcal{K}_0^{\prime}$, note that 
\begin{align}
\left.\frac{\partial D(l-p_1)}{\partial\sin\theta}\right\vert_{\sin\theta=0}=
-2lp_1\sin\varphi_1\cos\varphi_2 \frac{\partial D(l)}{\partial l^2}\,,
\end{align}
so that 
\begin{align}
 \mathcal{K}_0^{\prime} = -2lp_1\sin\varphi_1\cos\varphi_2 
\frac{[A(l^2)+A(p_2^2)]}{A^2(l^2)l^2+B^2(l^2)}\Delta(l-p_2) \frac{\partial D(l)}{\partial l^2}\,.
\end{align}

Note finally that, since in both the soft gluon and the quark-symmetric limits we have $p_1=p_2$, 
the  difference between the two depends on 
the value that $\theta$ will acquire in Eq.~\eqref{soft_gluon1}, namely $\theta=0$ or $\theta=\pi$, respectively.


\vspace{0.5cm}


\begin{thebibliography}{1}





\bibitem{Maris:2003vk} 
  P.~Maris and C.~D.~Roberts,
  Int.\ J.\ Mod.\ Phys.\ E {\bf 12}, 297 (2003).
	
\bibitem{Roberts:1994dr} 
  C.~D.~Roberts and A.~G.~Williams,
  Prog.\ Part.\ Nucl.\ Phys.\  {\bf 33}, 477 (1994).
	
\bibitem{Fischer:2003rp} 
  C.~S.~Fischer and R.~Alkofer,
  Phys.\ Rev.\ D {\bf 67}, 094020 (2003).
	
\bibitem{Aguilar:2010cn} 
  A.~C.~Aguilar and J.~Papavassiliou,
  Phys.\ Rev.\ D {\bf 83}, 014013 (2011).
	
\bibitem{Cloet:2013jya} 
  I.~C.~Cloet and C.~D.~Roberts,
  Prog.\ Part.\ Nucl.\ Phys.\  {\bf 77}, 1 (2014).
	
\bibitem{Maris:1999nt} 
  P.~Maris and P.~C.~Tandy,
  Phys.\ Rev.\ C {\bf 60}, 055214 (1999).
	
\bibitem{Bender:2002as} 
  A.~Bender, W.~Detmold, C.~D.~Roberts and A.~W.~Thomas,
  Phys.\ Rev.\ C {\bf 65}, 065203 (2002).
	
	
	
\bibitem{Williams:2015cvx} 
  R.~Williams, C.~S.~Fischer and W.~Heupel,
  Phys.\ Rev.\ D {\bf 93}, no. 3, 034026 (2016).
	
	
\bibitem{Eichmann:2016yit} 
  G.~Eichmann, H.~Sanchis-Alepuz, R.~Williams, R.~Alkofer and C.~S.~Fischer,
  Prog.\ Part.\ Nucl.\ Phys.\  {\bf 91}, 1 (2016).
	

\bibitem{Sanchis-Alepuz:2015qra} 
  H.~Sanchis-Alepuz and R.~Williams,
  Phys.\ Lett.\ B {\bf 749}, 592 (2015).


	
\bibitem{Bhagwat:2004hn} 
  M.~S.~Bhagwat, A.~Holl, A.~Krassnigg, C.~D.~Roberts and P.~C.~Tandy,
  Phys.\ Rev.\ C {\bf 70}, 035205 (2004).
	
\bibitem{Holl:2004qn} 
  A.~Holl, A.~Krassnigg and C.~D.~Roberts,
  Nucl.\ Phys.\ Proc.\ Suppl.\  {\bf 141}, 47 (2005).
	
\bibitem{Chang:2009zb} 
  L.~Chang and C.~D.~Roberts,
  Phys.\ Rev.\ Lett.\  {\bf 103}, 081601 (2009).
	
	
	
\bibitem{Williams:2014iea} 
  R.~Williams,
  Eur.\ Phys.\ J.\ A {\bf 51}, no. 5, 57 (2015).
  
\bibitem{Bender:1996bb} 
  A.~Bender, C.~D.~Roberts and L.~Von Smekal,
  Phys.\ Lett.\ B {\bf 380}, 7 (1996).
 
 
\bibitem{Hopfer:2013np} 
  M.~Hopfer, A.~Windisch and R.~Alkofer,
  PoS ConfinementX {\bf }, 073 (2012). 
		
	
\bibitem{Bhagwat:2004kj} 
  M.~S.~Bhagwat and P.~C.~Tandy,
  Phys.\ Rev.\ D {\bf 70}, 094039 (2004).
	
\bibitem{LlanesEstrada:2004jz} 
  F.~J.~Llanes-Estrada, C.~S.~Fischer and R.~Alkofer,
  Nucl.\ Phys.\ Proc.\ Suppl.\  {\bf 152}, 43 (2006).
	
\bibitem{Alkofer:2008tt} 
  R.~Alkofer, C.~S.~Fischer, F.~J.~Llanes-Estrada and K.~Schwenzer,
  Annals Phys.\  {\bf 324}, 106 (2009).
	
\bibitem{Matevosyan:2006bk} 
  H.~H.~Matevosyan, A.~W.~Thomas and P.~C.~Tandy,
  Phys.\ Rev.\ C {\bf 75}, 045201 (2007).
	
\bibitem{Aguilar:2013ac} 
  A.~C.~Aguilar, D.~Binosi, J.~C.~Cardona and J.~Papavassiliou,
  PoS ConfinementX {\bf }, 103 (2012).
	
\bibitem{Rojas:2013tza} 
  E.~Rojas, J.~P.~B.~C.~de Melo, B.~El-Bennich, O.~Oliveira and T.~Frederico,
  JHEP {\bf 1310}, 193 (2013).
	
	
\bibitem{Fischer:2006ub} 
  C.~S.~Fischer,
  J.\ Phys.\ G {\bf 32}, R253 (2006).
	
\bibitem{Salam:1963sa} 
  A.~Salam,
  Phys.\ Rev.\  {\bf 130}, 1287 (1963).
	
\bibitem{Salam:1964zk} 
  A.~Salam and R.~Delbourgo,
  Phys.\ Rev.\  {\bf 135}, B1398 (1964).
	
\bibitem{Delbourgo:1977jc} 
  R.~Delbourgo and P.~C.~West,
  J.\ Phys.\ A {\bf 10}, 1049 (1977).
	
\bibitem{Delbourgo:1977hq} 
  R.~Delbourgo and P.~C.~West,
  Phys.\ Lett.\ B {\bf 72}, 96 (1977).
	
\bibitem{Curtis:1990zs} 
  D.~C.~Curtis and M.~R.~Pennington,
  Phys.\ Rev.\ D {\bf 42}, 4165 (1990).

\bibitem{Bashir:1997qt} 
  A.~Bashir, A.~Kizilersu and M.~R.~Pennington,
  Phys.\ Rev.\ D {\bf 57}, 1242 (1998).

\bibitem{Kizilersu:2009kg} 
  A.~Kizilersu and M.~R.~Pennington,
  Phys.\ Rev.\ D {\bf 79}, 125020 (2009).


\bibitem{Ball:1980ay}
  J.~S.~Ball and T.~W.~Chiu,
  Phys.\ Rev.\  D {\bf 22}, 2542 (1980).

		
	
	
\bibitem{Aguilar:2014lha} 
  A.~C.~Aguilar, D.~Binosi, D.~Iba\~nez and J.~Papavassiliou,
  Phys.\ Rev.\ D {\bf 90}, no. 6, 065027 (2014).
  
  
\bibitem{Heupel:2014ina} 
  W.~Heupel, T.~Goecke and C.~S.~Fischer,
  Eur.\ Phys.\ J.\ A {\bf 50}, 85 (2014).
  
  
\bibitem{Braun:2014ata} 
  J.~Braun, L.~Fister, J.~M.~Pawlowski and F.~Rennecke,
  Phys.\ Rev.\ D {\bf 94}, no. 3, 034016 (2016).

\bibitem{Mitter:2014wpa} 
  M.~Mitter, J.~M.~Pawlowski and N.~Strodthoff,
  Phys.\ Rev.\ D {\bf 91}, 054035 (2015).



 
\bibitem{Skullerud:2002sk} 
  J.~Skullerud, P.~O.~Bowman and A.~Kizilersu,
  hep-lat/0212011.
	
\bibitem{Skullerud:2002ge} 
  J.~Skullerud and A.~Kizilersu,
  JHEP {\bf 0209}, 013 (2002).
	
\bibitem{Skullerud:2003qu} 
  J.~I.~Skullerud, P.~O.~Bowman, A.~Kizilersu, D.~B.~Leinweber and A.~G.~Williams,
  JHEP {\bf 0304}, 047 (2003).
	
\bibitem{Skullerud:2004gp} 
  J.~I.~Skullerud, P.~O.~Bowman, A.~Kizilersu, D.~B.~Leinweber and A.~G.~Williams,
  Nucl.\ Phys.\ Proc.\ Suppl.\  {\bf 141}, 244 (2005).
	
\bibitem{Lin:2005zd} 
  H.~W.~Lin,
  Phys.\ Rev.\ D {\bf 73}, 094511 (2006).
	
\bibitem{Kizilersu:2006et} 
  A.~Kizilersu, D.~B.~Leinweber, J.~I.~Skullerud and A.~G.~Williams,
  Eur.\ Phys.\ J.\ C {\bf 50}, 871 (2007).
	
\bibitem{Oliveira:2016muq} 
  O.~Oliveira, A.~Kizilersu, P.~J.~Silva, J.~I.~Skullerud, A.~Sternbeck and A.~G.~Williams,
  arXiv:1605.09632 [hep-lat]. 

\bibitem{Sternbeck:2017ntv} 
  A.~Sternbeck, P.~H.~Balduf, A.~Kizilersu, O.~Oliveira, P.~J.~Silva, J.~I.~Skullerud and A.~G.~Williams,
  arXiv:1702.00612 [hep-lat].
	
\bibitem{Davydychev:2000rt} 
  A.~I.~Davydychev, P.~Osland and L.~Saks,
  Phys.\ Rev.\ D {\bf 63}, 014022 (2001).
	
\bibitem{Gracey:2014mpa} 
  J.~A.~Gracey,
  Phys.\ Rev.\ D {\bf 90}, no. 2, 025014 (2014).


\bibitem{Bermudez:2017bpx} 
  R.~Bermudez, L.~Albino, L.~X.~Gutiérrez-Guerrero, M.~E.~Tejeda-Yeomans and A.~Bashir,
  Phys.\ Rev.\ D {\bf 95}, no. 3, 034041 (2017). 

\bibitem{Chetyrkin:2000fd} 
  K.~G.~Chetyrkin and T.~Seidensticker,
  Phys.\ Lett.\ B {\bf 495}, 74 (2000).
	
\bibitem{Chetyrkin:2000dq} 
  K.~G.~Chetyrkin and A.~Retey,
  hep-ph/0007088.


\bibitem{Bogolubsky:2007ud}
  I.~L.~Bogolubsky, E.~M.~Ilgenfritz, M.~Muller-Preussker and A.~Sternbeck,
  PoS {\bf LAT2007}, 290 (2007).
	


\bibitem{Bashir:2011dp} 
  A.~Bashir, R.~Bermudez, L.~Chang and C.~D.~Roberts,
  Phys.\ Rev.\ C {\bf 85}, 045205 (2012).

\bibitem{Chang:2012cc} 
  L.~Chang, C.~D.~Roberts and S.~M.~Schmidt,
  Phys.\ Rev.\ C {\bf 87}, no. 1, 015203 (2013).

\bibitem{Chang:2010hb} 
  L.~Chang, Y.~X.~Liu and C.~D.~Roberts,
  Phys.\ Rev.\ Lett.\  {\bf 106}, 072001 (2011).






\bibitem{Takahashi:1985yz} 
  Y.~Takahashi,
  ``Canonical Quantization and Generalized Ward Relations: Foundation Of Nonperturbative Approach,''
  Positano Symp.1985:0019.


\bibitem{Kondo:1996xn} 
  K.~I.~Kondo,
  Int.\ J.\ Mod.\ Phys.\ A {\bf 12}, 5651 (1997).

\bibitem{He:2000we} 
  H.~X.~He, F.~C.~Khanna and Y.~Takahashi,
  Phys.\ Lett.\ B {\bf 480}, 222 (2000).

\bibitem{Pennington:2005mw} 
  M.~R.~Pennington and R.~Williams,
  J.\ Phys.\ G {\bf 32}, 2219 (2006).



\bibitem{He:2006my} 
  H.~X.~He,
  Commun.\ Theor.\ Phys.\  {\bf 46}, 109 (2006).


\bibitem{Qin:2013mta} 
  S.~X.~Qin, L.~Chang, Y.~X.~Liu, C.~D.~Roberts and S.~M.~Schmidt,
  Phys.\ Lett.\ B {\bf 722}, 384 (2013).



\bibitem{Binosi:2016wcx} 
  D.~Binosi, L.~Chang, J.~Papavassiliou, S.~X.~Qin and C.~D.~Roberts,
  arXiv:1609.02568 [nucl-th].


\bibitem{Kizilersu:1995iz} 
  A.~Kizilersu, M.~Reenders and M.~R.~Pennington,
  Phys.\ Rev.\ D {\bf 52}, 1242 (1995).


\bibitem{Aguilar:2011ux} 
  A.~C.~Aguilar, D.~Binosi and J.~Papavassiliou,
  Phys.\ Rev.\ D {\bf 84}, 085026 (2011).
    
  
  
\bibitem{Aguilar:2013xqa} 
  A.~C.~Aguilar, D.~Iba\~nez and J.~Papavassiliou,
  Phys.\ Rev.\ D {\bf 87}, no. 11, 114020 (2013).

\bibitem{Aguilar:2012rz} 
  A.~C.~Aguilar, D.~Binosi and J.~Papavassiliou,
  Phys.\ Rev.\ D {\bf 86}, 014032 (2012).


\bibitem{Cornwall:1981zr} 
  J.~M.~Cornwall,
  Phys.\ Rev.\ D {\bf 26}, 1453 (1982). 
  


	
\bibitem{Aguilar:2008xm} 
  A.~C.~Aguilar, D.~Binosi and J.~Papavassiliou,
  Phys.\ Rev.\ D {\bf 78}, 025010 (2008).
	

\bibitem{Aguilar:2009nf} 
  A.~C.~Aguilar, D.~Binosi, J.~Papavassiliou and J.~Rodriguez-Quintero,
  Phys.\ Rev.\ D {\bf 80}, 085018 (2009).
 
 
\bibitem{Aguilar:2016lbe} 
  A.~C.~Aguilar, J.~C.~Cardona, M.~N.~Ferreira and J.~Papavassiliou,
  arXiv:1610.06158 [hep-ph]. 
 
 
\bibitem{Sternbeck:2005tk} 
  A.~Sternbeck, E.-M.~Ilgenfritz, M.~Muller-Preussker and A.~Schiller,
  Phys.\ Rev.\ D {\bf 72}, 014507 (2005). 

\bibitem{Nachtmann:1981zg} 
  O.~Nachtmann and W.~Wetzel,
  Nucl.\ Phys.\ B {\bf 187}, 333 (1981). 

\bibitem{Binosi:2014aea} 
  D.~Binosi, L.~Chang, J.~Papavassiliou and C.~D.~Roberts,
  Phys.\ Lett.\ B {\bf 742}, 183 (2015).


\bibitem{Grassi:1999tp} 
  P.~A.~Grassi, T.~Hurth and M.~Steinhauser,
  Annals Phys.\  {\bf 288}, 197 (2001).
 
 
\bibitem{Binosi:2002ez} 
  D.~Binosi and J.~Papavassiliou,
  Phys.\ Rev.\ D {\bf 66}, 025024 (2002).



  
\bibitem{Binosi:2009qm} 
  D.~Binosi and J.~Papavassiliou,
  Phys.\ Rept.\  {\bf 479}, 1 (2009).




\bibitem{Aguilar:2010gm} 
  A.~C.~Aguilar, D.~Binosi and J.~Papavassiliou,
  JHEP {\bf 1007}, 002 (2010).
 
 
 

 
 
 
\bibitem{Qin:2011dd} 
  S.~x.~Qin, L.~Chang, Y.~x.~Liu, C.~D.~Roberts and D.~J.~Wilson,
  Phys.\ Rev.\ C {\bf 84}, 042202 (2011).
  
  
\bibitem{Boucaud:2008gn} 
  P.~Boucaud, F.~De Soto, J.~P.~Leroy, A.~Le Yaouanc, J.~Micheli, O.~Pene and J.~Rodriguez-Quintero,
  Phys.\ Rev.\ D {\bf 79}, 014508 (2009).

  
\end{thebibliography}
\end{document}